\documentclass[twocolumn]{aastex631}
\usepackage[english]{babel}
\usepackage{appendix}

\def\asca{{\sl ASCA }}

\def\ros{{\sl ROSAT }}

\def\chandra{{\sl Chandra }}
\def\b1626{4U~1626-67~}
\def\tet01oric{\theta^1~Ori~C}
\def\tet01orib{\theta^1~Ori~B}
\def\tet01oria{\theta^1~Ori~A}
\def\tet01orie{\theta^1~Ori~E}
\def\tet01orid{\theta^1~Ori~D}
\def\tet02oria{\theta^2~Ori~A}
\def\tet02orib{\theta^2~Ori~B}

\def\ergcm{\hbox{erg cm$^{-2}$ s$^{-1}$ }}

\def\kms{\hbox{km s$^{-1}$}}

\def\Msun{$M_{\odot}$ }

\def\it{\sl}

\usepackage{amsmath}

\graphicspath{{./}{figures/}}

\shorttitle{X-Rays From the ONC}
\shortauthors{Schulz et al.}

\usepackage{xspace}
\newcounter{xion}     \newcommand{\eli}[2]  {\setcounter{xion}{#2}#1{~\sc\roman{xion}}}

\newcommand{\toneE}   {{$\theta^1\,$Ori~E}\xspace}
\newcommand{\toneC}   {{$\theta^1\,$Ori~C}\xspace}
\newcommand{\chan}    {{\it Chandra}\xspace}
\newcommand{\ks}      {\,\mathrm{ks}}
\newcommand{\mang}    {\,\mathrm{{\mbox{\AA}}}\xspace}

\begin{document}

\title{THE NATURE OF X-RAYS FROM YOUNG STELLAR OBJECTS IN THE ORION NEBULA CLUSTER
       - A Chandra HETGS Legacy Project}

\author{Norbert S. Schulz}
\affiliation{Department of Physics and Kavli Institute for Astrophysics and Space Research\\
Massachusetts Institute of Technology\\
Cambridge, MA 02139, USA}

\author{David P. Huenemoerder}
\affiliation{Department of Physics and Kavli Institute for Astrophysics and Space Research\\
Massachusetts Institute of Technology\\
Cambridge, MA 02139, USA}

\author{David A. Principe}
\affiliation{Department of Physics and Kavli Institute for Astrophysics and Space Research\\
Massachusetts Institute of Technology\\
Cambridge, MA 02139, USA}

\author{Marc Gagne}
\affiliation{West Chester University\\
West Chester, PA 19383, USA}

\author[0000-0003-4243-2840]{Hans Moritz G\"unther}
\affiliation{Department of Physics and Kavli Institute for Astrophysics and Space Research\\
Massachusetts Institute of Technology\\
Cambridge, MA 02139, USA}

\author[0000-0002-3138-8250]{Joel Kastner}
\affiliation{Center for Imaging Science, School of Physics \& Astronomy, and Laboratory for Multiwavelength Astrophysics, Rochester Institute of Technology, 54 Lomb Memorial Drive, Rochester, NY 14623, USA}

\author{Joy Nichols}
\affiliation{Harvard \& Smithsonian Center for Astrophysics\\
Cambridge, MA 02138, USA}

\author{Andrew Pollock}
\affiliation{Department of Physics and Astronomy\\
University of Sheffield
Sheffield S10 2TN, United Kingdom}

\author[0000-0003-3130-7796]{Thomas Preibisch}
\affiliation{Universit\"ats-Sternwarte M\"unchen,
Ludwig-Maximilians-Universität, \\ 
81679 M\"unchen, Germany}

\author{Paola Testa}
\affiliation{Harvard \& Smithsonian Center for Astrophysics\\
Cambridge, MA 02138, USA}

\author{Fabio Reale}
\affiliation{University of Palermo\\
90133 Palermo, Italy}

\author{Fabio Favata}
\affiliation{ESA European Space Research and Technology Centre (ESTEC)\\
Keplerlaan 1, 2201 AZ Noordwijk, The Netherlands}
\affiliation{INAF - Osservatorio Astronomico di Palermo\\
Piazza del Parlamento 1, 90134 Palermo, Italy}

%\author[0000-0002-7584-4756]{Jun Yang}
%\affiliation{Department of Physics and Kavli Institute for Astrophysics and Space Research\\
%Massachusetts Institute of Technology\\
%Cambridge, MA 02139, USA}

\author{Claude R. Canizares}
\affiliation{Department of Physics and Kavli Institute for Astrophysics and Space Research\\
Massachusetts Institute of Technology\\
Cambridge, MA 02139, USA}

\begin{abstract}
The Orion Nebula Cluster (ONC) is the closest site of very young ($\sim$ 1 Myrs) massive star formation
The ONC hosts more than 1600 young and X-ray bright stars with masses ranging from $\sim$ 0.1 
to 35 \Msun. The \chandra HETGS Orion Legacy Project observed the ONC 
with the \chandra high energy transmission grating spectrometer (HETGS) for $2.1\,$Ms. We describe the spectral extraction and
cleaning processes necessary to separate overlapping spectra. We obtained 
36 high resolution spectra which includes a high brilliance X-ray spectrum of $\theta^1$ Ori C with over 
100 highly significant X-ray lines. The lines show Doppler broadening between 300 and $400~\kms$.
Higher spectral diffraction orders allow us to resolve line components of high Z He-like triplets in $\theta^1$ Ori C with unprecedented spectral
resolution. Long term light curves spanning $\sim$20 years show all stars to be highly variable, including the massive stars.
Spectral fitting with thermal coronal emission line models reveals that
most sources show column densities of up to a few times $10^{22}\,$cm$^{-2}$ and high coronal temperatures
of 10 to 90 MK. We observe a bifurcation of the high temperature component where some stars show
a high component of 40 MK, while others show above 60 MK indicating heavy flaring activity. Some lines are resolved
with Doppler broadening above our threshold of $\sim200~\kms$, up to $500~\kms$.
This data set represents the largest collection of HETGS high resolution X-ray spectra from young 
pre-MS stars in a single star-forming region to date.
\end{abstract}

\section{Introduction}

The Orion Nebula Cluster (ONC) is a very young star forming region hosting a large number
of young stellar objects in terms of mass, age, and evolutionary stages.
The cluster is part of the Orion A molecular cloud hosting a hierarchical structure
of ongoing star formation cells~\citep{bally2000}. The part of this region we generally refer
to as the ONC is a somewhat older formation bubble located at the foreground of the main molecular cloud.
Two very massive stars - $\theta^1$ Ori C and $\theta^2$ Ori A - are members of the Orion Trapezium Cluster at the 
core of the ONC with $\theta^1$ Ori C being the main source of illumination and ionization of the Orion Nebula (M42).
The ONC also hosts a large assembly of young stars with about 80$\%$ of
its members being younger than a few Myrs. With over 3000 stars in the 
vicinity of the Orion Trapezium the average stellar density amounts to about
250 stars per pc$^{3}$ within a radius of about 3 pc \citep{hillenbrand1997}. The ONC is the nearest site of
massive star formation rich in a low- and intermediate mass pre-main sequence (PMS)
stellar population as well as early-type zero-age main sequence (ZAMS) stars. It is well
studied in the optical and infra-red bands with about 1600 sources classified to some limited 
extent through spectroscopic and photometric measurements \citep{hillenbrand1997, hillenbrand2013} and over 2000 stars 
being observed in the IR band with 2MASS \citep{skrutskie2006} and ground based surveys ~\citep{muench2002, robberto2010, manara2012}.

The ONC also has a long history of X-ray observations. From its first discovery with 
\emph{Uhuru} \citep{giacconi1972} identified as a bright X-ray source 3U0527-05 to the 
realization that this is a more extended emission region containing X-rays from stellar coronae
around young T Tauri stars \citep{denboggende1978, feigelson1981, gagne1995}, decades 
of observations established the ONC as one of the richest X-ray emitting star forming clusters.
However, while most of these studies were severely limited by low angular resolution of 
their satellite telescopes, \ros in the 1990's came in best with 5 arcsec, a true breakthrough
came with the launch of \chandra in 1999 which then offered an angular resolution of 0.5 to 2 arcseconds
over a few arcmin field of view. The \chandra Orion Ultradeep Project (COUP, \citealt{feigelson2005}) took 
full advantage of this superb observing capability and observed the ONC for nearly 10 days total
to detect 1616 X-ray sources, measure column densities,
source fluxes, and basic X-ray spectral and photometric parameters \citep{getman2005}. Many
X-ray surveys of other young stellar clusters were performed with \chandra, examples are RCW38 \citep{wolk2006},  
30 Doradus \citep{townsley2006}, NGC 6357
\citep{wang2007}, M17 \citep{broos2007}, NGC 2244 \citep{Wang2008} or recently in the 
Tarantula Nebula \citep{crowther2022}. Perhaps the
most notable survey is the large Chandra Carina Complex Project, which detected over 14\,000 X-ray sources,
with a large number of multi-wavelength counterparts \citep{townsley2011, broos2011, gagne2011, feigelson2011, preibisch2011}.

Young, low-mass (0.1 \Msun to about 2 \Msun{}) pre-main sequence (PMS) stars are brighter in X-rays than their more 
evolved counterparts on the main sequence. The ratio of X-ray to bolometric luminosity in these stars lies between
10$^{-4}$ and 10$^{-3}$, close to the saturation threshold~\citep{vilhu1984,vilhu1987,Wright2011}. Besides coronal activity, 
accretion and outflows can also contribute X-ray flux for those stars still surrounded by a proto-planetary 
disk \citep[for a review, see][]{schneider2022}. Those stars are called classical T~Tauri stars (CTTS). X-rays from shocks in outflows are very soft and orders of magnitude fainter than 
coronal emission \citep{guedel2011}; they can generally only be seen in near-by stars with little absorption 
where the jet is spatially resolved. One of the first detections of soft X-rays from shocks at the base of an outflow was an 
Orion proplyd using the COUP dataset \citep{kastner2005}. 
Another source of X-rays is the accretion shock itself. The disk does not reach down to the star, but instead mass falls onto the 
stellar surface along the magnetic field lines. It is accelerated to free-fall velocities and forms a strong shock at the stellar 
surface. This shock heats the infalling gas to X-ray emitting temperatures \citep{lamzin1998,guenther2007,hartmann2016}. The density in the 
shock is high enough that it alters the line ratios in the He-like triplets, which are resolved in high-resolution X-ray grating 
spectroscopy \citep[e.g.][]{kastner2002, kastner2004,testa2004,schmitt2005,guenther2006,argiroffi2007,brickhouse2010,argiroffi2012}. 
However, it is not clear if it is actually the shock itself that is observed \citep{reale2013,reale2014}, or if the depth of the shock in the photosphere 
and the outer layers of an inhomogenous accretion column hide the shock from view \citep{sacco2010,schneider2018,espaillat2021}, 
and the observed line-ratios would be a secondary effect, formed where cooler and denser plasma flows up into the corona as seen in
simulations \citep{orlando2010,orlando2013}.

Older weak-lined T Tauri stars (WTTS) do not show accretion and thus have coronal line ratios in their 
He-like triplets, e.g., in the WTTS \object{HID 98890} \citep[e.g.,][]{kastner2004}. \citet{telleschi2007} also
showed that many CTTS have hard spectra with substantial emissions up to 10~keV, far beyond the reach of accretion shock heated plasma.
Yet, in the accretion phase the stars accrete not only mass, but also angular momentum; young stars, CTTS and WTTS, thus rotate faster 
than their older main-sequence counterparts, which explains the saturated level of coronal activity.
This fact is often used to identify young stars in a dense field, e.g., \citet{pillitteri2013} use X-ray observations 
in the Orion A cloud south of the ONC to find young, but disk-less cluster members.

Performing high spectral resolution X-ray studies of very young stellar clusters is challenging. The \chandra High Energy 
Resolution Transmission Grating Spectrometer (HETGS) disperses the image of a point source across the field of view 
\citep[see][]{canizares2000}. 
This works well for isolated objects, but is susceptible to confusion from intersecting and overlapping spectra in crowded fields, 
such as young stellar associations. 
HETGS spectra of the close by TW Hydra association were easy to obtain because the member stars are sufficiently well 
separated in individual pointings \citep{kastner2002, kastner2004, huenemoerder2007}. Stars of the Cygnus OB2 association fit 
into one single pointing, but they are still sufficiently well separated to prevent serious confusion \citep{waldron2004}.

 The ONC is the nearest massive star forming cluster
at a distance of $\simeq 400$~pc \citep{menten2007,kounke2017,kuhn2019,maizapellaniz2022}. Its brightest sources were 
a focus early in the \chandra mission,
involving $\theta^1$ Ori A, C and E \citep{schulz2003, gagne2005, huenemoerder2009}, and $\theta^2$ Ori A 
\citep{schulz2006, mitschang2011}. \cite{schulz2015} used an early set of \chandra HETG observations to 
study 6 bright PMS stars in the near environment of the Orion Trapezium at the core of the ONC.
Here significant confusion between overlapping spectra was encountered.
That study specified the limitations of high angular resolution as offered by the \chandra optics
and dispersive high resolution spectroscopy offered by the HETGS. In the ONC field of view the
closest separation within bright sources is between 5 to 8 arcsec which appeared to make
a deep high resolution study feasible. However, it also indicated that even though the angular resolution 
of \chandra is 0.5 arcsec, dispersive studies of PMS stars separated by less then 3-5 arcsec are not feasible. 
The study by \citet{huenemoerder2007} of Hen 3-600 shows this limitation well for a 1.5 arscec binary.
This excludes all clusters more distant than the ONC.

In this pilot paper we describe our observation of the ONC
with the \chandra HETGS in order to obtain more than 3 dozen high resolution
X-ray grating spectra of ONC member stars. The data described in this pilot paper
are made public and we anticipate several science publications to follow by the authors and
the science community. We present observations, spectral
confusion cleaning procedures, a set of final spectra bearing a total number of
counts and exposure time after spectral cleaning  and a first in depth analysis
of X-ray properties of massive, intermediate mass stars and low-mass PMS stars in the ONC 
for which we have sufficient spectral data. Any follow-up paper then should refer to
this pilot paper and the official data release site for a full description of the data.

\section{Observations and Data Reduction}

\subsection{The \chandra HETGS}
The \chandra HETG assembly consists of an array of periodic gold microstructures that can be interposed in the converging X-ray beam 
just behind the \chandra High Resolution Mirror Assembly. When the telescope observes a point source with the gratings in place, a 
fraction of the X-rays are dispersed, according to wavelength, to either side of the point source zeroth-order image. The zeroth 
order image and the dispersed +/- first and less prominent higher orders are detected at the focal plane by the linear array
of CCD detectors, ACIS-S. Thus the whole system of mirror, gratings and detector constitute a slitless spectrometer, the HETGS
\citep{canizares2000}.
The HETG assembly has two different grating types, designated MEG and HEG, optimized for medium and high energies, respectively. 
The gratings are mounted so that the dispersed +/- spectra of the MEG and HEG are offset from one another by an angle of 10 degrees, 
forming a shallow "X" in the focal plane with the zeroth order image at its center (Fig. \ref{fig:confusion}).

The HETGS provides spectral resolving powers of $\lambda/ \Delta \lambda = 100-1000$ in its first orders for point sources, 
corresponding to a line FWHM of about 0.02\AA\ for MEG and 0.01\AA\ for HEG, and effective areas of 1-180 cm$^{2}$  
over the wavelength range of 1.2-30\AA\ (0.4-10~keV). Multiple overlapping orders are separated using the 
moderate-energy resolution of ACIS-S.

\begin{figure*}[t]
\centering
\includegraphics[angle=0,width=16.2cm]{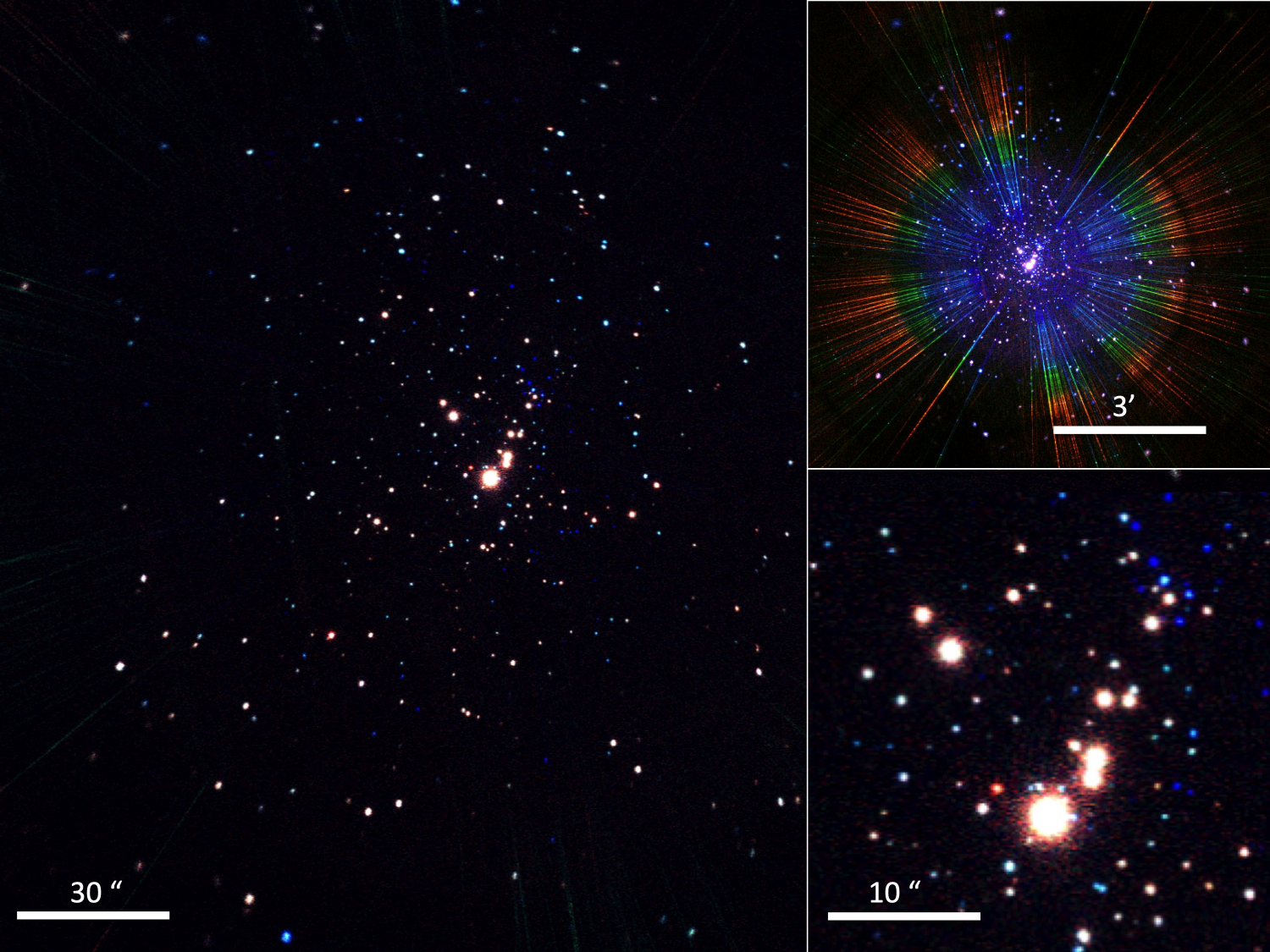}
\caption{Merged zero order image over the entire exposure using a three color (rgb) scheme reflecting the stars energy spectra. 
The main image is shown with a 30 arcsec scale covering about 60\% of the entire captured ACIS-S field of view. The dispersive
HETG 1st and higher order dispersion events of the brightest star $\theta^1$ Ori C were removed. 
The top right inset shows a wider view for 3 armin with all dispersion streaks included. The most prominent one are from $\theta^1$ Ori C.
The bottom right inset shows a zoomed version of the Orion Trapezium region, which includes about 10 of the brightest
stars in the region and for which we have most significant HETG 1st order spectra. 
%A click on the merged image shows a
%movie of how the stars vary in flux with time (on-line version only).
}
\label{fig:hetgcluster}
\end{figure*}

\subsection{HETGS Observations}

The data contains a set of 70 observations of the ONC with the HETG aimed at the central star 
of the Orion Trapezium $\theta^1$ Ori C, obtained by the Chandra X-ray Observatory, contained in the Chandra Data Collection (CDC) 192~\dataset[doi:10.25574/cdc.192]{https://doi.org/10.25574/cdc.192}.
The total amount of the exposure is 2,086.14 ks taken over
a period of about 20 years. The top right inset of Fig~\ref{fig:hetgcluster} shows the merged image of all
observations
over the most effective field of view summed over all roll angles. Nearly all visible dispersive HETG streaks are due 
the three brightest sources in the field, $\theta^1$ Ori C, $\theta^1$ Ori E, and MT Ori.
The observations are divided into two observation periods; one taken over six years after 
the launch of \chandra in 1999 up to the year 2007 amounting to 470.96 ks and a second one  
during the years 2019 and 2020 amounting to a total of 1615.18 ks all summarized in Tab.\ref{tab:obs} . 
%The latter suite is separated into two 
%periods before (Tab.\ref{tab:obs2}) and after sun block (Tab.\ref{tab:obs3}) of the Orion region.

The first period observations used
the full array of ACIS-S charge-coupled devices (CCDs). This means for these data full access of the \chandra wavelength
band is available from 1.70 \AA\ to 30 \AA. These observations also provide the bulk of X-rays above 16 \AA\ due to
progressing ACIS contamination at later stages in the \chandra mission.

\startlongtable
\begin{deluxetable}{rccccc}\label{tab:obs}
\tablecaption{CHANDRA HETGS Observations}
  \tablehead{
  \colhead{Obsid}&
	 \colhead{Exp.}&
	 \colhead{Date} &
  \colhead{Time} &
  \colhead{CCDs} &
  \colhead{MJD} 
  \\
  &
  \colhead{ [ks]} &
  \colhead{ [UT]} &
  \colhead{ [UT]} &
  &
  \colhead{ [d]} 
  \\
  }
  \startdata
       3  & 49.62  & 1999-10-31 & 05:47:21 & 6 & 51482.2 \\ 
       4  & 30.92  & 1999-11-24 & 05:37:54 & 6 & 51506.2 \\
    2567  & 46.36  & 2001-12-28 & 12:25:56 & 6 & 52271.5 \\
    2568  & 46.34  & 2002-01-19 & 20:29:42 & 6 & 52324.9 \\
    7407  & 24.64  & 2006-12-03 & 19:07:48 & 6 & 54072.8 \\
    7408  & 24.98  & 2006-12-19 & 14:17:30 & 6 & 54075.5 \\
    7409  & 27.09  & 2006-12-23 & 00:47:40 & 6 & 54088.6 \\
    7410  & 13.10  & 2006-12-06 & 12:11:37 & 6 & 54092.0 \\
    7411  & 24.64  & 2007-07-27 & 20:41:22 & 6 & 54308.9 \\
    7412  & 25.20  & 2007-07-28 & 06:16:09 & 6 & 54309.3 \\
    8568  & 36.08  & 2007-08-06 & 06:54:08 & 6 & 54318.3 \\
    8589  & 50.71  & 2007-08-08 & 21:30:35 & 6 & 54320.9 \\
    8895  & 24.97  & 2007-12-07 & 03:14:07 & 6 & 54419.4 \\
    8896  & 22.66  & 2007-11-30 & 21:58:31 & 6 & 54434.8 \\
    8897  & 23.65  & 2007-11-15 & 10:03:16 & 6 & 54441.1 \\
   23008 &  47.43   &   2019-11-27  & 12:07:33  & 4 & 58814.5\\ 
   22893 &  24.73   &   2019-12-02  & 17:18:23  & 5 & 58819.7\\
   22994 &  24.73   &   2019-12-05  & 09:22:57  & 4 & 58822.4\\
   23087 &  39.54   &   2019-12-08  & 16:56:56  & 4 & 58825.7\\
   22904 &  36.58   &   2019-12-10  & 17:49:59  & 4 & 58827.7\\
   23097 &  35.88   &   2019-12-11  & 12:12:24  & 4 & 58828.5\\
   22337 &  37.66   &   2019-12-13  & 04:25:33  & 4 & 58830.2\\
   23006 &  24.73   &   2019-12-14  & 06:35:20  & 5 & 58831.3\\
   22343 &  24.73   &   2019-12-15  & 20:04:15  & 4 & 58832.8\\
   23003 &  24.74   &   2019-12-21  & 05:12:39  & 4 & 58838.2\\
   23104 &  24.73   &   2019-12-21  & 21:47:04  & 5 & 58838.9\\
   22336 &  25.59   &   2019-12-22  & 11:01:50  & 4 & 58839.5\\ 
   23007 &  37.41   &   2019-12-24  & 23:12:06  & 4 & 58842.0\\
   22339 &  31.64   &   2019-12-26  & 02:06:17  & 4 & 58843,1\\
   22892 &  30.66   &   2019-12-26  & 22:46:53  & 4 & 58843.9\\
   22995 &  38.74   &   2019-12-27  & 14:29:16  & 4 & 58844.6\\
   22338 &  39.15   &   2019-12-30  & 06:02:12  & 4 & 58847.3\\
   22334 &  24.73   &   2019-12-31  & 09:17:51  & 4 & 58849.3\\
   23000 &  42.50   &   2020-01-01  & 07:04:24  & 4 & 58851.7\\
   22996 &  26.70   &   2020-01-03  & 00:38:17  & 5 & 58852.4\\
   23114 &  37.56   &   2020-01-03  & 16:46:28  & 4 & 58855.0\\
   23115 &  29.67   &   2020-01-04  & 10:02:01  & 4 & 58856.5\\
   22335 &  29.67   &   2020-01-06  & 23:19:34  & 4 & 58859.55\\
   23005 &  24.73   &   2020-01-08  & 10:14:19  & 5 & 58941.05\\
   23120 &  39.54   &   2020-01-11  & 12:12:26  & 4 & 58941.6\\
   23012 &  10.81   &   2020-04-01  & 23:58:25  & 6 & 58943.9\\
   23206 &  17.71   &   2020-04-02  & 13:57:51  & 6 & 58944.4\\
   23207 &  14.75   &   2020-04-04  & 12:21:33  & 6 & 58948.4\\
   23208 &  14.75   &   2020-04-05  & 08:54:30  & 6 & 58951.0\\
   23011 &  51.69   &   2020-04-21  & 18:33:18  & 5 & 58960.8\\
   22341 &  32.12   &   2020-04-29  & 09:07:29  & 5 & 58968.4\\    
   23233 &  34.59   &   2020-05-01  & 13:36:01  & 5 & 58970.6\\
   23010 &  25.72   &   2020-07-27  & 11:07:30  & 4 & 59057.5\\
   23001 &  25.62   &   2020-07-28  & 05:42:17  & 6 & 59058.2\\  
   23009 &  25.01   &   2020-07-28  & 23:57:16  & 6 & 59059.0\\
   22340 &  25.62   &   2020-10-14  & 17:14:19  & 6 & 59136.7\\   
   24832 &  27.59   &   2020-10-15  & 05:57:17  & 6 & 59137.3\\
   22997 &  26.60   &   2020-10-15  & 18:40:16  & 6 & 59137.8\\
   24834 &  26.91   &   2020-10-16  & 07:35:44  & 5 & 59138.3\\
   22342 &  34.50   &   2020-10-20  & 03:07:56  & 6 & 59142.1\\  
   24842 &  29.57   &   2020-10-21  & 01:56:20  & 6 & 59143.1\\
   22993 &  24.63   &   2020-10-23  & 05:55:39  & 6 & 59145.3\\
   22998 &  23.09   &   2020-11-01  & 04:40:26  & 5 & 59154.2\\
   22999 &  35.58   &   2020-11-08  & 07:34:20  & 5 & 59161.3\\
   24830 &  26.52   &   2020-11-22  & 07:56:09  & 4 & 59175.3\\
   24622 &  24.56   &   2020-11-23  & 09:17:18  & 4 & 59176.4\\
   24873 &  24.74   &   2020-11-24  & 03:01:16  & 4 & 59177.1\\
   24874 &  25.72   &   2020-11-24  & 15:24:52  & 4 & 59177.6\\
   24829 &  26.46   &   2020-11-27  & 14:58:09  & 4 & 59180.6\\ 
   24623 &  24.74   &   2020-11-29  & 13:30:17  & 4 & 59182.6\\
   24624 &  29.67   &   2020-12-09  & 22:23:51  & 4 & 59192.9\\
   23002 &  30.66   &   2020-12-10  & 13:20:39  & 4 & 59193.6\\
   23004 &  32.14   &   2020-12-12  & 01:34:32  & 4 & 59195.1\\    
   24831 &  30.66   &   2020-12-25  & 05:12:20  & 4 & 59208.2\\
   24906 &  28.60   &   2020-12-25  & 21:09:10  & 4 & 59208.9\\            %
\enddata
\end{deluxetable}
%\endlongtable

The second period of observations happened about 13 years later, after the observing conditions
of the satellite had changed. Progressing contamination of the focal plane CCD array optical blocking filter 
effectively blocks soft X-rays below 1 keV ($>12.3485\,$\AA). In addition, thermal constraints
due to deteriorating thermal protection of the spacecraft requires reducing the number of CCD devices activated during 
observations. 
%We added a column in Tab.~\ref{tab:obs2} and Tab.~\ref{tab:obs3} listing the number of 
%CCD devices active during the observation. 
With 6 CCDs we have the full wavelength band available;
with 5 CCDs this still holds, but we lose some exposure above about 24 \AA; with 4 CCDs we lose exposure
above about 18 \AA. This is not an additional limitation, however,  as the progressive ACIS filter contamination 
blocks most of the exposure above 16~\AA{} anyway.

\subsection{Spectral Extraction}

For most data preparation and spectral analysis we used the 
Interactive Spectral Interpretation System (ISIS)~\citep{houck2000}.
To uniformly process the many observations each with multiple objects of
interest in a crowded field, we modified 
the standard procedures of the CIAO software \citep{fruscione2006}. Events
were rerun through standard event processing to update bad pixel maps
and to ``destreak'' bad events on CCD\_ID 8 (ACIS-S4). We then reran
\emph{acis\_process\_events} to re-create a Level 1 event file
identical to what is done in standard processing. Since we have many
observations with an ensemble of sources of interest in a crowded field, we
matched and updated the world-coordinate-system (WCS).
This is so that we can run source spectral extractions using {\it a
priori} source celestial coordinates from COUP \citep{getman2005}.
This avoids small position uncertainties in zeroth order detection due 
to low exposure or confusion by dispersed spectra. We then simply skip the 
detection step and map the celestial coordinates to sky
pixel for each observation using the WCS. In order to provide the WCS
registration, we ran a CIAO source detection program, \emph{wavdetect}, on the central region over an
8 arcmin radius for several spatial scales. For that we used a PSF-map 
which we created using \emph{mkpsfmap} at 2.3 keV for an enclosed counts fraction of 0.9.
We then applied \emph{wcs\_match} to fit the
rotation and translation of the coordinate system of each ObsID relative
to COUP, and updated all Level 1 event files and corresponding aspect
solution files with these solutions. Spectral extraction then followed the usual CIAO steps
but with narrower than default cross-dispersion extraction regions 
($2\,$ arcsec full width instead of the default $4$) 
to reduce the overlap of crossing HEG or MEG orders from different sources. 
This does not change the overall spectral extraction process, but reduces
the ambiguity about from which source an event originates in the extraction mask.
The aperture efficiency is reduced a bit, by about $5\%$ at $6\mang$, and by 
about $8\%$ at $12\mang$, but this was seen to significantly reduce the number of contaminated sources at a small loss of
signal.

Responses were made in the usual way for each source extraction, via
the CIAO commands \emph{mkgrmf}  and \emph{mkgarf}. While ARFs depend 
critically on source position and observation details (such as the aspect history), 
RMFs do not. The RMFs depend on the spectral extraction region width which we chose to be the same for all 
sources and observations. Thus, there are only four unique RMFs for
HEG and MEG $\pm$1 orders for all sources.

\subsection{Confusion Analysis}\label{sec:confusion} 

The region of the sky observed by the HETGS includes more than 1000 known X-ray sources (Fig~\ref{fig:hetgcluster}) and the majority of 
these are present in the field of view of individual epochs. The HETGS instrument disperses light from each X-ray source in a 
characteristic, shallow 'X' shape on the ACIS-S detectors\footnote{\url{https://cxc.harvard.edu/proposer/POG/}}. 
The non-dispersed (0th order) events are located at the right ascension (RA) and declination (DEC) of the X-ray source in the sky.  
The first, second and third order events for each source are dispersed by an angle given by the dispersion equation. The orders 
overlap along a line, one pair for the +/- HEG and one for the +/- MEG. While every X-ray source in an HETGS field of view 
has its light dispersed 
in the characteristic X-shaped pattern, only those sources that are sufficiently bright will disperse enough events to yield meaningful spectra.

\begin{figure*}[t]
\centering
\includegraphics[angle=0,width=18cm]{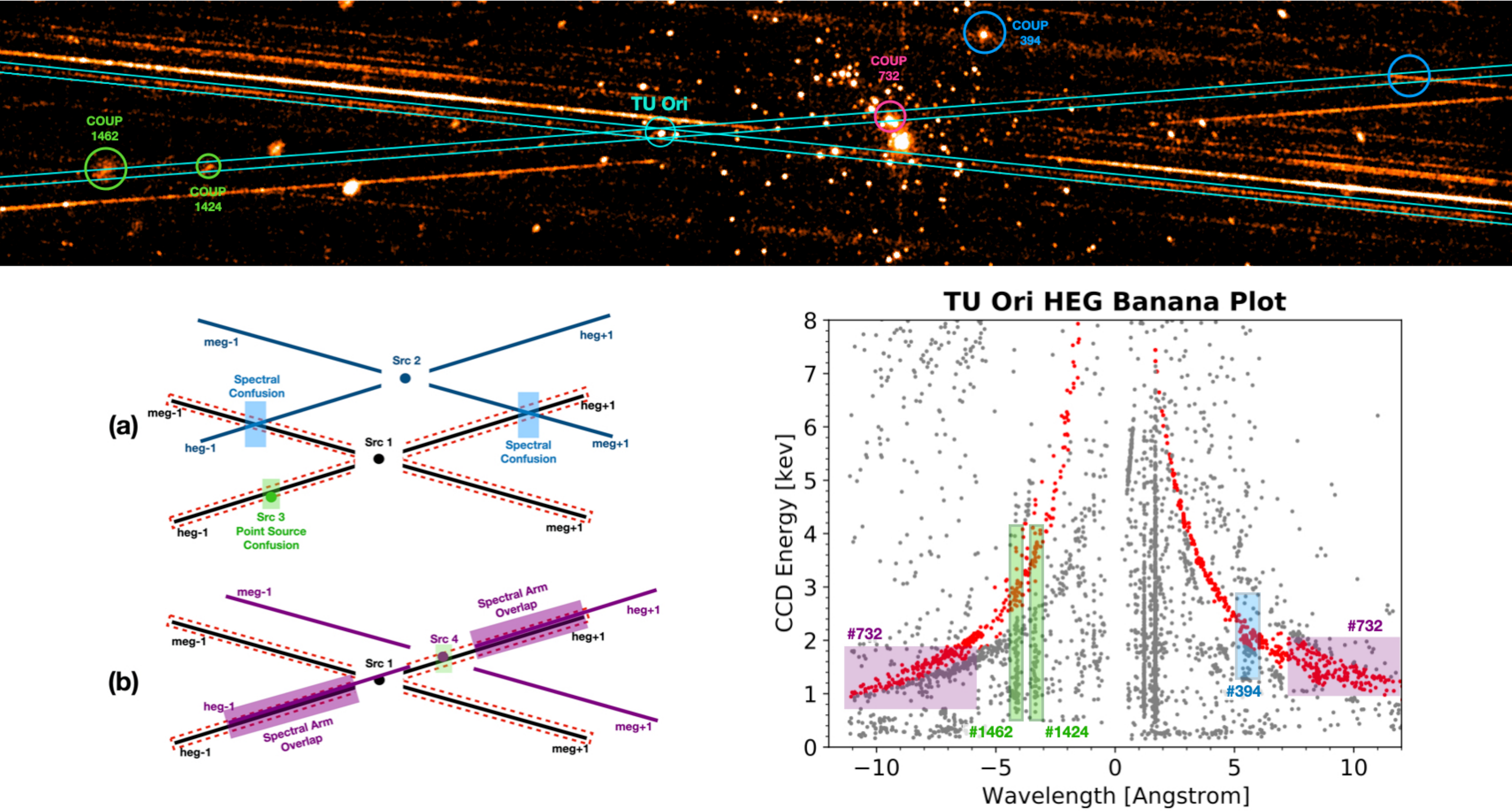}
\caption{Top: An example HETG observation (obsid 3) demonstrating the need to account for confusion 
when extracting spectra in the ONC dataset. An example dispersed spectrum of TU Ori is 
displayed (cyan rectangle) with sources of confusion highlighted 
with circles (green-point source confusion, blue- spectral confusion, magenta-spectral arm overlap). 
Left: An illustration (not to scale) demonstrating (a) point source (green) and spectral (blue) 
confusion and (b) spectral arm confusion (magenta). The black X labeled Src 1 corresponds to the source 
intended for spectral extraction with the 
red dashed box corresponding to dispersed events. Specific locations in the extracted spectra where confusion can occur are identified with 
colored boxes. Right: ACIS order sorting banana plot showing confused events from different sources in the field erroneously being assigned 
to the spectra of TU Ori. Red dots indicate events that standard CIAO processing assigns to the extracted 
source (TU Ori) while other events, whose ccd-resolved energy do not match the expected wavelength of TU Ori, are 
not included in the standard CIAO source extraction. Colored numbers represent the COUP number of the source causing confusion for this case. Examples where standard CIAO processing has the potential to erroneously 
include events from other sources in the extracted spectrum of TU Ori are shown as red dots within the colored boxes.  }
\label{fig:confusion}
\end{figure*}

HETGS observations of crowded fields, where multiple bright point sources cast their X-shaped patterns on the CCDs, suffer 
from event confusion, a scenario where events from two (or more) astrophysical sources could arrive at the same location on 
the detector and be erroneously assigned with standard CIAO processing (Fig. \ref{fig:confusion}, Top). The relative 
locations of the dispersed spectra for each source depend on 
the roll angle of the observation. Dispersed spectra roll with the spacecraft, but zeroth order sky positions do not. Hence, the 
relative positions of spectra change with roll and every epoch in the ONC HETGS dataset will have unique sources of confusion 
(Fig~\ref{fig:hetgcluster}). To identify and account for all the potential sources of confusion when extracting spectra, 
we created a custom Python program called \emph{CrissCross} which utilizes the fixed geometry of the X-shaped spectral dispersion 
region and the known location of X-ray sources in the field of view to produce un-confused spectra. 
While the details of \emph{CrissCross} will be published in a forthcoming paper (Principe et al.\ in prep), 
we summarize its utility here.

 In the ONC HETGS dataset, 
there are three primary causes of confusion when assigning events to a specific source for spectral extraction: (a) 0th order 
(non-dispersed) point sources falling on a extracted sources spectral arm, (b) dispersed events from one source intersecting 
the arm of an extracted source, and (c) a bright source whose 0th order lands on or near another sources spectral arm dispersing its 
events along the same location on the CCD (Fig.~\ref{fig:confusion}, left). 
The location where confusion occurs in the spectrum of an extracted source is straightforward to calculate
using the location on the CCD of the confuser and the well-calibrated energy to dispersion distance relation for HEG and MEG.

Standard CIAO processing already mitigates some portion of confusion by utilizing ACIS order sorting (Fig.~\ref{fig:confusion}, right). 
When events are assigned to 
a specific source during spectral extraction, the CCD-resolved event energy is compared to the expected energy of the event based 
on its dispersion distance (i.e., the distance from the 0th order in the dispersion direction). If these energies do not match, 
within an energy range based on the spectral energy resolution of the CCD, then events from a confusing source will automatically 
be rejected from the extracted spectrum, effectively removing confusion. However, in a region with a large number of X-ray sources 
like the ONC, there are often cases where the CCD-resolved energy of confusing events happens to match the expected energy of 
dispersed events during spectral extraction. In these cases ACIS order sorting will erroneously assign events from a confusing source to the 
extracted spectrum. Therefore, we use \emph{CrissCross} to identify scenarios where this confusion occurs so that 
we can account for this during spectral fitting. 

\emph{CrissCross} is run for each observation and ultimately identifies all three sources of confusion for every source of 
interest (e.g., Table \ref{tab:master}). In order to achieve this goal \emph{CrissCross} runs through multiple steps starting 
with building a source list of all detected point sources and an estimation of their brightness in terms of counts per observation.  
This is achieved with \emph{wavdetect} which identifies sources with a Mexican-Hat Wavelet source 
detection algorithm. However, \emph{wavdetect} is not designed to be run on grating observations where HETG dispersed events are 
often misidentified as point sources. Nevertheless, the \emph{wavdetect} tool still correctly identifies point sources and we cross match 
all \emph{wavdetect} sources to the list of known COUP sources (\citealt{feigelson2005}). If a detected source is within 3 arcseconds 
of a COUP source then it is recognized as a valid source. If more than one \emph{wavdetect} source is detected within 3 arcseconds 
of a known COUP source, the closest source is assigned to the COUP source. The majority of cluster members are near the center of
the field of view where 0th order events dominate (Fig~\ref{fig:hetgcluster}) and thus their detection is not 
affected by dispersed events.  Off-axis COUP point sources were also accurately matched. The location of 0th order point sources 
and the estimated number of counts for each source provided by \emph{wavdetect} is used to calculate the location of every
dispersed spectrum in each field of view.  All three primary causes of confusion are then identified for every source in 
Table \ref{tab:master}. The ONC HETGS observations were carried out with ACIS-S while the COUP project used ACIS-I. Since the ACIS-S array 
covers a larger area of the sky, there are 27 X-ray sources detected in the HETGS ONC observations that were outside of the field 
of view of the COUP. Regardless of whether or not these sources represent young stars in the ONC, we include these objects when 
considering spectral confusion of the bright HETGS sources. All of these X-ray sources have 2MASS counterparts.

Point source confusion occurs when a 0th order point source is detected on or near an HEG or MEG arm of an extracted source within 
some margin. Since the Chandra PSF increases  in radius as a function of distance off-axis (i.e., distance from the optical axis, or aimpoint), 
the margin used to initially determine whether a point source is a confuser also depends on off-axis angle. A point source located 
within 3 arcminutes of the aimpoint is initially  considered a potential confusing source if its centroid is located within 8 pixels 
($\sim$4 arcsec) of the dispersed arm in the cross dispersion direction (perpendicular to the arm on the CCD). 
If a source is considered confusing, the energy and number of events within the fraction of the PSF that overlaps with
the spectral arm of the extracted source is estimated. The number of 0th order counts in the same energy range for the source intended 
for spectral extraction is also determined. If the confusing source contributes more than 10\% of the counts in the specific energy 
range where confusion occurs then it is considered a genuine case of point source confusion. 

%This distance is adjusted for off-axis cases allowing larger cross-dispersion distances to be considered for potential confusers. 
%For off-axis cases between 3-6 arcmin and greater than 6 arcmin a cross-dispersion distance of 16 and 24 pixels, respectively are used.
%In all cases points sources are required to have at least 3 counts to be considered a potentially confusing source. --
%I'm not sure this is relevant enough to include

Spectral confusion occurs when the dispersed spectrum of a confusing source intersects with the dispersed spectrum of the source intended 
for spectral extraction.  In most cases, this type of event confusion is already removed with ACIS order sorting under standard CIAO processing.
However, if the location where the two spectra intersect corresponds to the same energy in both spectra (i.e., the confusing events are 
within the order sorting energy range of the extracted spectrum) then genuine confusion will occur and the confusing events could be erroneously 
assigned to the extracted source's spectrum. \emph{CrissCross} identifies these cases and determines the number of counts 
in both the confuser and 
extracted sources 0th orders in the same energy range. After accounting for the different efficiencies between HEG and MEG spectral arms, 
if the ratio of 0th order confuser counts to 0th order extracted counts is greater than 15\% it is considered a 
genuine source of spectral confusion.

%If the ratio of 0th order confuser counts to 0th order extracted counts is greater than 7.5\% for a case of HEG spectral confusion or 
%greater than 30\% for MEG spectral confusion, then the instance is considered genuine confusion.  The different ratios account for 
%different efficiencies for HEG and MEG (i.e., an MEG arm confusing an HEG arm will have 2x the counts).

The final primary cause of confusion in the ONC HETGS dataset comes from spectral arm confusion. Cases of spectral arm confusion occur 
when a bright 0th order point source (e.g., a source bright enough to disperse many events in the 1st order) falls on or near the 
spectral arm of a source intended for extraction. Identifying potential cases of spectral arm confusion begins by identifying 0th order
point sources with more than 50 counts that fall within a specific cross-dispersion distance of the intended source for spectral extraction. 
As is the case with point source confusion, we consider off-axis angle when determining an appropriate cross dispersion distance for 
potential confusion. A single on-axis source will have a cross-dispersion width of about $\sim$4 arcsec (8 pixels). As the PSF gets 
larger farther off-axis, the cross-dispersion distance used to identify confusing sources is increased based on the off-axis 
locations of both the confusing and the intended source for spectral extraction. 

Unlike other sources of confusion, spectral arm confusion has the potential to contaminate the entire HEG or MEG arm of the source intended 
for extraction. If the 0th order location of the two sources are close enough in the dispersion direction, many of the confusing spectral 
events can fall in an energy window that the extracted source is expecting (i.e., ACIS order sorting would erroneously assign events from 
the confused source to the extracted source). For every potential arm confusing case, \emph{CrissCross} uses the distance between two 
0th orders in the dispersion direction to evaluate the boundaries in energy space within a spectrum where a standard spectral extraction 
would have erroneously included events from the confused source. These spectral regions are then flagged as confused and accounted for 
in spectral fitting (Section \ref{sec:cleaning}).

The three causes of confusion were determined for every source in Table \ref{tab:master} on a per-epoch basis and collated into a master table
to be used in spectral cleaning (Section \ref{sec:cleaning}). The reduction and analysis of the high resolution X-ray spectra 
in Tab.\ref{tab:master} from the
70 HETGS observations of the ONC represents a very large dataset with tens of thousands of potential instances of confusion over all the 
individual spectra. Many instances of confusion were checked by eye but it is not feasible to check them all. Therefore, conservative 
parameter values were chosen with \emph{CrissCross} to err on the side of removing some genuine source events in an effort to ensure 
confusion events are not included in our final spectral extractions.  This provides a first set of quality spectra for analysis. 

\subsection{Spectral Cleaning Process} \label{sec:cleaning}

The spectral extraction results in standard products for data analysis for all
sources over the entire exposure. This includes a PHA file containing 
binned spectra, and their corresponding ARFs and RMFs.  We
did not extract backgrounds adjacent to spectra, since the
``background'' will be largely due to confusing sources, both zeroth
orders and dispersed spectra as described in Sec.\ref{sec:confusion}.
For this analysis we combine the single source spectra
(i.e. PHA files) to one merged spectrum but ignore the confused regions.
To do this, we load all the spectra for a given source, then apply the confusion 
information which defines the regions to be ignored in each order of each spectrum.
The confusion analysis described in Sec.~\ref{sec:confusion} produced a confusion table
which contains all locations where cluster stars interfere with each other 
either via zero order overlaps with grating arms spatially or where grating arms
overlap with each other spatially \emph{and} in PHA space. In standard analysis of a single, 
isolated source, the PHA is used to sort the grating orders. In a multi-source confused situation as
we encounter in the ONC, PHA space also has to sort out orders from other confusing sources.
The application of the information from the confusion table is straightforward for the 
zero order point source overlaps, but somewhat subjective when it comes to confusion due to
spectral arm overlaps. Here we defined a parameter, which is basically the 
zero-order flux ratio of the involved sources, that controls how low of an interfering overlap
we allow with respect to contributing flux. The farther below unity this parameter is chosen to be,
the more overlapping flux is excluded. This has to be done manually by adjusting this parameter
until the HEG and MEG positive and negative first order fluxed spectra agree within their statistical 
uncertainties. Here it is mandatory that \emph{all} four spectral arms agree.
This then defines the exclusion criteria, i.e., the `ignore' ranges in each spectral
histogram. 

Tab.~\ref{tab:master} lists the total number of counts in the added HETG 1st orders after that cleaning procedure was applied
and an effective exposure. The effective exposure shows how much of the original ~2 Msec exposure remained for each source.
In theory for bright sources such as $\theta^1$ Ori A, C, E and MT Ori there should be little arm
confusion. It turned out that this was only true for $\theta^1$ Ori C mostly because it is so much
brighter than any of the other sources. The other three sources suffered significant losses due to unfortunate
observation roll angles which resulted in the situation that they confused each other. Here 
$\theta^1$ Ori E interfered with $\theta^1$ Ori C and A. The latter source suffered the most as it 
overlapped with three very bright sources, $\theta^1$ Ori C, E and MT Ori. This situation was anticipated 
and minimized during observation planning by selecting more favorable roll angles. We also had 
over half a dozen cases where overlaps were so severe
that at this point we could not recover any reasonable flux in the 1st orders. 
We note, that the method we apply here is likely over-cleaning the spectra, i.e. future refinements 
may improve these numbers, even recover 1st order counts in those sources that have zero counts and zero effective
exposures in the present analysis.

In order to compare the resulting spectra with spectral models, the models must also ignore the same regions 
in the responses and sum to the cleaned observed counts. The rigorous way to do this would be to zero the
corresponding channel range in the response matrix.  However, since the response matrices for HETG dispersed orders are nearly 
diagonal, and since regions are randomly distributed throughout the count spectra, it is easier
to modify the ARF in the same way as the counts. We can thus, for each order and grating type, add the 
counts, add the ARFs with exposure weighting, and use the RMF as is, to provide a merged set of 
data products suitable for further analysis. These data products, i.e. the cleaned merged spectra and 
their corresponding ARFs and RMFs, are available to the public and can be downloaded from the \chandra archive
contributed data page\footnote{\url{https://cxc.harvard.edu/cda/contributedsets.html}} and 
alternatively from \emph{Zenodo at \url{https://doi.org/10.5281/zenodo.10853416}}.

\section{Source Detection and Master Source List}

\subsection{0th Order Source Detection}

The main field of view of Fig.~\ref{fig:hetgcluster} shows the merged zero order image of the ONC as observed
with the \chandra HETG. We ran \emph{wavedetect} on that field of view and compared the resulting source list with 
the COUP source list~\citep{getman2005}. Some of the sources
in the COUP list were not detected, even though based on their brightness during the COUP campaign, they should have been detectable.
This emphasizes the extreme flux variability young stars exhibit in X-rays.

\subsection{1st Order Source List}

We have accumulated a final master source list that emerged after all cleaning procedures. Of the 45 sources we found to be bright
enough to produce good 1st order spectra and which are shown in Tab.~\ref{tab:master}, 36 sources survived the cleaning process
described in Sec.~\ref{sec:cleaning} with well above several 1000 1st order counts. One source, 
$\theta^1$ Ori C remained with over 2 Ms exposure after cleaning and 25 sources have between 1 and  2 Ms exposures. 
The smallest exposure is for COUP 662 with 750 ks. 24 sources yield over 10\,000 1st order counts, 11 sources have more than 5000 
counts, only 2 sources are below that number. Nine sources were excluded because their spectra had less than a 
few 100 counts left after cleaning. These sources are fainter than the rest and we anticipate that future improvements  
in the cleaning procedure may recover some more counts. Tab. \ref{tab:master} provides the number of the final number of counts
after cleaning, the fraction of counts between the final spectra divided by the original number of counts as a figure of  merit towards
the cleaning process f$_{cl}$. It also states the final remaining exposure after cleaning process. 

\begin{deluxetable*}{lcccccccccc}\label{tab:master}
\tablecaption{HETGS 1ST Order Master Source Table (Data available in machine readable table (MRT))}
  \tablehead{
   	 \colhead{Star}&
 	 \colhead{RA}&
      \colhead{DEC} &
 	 \colhead{Primary} &
 	 \colhead{T$_{eff}$} &
      \colhead{Mass} &
      \colhead{log(age)} &
      \colhead{COUP} &
      \colhead{1st order} &
      \colhead{f$_{cl}$} &
      \colhead{fin. exp}
      \\
      &
    \colhead{ [h m s]}&
    \colhead{ [d m s]} &
    \colhead{ Spec. Type} &
    \colhead{ [kK] } &
    \colhead{ [$M_\odot$] } &
    \colhead{ [yr] }  &
    \colhead{ [$\#$]} &
    \colhead{ [counts] } &
    \colhead{} &
    \colhead{ [ks] }
    }
 \startdata
$\theta^1$ Ori C   & 5 35 16.46 & -5 23 22.8  &  O7V   &    44.6     &   35    &          &  809 &  1033433 & 1.00 & 2085 \\
$\theta^2$ Ori A   & 5 35 22.90 & -5 24 57.8  & O9.5IV  &    30.9     &   25    &          & 1232 &    19573 & 0.85 & 1445 \\
$\theta^1$ Ori A   & 5 35 15.83 & -5 23 14.3  &  B0.5Vp   &    28.8     &   15    &          &  745 &    71578 & 0.48 & 1276 \\
$\theta^1$ Ori B   & 5 35 16.14 & -5 23 06.8  & B3V  &             &    7     &          &  778 &        0 &  0 &    0 \\
$\theta^1$ Ori E   & 5 35 15.77 & -5 23 09.9  & G2IV &    14.8     &   2.8   &          &  732 &   131865 & 0.67 & 1592 \\
$\theta^1$ Ori D   & 5 35 17.26 & -5 23 16.6  &  B1.5Vp   &    32.4     &   16  & $<$6.39   &  869 &        0 &  0 &    0 \\
$\theta^2$ Ori B   & 5 35 26.40 & -5 25 00.8  &  B0.7V    &    29.5     &   15  &  $<$6.30   & 1360 &        0 &  0 &   0 \\
MV Ori       & 5 35 18.67 & -5 20 33.7  &  F8-G0    &    5.24     &   2.72  &   6.17   &  985 &  17368 & 0.74 & 1189 \\
TU Ori       & 5 35 20.22 & -5 20 57.2  &  F7-G2    &    5.90     &   2.43  &   5.55   & 1090 &  9813 & 0.53 &  1027 \\
V2279 Ori    & 5 35 15.93 & -5 23 50.1  &  G4-K5    &    5.24     &   2.37  &   6.12   &  758 &  16545 & 0.27 & 1058 \\
V348 Ori     & 5 35 15.64 & -5 22 56.5  &  G8-K0    &    5.24     &   2.33  &   6.23   &  724 &  34731 & 0.43 & 1236 \\ 
V1399 Ori    & 5 35 21.04 & -5 23 49.0  &  G8-K0    &    5.11     &   2.28  &   6.17   & 1130 &  32765 & 0.63 & 1816 \\
V1229 Ori    & 5 35 18.37 & -5 22 37.4  &  G8-K0    &    5.24     &   2.22  &   6.14   &  965 &  28267 & 0.55 & 1349 \\
V2299 Ori    & 5 35 17.06 & -5 23 34.2  &  K0-K7    &    5.11     &   2.08  &   6.27   &  855 &  10640 & 0.23 &  905 \\
LR Ori       & 5 35 10.51 & -5 26 18.3  &  K0-M0    &    5.24     &   2.05  &   6.43   &  387 &  9549 & 0.73 & 1193 \\
2MASS3       & 5 35 17.22 & -5 21 31.7  &  K4-K7    &    4.68     &   1.97  &   5.56   &  867 &   7024 & 0.50 &  942 \\
MT Ori       & 5 35 17.95 & -5 22 45.5  &  K2-K4    &    4.58     &   1.99  &   5.39   &  932 & 150965 & 0.84 & 1701 \\
LU Ori       & 5 35 11.50 & -5 26 02.4  &  K2-K3    &    4.78     &   1.86  &   6.07   &  430 &  13386 & 0.77 &  1259 \\
V1338 Ori    & 5 35 20.17 & -5 26 39.12 &  K0-G4    &    5.25     &   1.83  &   6.32   & 1087 &   0 &  0 & 0 \\
Par 1841     & 5 35 15.18 & -5 22 54.53 &  K6-G4    &    5.25     &   1.83  &   6.74   &  682 &      0 &  0 &   0 \\
V1333 Ori    & 5 35 17.00 & -5 22 33.0  &  K5-M3    &    4.95     &   1.68  &   6.32   &  854 &  13484 & 0.31 &  918 \\
V2336 Ori    & 5 35 18.70 & -5 22 56.8  &  K0-K3    &    4.79     &   1.65  &   6.50   &  993 &      0 &  0 &   0 \\
Par 1842     & 5 35 15.27 & -5 22 56.8  &  G7-G8    &    5.56     &   1.56  &   6.62   &  689 &  15783 & 0.19 & 941 \\
V1330 Ori    & 5 35 14.90 & -5 22 39.2  &  K5-M2    &    4.58     &   1.47  &   5.88   &  670 &  21357 &  0.41 & 1314 \\
Par 1837     & 5 35 14.99 & -5 21 59.93 &   K3.5    &    4.58     &   1.47  &   6.30   &  669 &   6956 & 0.54 & 1096 \\
Par 1895     & 5 35 16.38 & -5 24 03.35 &  K4-K7    &    4.00     &   0.91  &   5.59   &  801 &   5724 &  0.28 & 838 \\
V1279 Ori    & 5 35 16.76 & -5 24 04.3  &  M0.9e    &    4.20     &   0.91  &   5.84   &  828 &  13683 &  0.64 & 1251 \\
V491 Ori     & 5 35 20.05 & -5 21 05.9  &  K7-M2    &    3.99     &   0.74  &   5.92   & 1071 &  18586 &  0.78 & 1380 \\
Par 1839     & 5 35 14.64 & -5 22 33.70 &   K7      &    3.99     &   0.74  &   5.30   &  648 &   6382 &  0.27 & 877 \\
LQ Ori       & 5 35 10.73 & -5 23 44.7  &  K2       &    3.90     &   0.70  &   3.99   &  394 &  34093 & 0.84 & 1617 \\
V1326 Ori    & 5 35 09.77 & -5 23 26.9  &  K4-M2    &    3.90     &   0,64  &   5.76   &  343 &  17530 & 0.68 & 1402 \\
COUP 1023    & 5 35 19.21 & -5 22 50.7  &  K5-M2    &    4.40     &   0.62  &   6.36   & 1023 &  6119 & 0.39 &  815 \\
V495 Ori     & 5 35 21.66 & -5 25 26.5  &   M0      &    3.80     &   0.58  &   6.43   & 1161 &  13126 & 0.83 & 1453 \\
V1527 Ori    & 5 35 22.55 & -5 23 43.7  &   M0      &    3.80     &   0.57  &   6.43   & 1216 &      0 &   0 &  0 \\
V1228 Ori    & 5 35 12.28 & -5 23 48.0  &  K1-M0    &    3.80     &   0.56  &   5.95   &  470 &   9440 &  0.36 & 1133 \\
V1501 Ori    & 5 35 15.55 & -5 25 14.15 &  K4-M1    &    3.80     &   0.55  &   4.65   &  718 &  16384 & 0.87 & 1564 \\
2MASS4       & 5 35 23.81 & -5 23 34.3  &   M1e     &    3.72     &   0.47  &   6.21   & 1268 &      0 &  0 &  0 \\
V1496 Ori    & 5 35 13.80 & -5 22 07.02 &   K2e     &    3.43     &   0.39  &   5.16   &  579 &  6425 &  0.49 & 1040 \\
2MASS1       & 5 35 09.77 & -5 21 28.3  &   M3.5    &    3.31     &   0.28  &   6.52   &  342 &   13960 & 0.90 & 1581 \\
COUP 450     & 5 35 11.80 & -5 21 49.3  &   M4.4    &    3.16     &   0.22  &   6.47   &  450 &  24771 & 0.85 & 1642 \\
Par 1936     & 5 35 19.30 & -5 20 07.9  &   K2      &    4.95     &    1.4  &   6.78   & 1028 &   4301 & 0.55 &  959 \\  
V1230 Ori    & 5 35 20.72 & -5 21 44.3  &   B1      &   18.6      &    6.4  &          & 1116 &  24363 & 0.83 &  1507 \\
COUP 662     & 5 35 14.90 & -5 22 25.41 &           &             &         &          &  662 &   4026 &  0.33 & 750 \\
JW 569       & 5 35 17.95 & -5 25 21.24 &   M3.5    &    3.16     &   0.1   &          &  936 &      0 &   0 &  0 \\
V1398 Ori    & 5 35 13.45 & -5 23 40.43 &   M0      &             &         &          &  545 &   7068 & 0.43 &  980 \\
\enddata
\end{deluxetable*}          
 
 As expected, all bright sources were detected by COUP \citep{feigelson2005} and Tab.~\ref{tab:master} provides the COUP numbers of the object as 
 well as the coordinates as provided by COUP \citep{getman2005}. The table also provides some physical parameters describing each source which
 were collected from previous optical studies~\citep{hillenbrand1997, herbig2006, dario2010, hillenbrand2013, maizapellaniz2022}. 
 In fact the table itself is approximately sorted by modeled
 stellar masses, even though for some stars we could not find model predictions. 

All of the early (O and B) type stars in our sample are known to be
multiple systems 
\citep[see][and references therein]{petr1998, preibisch1999, grellmann2013, karl2018}.
Table 4 lists only the properties of the primary component,
but a summary of the companion properties is provided
in Tab.~\ref{tab:theta1Ori}.

\begin{deluxetable}{llcrr}\label{tab:theta1Ori}
\tablecaption{Multiplicity and components of $\theta^1$~Ori and $\theta^2$~Ori}
  \tablehead{
 	 \colhead{Star}&
 	 \colhead{Comp.}&
      \colhead{SpT} &
 	 \colhead{Mass} &
 	 \colhead{Separation} 
      \\
    \colhead{ } &
    \colhead{ } &
    \colhead{ } &
    \colhead{ [$M_\odot$] } &
    \colhead{ [AU] } 
    }
 \startdata
$\theta^1$ Ori A --  & 1 &  B0.5Vp &  15  &    \\
                   & 2 &         &  $\approx 4$  &  100  \\
                   & 3 &         &  $\approx 2.6$ &  0.71 \\ \hline
$\theta^1$ Ori B --  & 1 &  B3V    &   7  &   \\
                     & 2 &         & $\approx 4$  & 382 \\
                     & 3 &         &  $\approx 3$  & 49 \\
                  & 4 &         &  $\approx 1$  & 248 \\     
                  & 5 &  & $\approx  2$ &  0.12 \\  
                  & 6 &         &  $\approx 2$  &   5  \\  \hline
$\theta^1$ Ori C --  & 1 &  O7V    &  35  &  \\ 
                     & 2 &        & 9  &  18.1 \\
                     & 3 &        & $\approx 1$  & 0.41 \\ \hline
$\theta^1$ Ori D --  & 1 &  B1.5Vp &  16  &       \\ 
                     & 2 &          &   $\approx 1$   &  580 \\
                     & 3 &         & $\approx 6$ & 0.77 \\ \hline
$\theta^1$ Ori E --  & 1 &  G2IV   & 2.8  &   \\ 
                     & 2 &  G0IV-G5III        & 2.8  & 0.09 \\ \hline \hline
$\theta^2$ Ori A --  & 1 &  O9.5IV &  $\simeq 25$  &  \\ 
                    & 2 &     & $\approx 10$  & 0.42 \\
                    & 3 &     & $\approx 10$ &  157  \\ \hline
$\theta^2$ Ori B --  & 1 &  B0.7V  &  15  &     \\
                    &  2 &     &  $\approx 1.6$  &  40 \\
\enddata
\tablerefs{References: \citet{preibisch1999},
 \citet{kraus2009}, \citet{grellmann2013},
\citet{karl2018},
\citet{maizapellaniz2022}}
\end{deluxetable}     

\subsection{Gaia Distances of the ONC and Our Stellar Sample}                                                        
Thanks to Gaia parallaxes, the distance to the Orion Nebula Cluster is 
very well known today. In the recent study \citep{maizapellaniz2022}
based on the Gaia DR3 data, a distance of $D = (390 \,\pm\, 2)$~pc 
was determined for a sample of astrometrically selected cluster members.

Although it is highly likely that the X-ray selected stars in our
Master Source List are ONC members, the
X-ray detection alone does not immediately prove that this star
is actually a young star in the Orion Nebula Cluster; there may be
some level of contamination by foreground and background objects.

In order to check this, we
obtained the parallaxes for the stars in our Master Source List
from the Gaia DR3 archive.
Parallaxes were found for 43 of the 45 stars in our Master Source List;
the two exceptions are COUP~450 and COUP~662.
We performed the bias-correction of the parallaxes with the algorithm 
described in \citet{lindegren2021}.

All parallaxes are approximately in the expected range for ONC members around $\varpi \approx 2.5$~mas
and there are no immediately obvious foreground or background objects in the sample.
However, the parallaxes show (of course) some scatter, and there are four stars (V2299~Ori,
V1279~Ori, LQ~Ori, and Par~1936) for which the
3$\sigma$  uncertainty range for their parallax (i.e., $\varpi \pm 3 \,\sigma_\varpi$)
does not include the expected value,
which, in principle, qualifies them as ``outlier candidates''.
However, in all four cases the
``Renormalised Unit Weight Error'' (RUWE) associated to the Gaia data of these stars
is high ($> 1.4$).
The RUWE value is a goodness-of-fit statistic
describing the quality of the astrometric solution
\citep[see][]{lindegren2018}, and
RUWE values above 1.4 indicate a low reliability of the astrometric parameters
\citep{fabricius2021}.

We determined the most likely distance to the sample of stars in our Master Source List
with a Bayesian inference algorithm, employing the program \textit{Kalkayotl}
 \citep{olivares2020}.
\textit{Kalkayotl} is a free and open code that
uses a Bayesian hierarchical model to obtain samples of the
posterior distribution of the cluster mean distance by means of a Markov
chain Monte Carlo (MCMC) technique implemented in PyMC3.
\textit{Kalkayotl} also takes the parallax spatial correlations into account,
which improves the credibility of the results, and
allows to derive trustworthy estimates of
cluster distances up to about 5~kpc from Gaia data \citep{olivares2020}.

We used \textit{Kalkayotl} version 1.1.
For the prior, we used the implemented  Gaussian model with a mean distance of
$D_{\rm prior} = (390 \pm 10)$~pc and a cluster scale
of $S_{\rm prior} = 10$~pc. The calculations were done in distance space, and
the reported uncertainties for the inferred mean distances
are the central 68.3\% quantiles (corresponding to the ``$\pm 1 \sigma$
range'' for a Gaussian distribution).

For the complete sample of 43 stars with parallaxes we obtained
a distance of 396.5~pc with an uncertainty range of  $[ 391.8 \, , \, 401.2 ]$~pc.
Excluding the above mentioned  four ``outlier candidates'', the result
changes only very slightly to
 395.9~pc with an uncertainty range of  $[ 392.9 \, , \, 398.9 ]$~pc.
These distance values for our sample are well consistent with the above mentioned distance
determination for the ONC.
 
\section{Global HETG Properties}

\subsection{Zeroth Order Light Curves}

\begin{figure}[htb]
\centering
\begin{interactive}{animation}{tst_14.mp4}
\includegraphics[angle=270,width=1.\columnwidth]{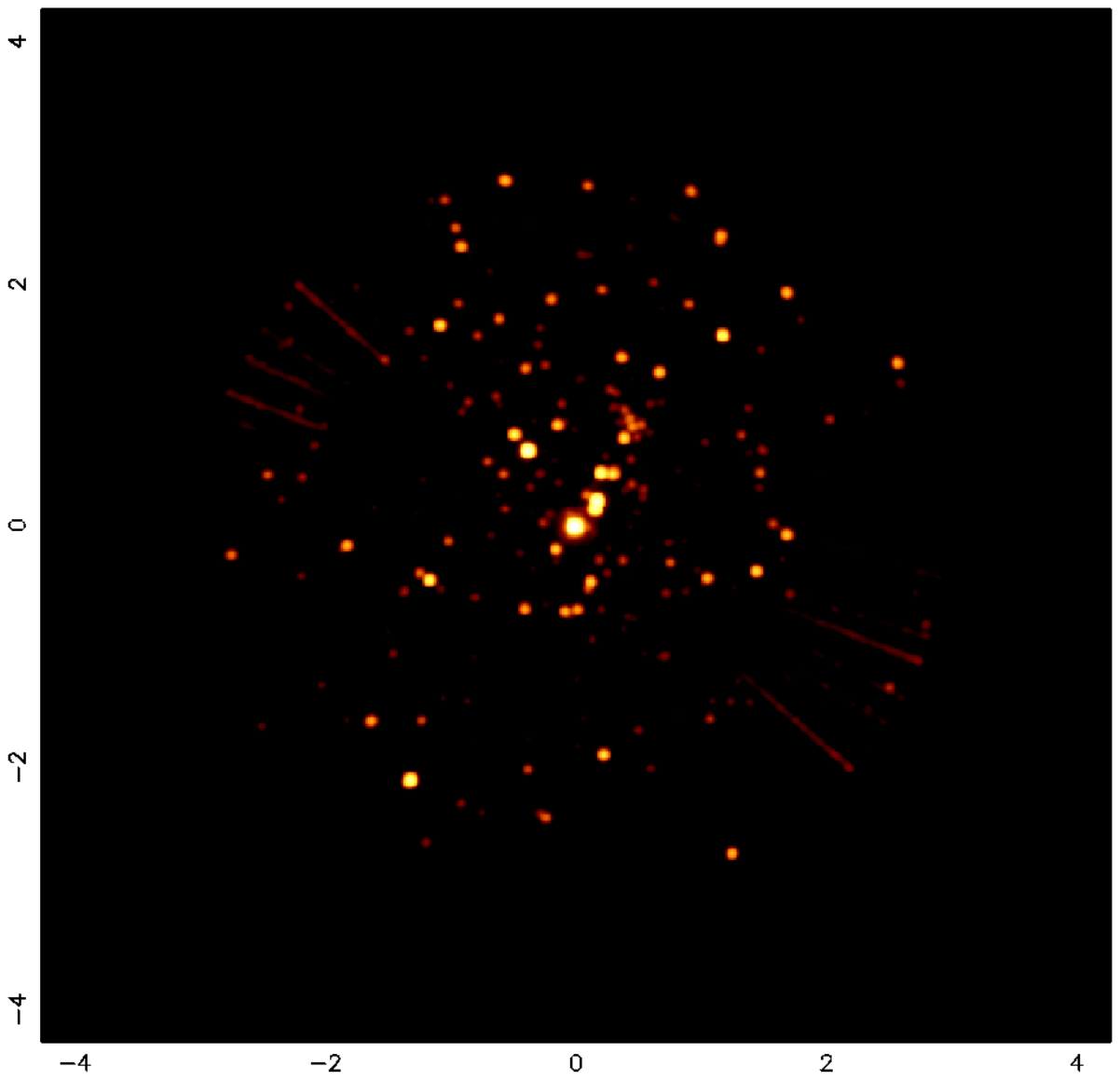}
\end{interactive}
\caption{Example image of the Orion Nebula Cluster for approximately 3' around $\theta^1 Ori C$.  This image represents 10\% of the exposure time on this field during the 2018-2019 campaign, whereas the associated video includes all of the exposure time on this field.  The video, visible in the HTML version of the paper, is composed of frames with 1 ksec of the exposure time in sequential order, organized into a movie to highlight the remarkable short-term variability of the sources in this region. The video runs 16 min, 40 sec at normal speed. In many cases, a source varies from bright to not detectable in the space of 1 frame (1 ksec).}
\label{fig:hetgmovie}
\end{figure}

The field of the Orion VLP observations includes a wealth of sources that vary in brightness with time.  
Many of the sources are late-type stars
that can flare.  Fig.~\ref{fig:hetgmovie} (on-line version only) shows a video that gives a full 
appreciation of variability in this field by watching the zeroth orders of the spectra as they change with time.  The video was created from 
the merged evt2 event file of all 70 obsids, split equally into 1000 frames. Therefore, each frame is a subsample of the total exposure time. 

We have examined the lightcurve of the zeroth order of each source listed in Tab.~\ref{tab:master}, searching for 
variability. The light curves were binned into 1 ksec bins for each obsid individually, then concatenated for each source.  Sources 
for which the zeroth order was confused by an overlaying spectral arm of another source in an obsid, as determined by method described
in Sec. \ref{sec:confusion}, were eliminated from the variability analysis.

We investigated the variability of the zeroth order of the spectrum for each of the 45 sources using the Gregory-Lorado variability index.  
The variability index is determined using the algorithm of \citet{gregory1992}, 
as implemented in CIAO as {\it glvary}, and is based on the 
probability that the count rate of the source is not constant during the observation, using a comparison of binned event arrival times.  
This index is normally used only within an individual observation, but can also be used for merged data if 
the Good Time Intervals are properly handled. According to \citet{rots2012}, if the source has a
variability index of 0-3, it is not considered variable within the observation.  A variability index of 8 or above is definitely variable.  
To examine the variability of each source, a merged file of all non-confused  observations  was created (see Table \ref{tab:obs} for a list of observations). 
{\it glvary} was used to evaluate 
the variability index for this set of non-confused observations for each source.  We find that all sources are definitely variable 
with a variability index of 9-10, except for COUP 1023 which is possibly variable and $\theta$ Ori D and V1527 Ori which are not variable.
An example of the zeroth order light curves produced by the merged observation files is shown in Fig. \ref{plt:lq}. The remainder of the zeroth order light curves appear in Appendix A.  The time gaps 
between the individual observations have been eliminated in these plots and the light curves display the data as if they were one long 
continuous observation for each source. Data for confused zeroth orders are not included in the light curves.

%\begin{figure}[htb]
%\centering
%\includegraphics[angle=0,width%=0.95\columnwidth]%{/Plots/V1230_Ori_1ks_lc_cat_no%bad.pdf}
%\caption{Concatenated light curve for 2019-2020 V1230 Ori observations, each in 1 ksec bins.  Time on x-axis is cumulative exposure since the start of the first observation plotted. 
%Data for obsids where %confusion affects the zeroth %order have been eliminated in %the plot. }
%\label{plt:v1230}
%\end{figure}

\begin{figure}[htb]
\centering
\includegraphics[angle=0,width=0.95\columnwidth]{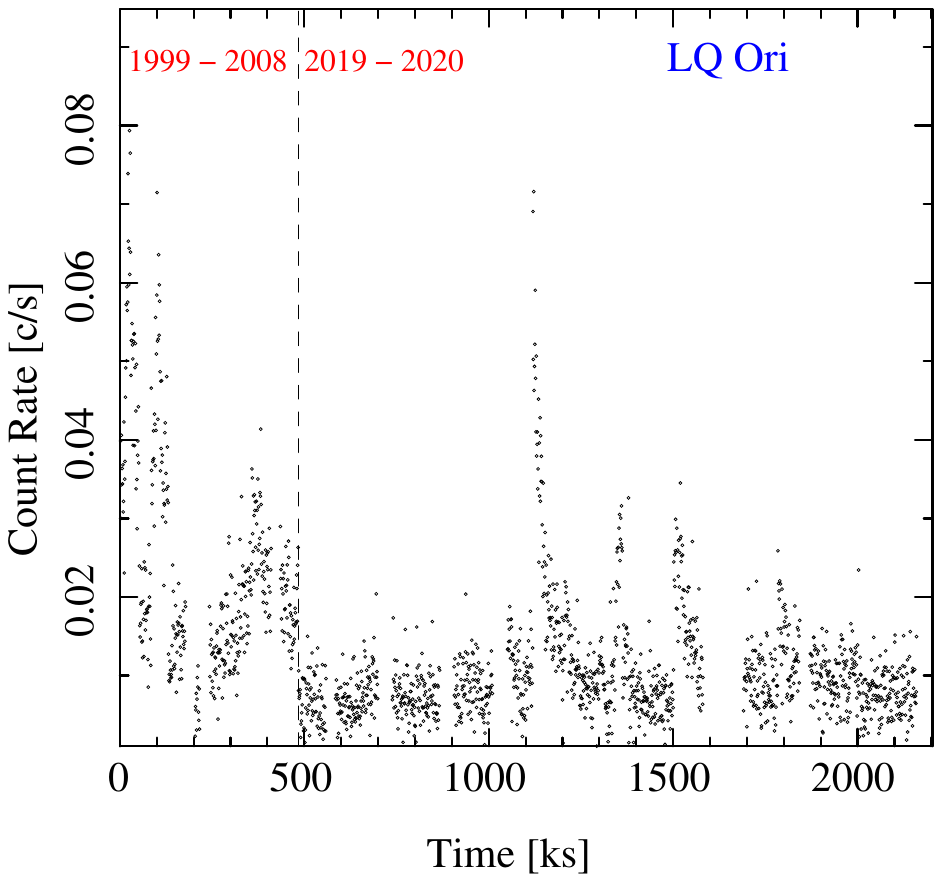}
\caption{Concatenated light curve for LQ Ori observations, each in 1 ksec bins.  Time on x-axis is cumulative observing time since the beginning of the first observation.  Data for obsids where confusion affects the zeroth order count rate have been eliminated in the plot.}
\label{plt:lq}
\end{figure}

%In addition to the primary 45 sources, the variability index of about %1600 additional sources in the fields are being calculated.

The analysis of flares in later type stars is an important component of the Orion VLP program. The ultimate goal is to analyze the high 
resolution spectra near the times of flares to obtain detailed information about the spectral parameters both before and after the flares. 
A follow-on paper will identify timing and other parameters of flares using the zeroth order light curves presented here.
%in detail the method for eliminating confused zeroth order sources and %give quantitative information on 
%the variability and flare-identification methods used.

\subsection{HETG 1st Order Spectra and Background}\label{sec:specbkg}

\begin{figure*}[t]
\centering
\includegraphics[angle=90,width=18cm]{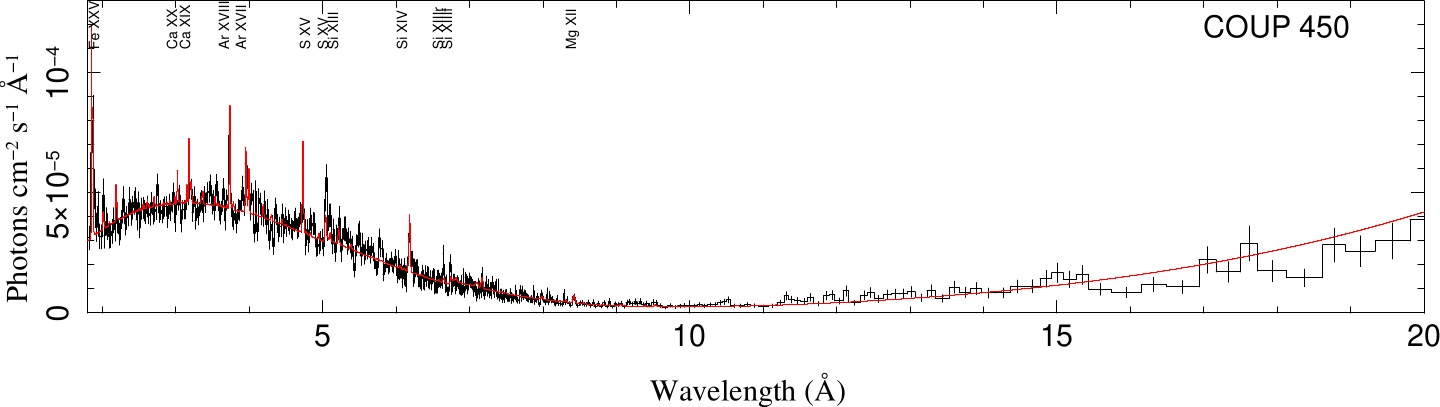}
\caption{Absorbed one-component plasma fit with a model background for COUP 450. The background has a 
power law shape and becomes noticeable above 10 \AA~ and dominant above 16 \AA; it is primarily due to local instrumental background.}
\label{fig:background}
\end{figure*}

The sample of 36 sources that passed the cleaning process contains four massive ($>6$ \Msun) stars, about
a dozen intermediate mass ($\sim 2-3$ \Msun) stars, and about twenty low-mass ($< 2$ \Msun ) stars 
(see Table \ref{tab:master}). The modeling of the spectra and the X-ray line emission is done in various steps. 
One item is selective bandpass. The bright sources as observed in the early phases of the \chandra mission also 
have low absorption and provide significant flux above 16~\AA. Observations in cycles later than Chandra Cycle 16
have too much contaminant absorption to allow for much flux above 16~\AA. Thus we allow a wider bandpass for 
bright sources analysed in the early \chandra Cycles up to 22~\AA, while limiting the bandpass for sources otherwise 
to 16~\AA. The model spectra apply the Astrophysical Database Emission Database (APED) to fit collisionally ionized 
emissions to the spectra. The number of temperature components mostly depends on the need to cover the available 
wavelength range but also depends on the strength of the recorded X-Ray continuum. As for the fitting procedure we
applied a number of APED temperature components plus background. All models have thermal line broadening
applied and are folded forward through the instrument responses, then fitted to the data applying
appropriate statistics.

Most of the stars in the sample are fainter than a few $10^{-13}$ \ergcm  and thus require the inclusion of 
an X-ray background which becomes significant at soft X-rays. This background consists mainly of an  
HETG/ACIS-S instrumental component\footnote{\raggedright For details, see 
the \chan Proposers' Observatory Guide \S8.2.3 (\url{https://cxc.harvard.edu/proposer/POG/html/chap8.html\#tth_sEc8.2.3}) 
and memo referenced therein.} with
some contribution of a flat diffuse stellar background from weak off-axis sources from the outer regions of the ONC cluster.
Given that we have so many roll angles involved in the available 70 observations, this background should be fairly isotropic for
all sources. The sample contains over half a dozen of absorbed sources where we can directly determine this background
contribution. Figure~\ref{fig:background} shows the example of COUP 450. It is heavily absorbed and the hard X-ray bandpass
below 9~\AA{} is fitted by a single APED temperature function, while the soft part directly shows this background. It is
a powerlaw of photon index 6.5 with a normalization of 6.068$\times10^{-5}$ photons cm$^{-2}$ s$^{-1}$. We tested this
function with half a dozen absorbed sources with powerlaw parameters agreeing within 5$\%$. We then added this powerlaw 
to every spectral fit procedure.  This rising tail beyond 13~\AA~ is well predicted by the empirically measured instrumental background.

\subsection{Massive Stars}

There are four massive stars in the sample, the two most massive are $\theta^1$ Ori C (O7~V) and $\theta^2$ Ori A (O9.5~IV), plus
two less massive stars $\theta^1$ Ori A and V1230 Ori. Even though all of these stars are bright with respect to the 
HETG background, we include this background in all the fits. Except for V1230 Ori, some early \chandra HETG results have been
published before on all the other massive  stars ~\citep{schulz2000, schulz2003, gagne2005, mitschang2011}. Here we assess
how the new 2.2 Msec data can serve to provide new insights.  

\subsubsection{$\theta^1\,$Ori~C}

\begin{figure*}[t]
\centering
\includegraphics[angle=-90,width=18cm]{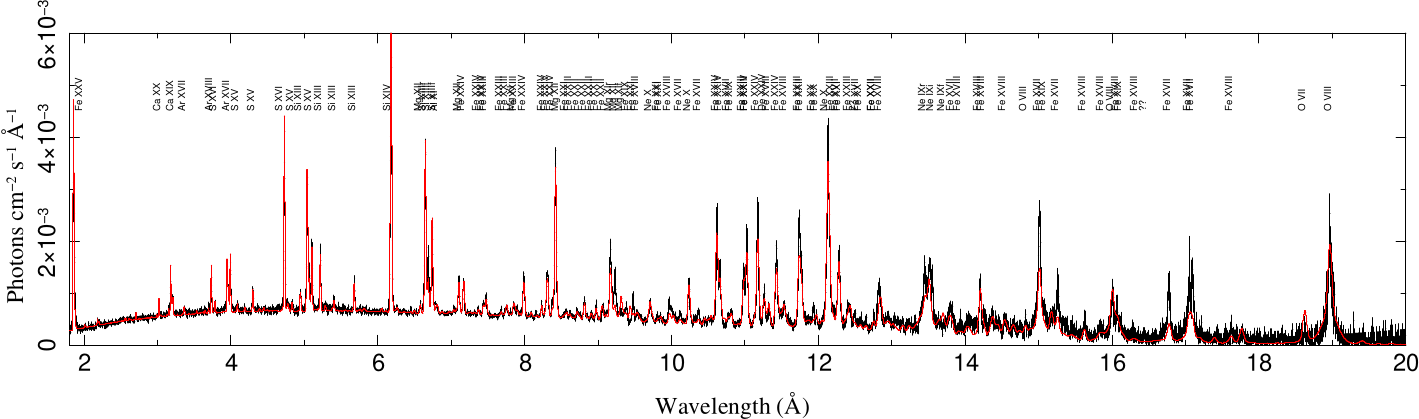}
\caption{The broadband 2.1 Msec spectrum of $\theta^1$ Ori C with line labels. The spectrum shows over 100 detected lines at 
high signal to noise.}
\label{fig:tet1oric}
\end{figure*}

The most massive component of the Trapezium cluster is the triple system $\theta^1$ Ori~C, comprised of a $\sim 33 M{\odot}$ oblique 
magnetic rotator $\theta^1$ Ori~C1,
a $\sim 1\;M_{\odot}$ star C3 at only $\approx 0.04$~AU \citep[][and references therein]{gravity2018}, and
a $\sim 10\;M_{\odot}$ star C2 at $16.7$~AU, with an orbital period of 11.26~years \citep{rzaev2021}.

The cleaning procedure left about 95$\%$ of the exposure for $\theta^1$ Ori C intact, yielding a total exposure time
of 2.085 Msec in 68 OBSIDs. The X-ray source is very bright with an
average unabsorbed $0.5-8.0$~keV X-ray flux of $4.0\times10^{-11} {\rm ~erg~s}^{-1}~{\rm cm}^{-2}$,
and an average X-ray luminosity $L_{\rm X} \approx 7.2\times10^{32} {\rm ~erg~s}^{-1}$ at 395.5 pc.
The high-signal-to-noise HEG and MEG spectra were analyzed using a bin size as low as 0.005 \AA\ over
the 1.65~\AA\ to 23~\AA\ bandpass. While still very good, count statistics decline towards larger wavelengths 
due to interstellar absorption, and the worsening low-energy response of the ACIS-S detector.
In fact, including data sets obtained after 2007 does not improve signal-to-noise above 16 \AA. 
Fig.~\ref{fig:tet1oric} shows the combined, first-order HEG/MEG spectrum of $\theta^1$ Ori~C,
exhibiting hundreds of X-ray lines in over seventy individual line complexes. 

Each data bin of the spectrum of $\theta^1$ Ori C has sufficient counts to allow for 
the application of Gaussian statistics with a $\chi^2$ minimisation process. We performed a fit using a multi-APED temperature model,
which was successfully used in previous analyses by \citet{schulz2003} and \citet{gagne2005}, however
on only about 10$\%$ of the exposure and selected orbital phases. The fit here involved 5 APED temperatures and resulted in a 
reduced $\chi^2_{\nu}$ of 2.97. This fit is shown in Fig.~\ref{fig:tet1oric}.
While the fit appears good with respect to the continuum, the $\chi^2_{\nu}$ shows it is not as 
there are significant residuals with respect to the line fits. These residuals require a more detailed modeling approach,
which has to involve more realistic line profiles.  
A detailed line-by-line analysis of the phase-resolved X-ray spectra will be presented by
Gagn\'e et al. (2024, in preparation). In addition, numerical 3D modeling of the magnetically confined wind shocks will be presented by
Subramanian et al. (2024, in preparation).
%Even though a simple broadband fit like this 
%does not provide a perfect description to the entire data set, it provides an adequate assessment of
%the underlying continuum and average X-ray temperatures which range between 8 and 32 MK, basically similar to results reported by \cite{schulz2003}.
%Residuals} indicate that more detailed line profile analysis will be necessary. However it is interesting
%to note that this fit of the long exposure spectrum did not require overly hot temperature components, we stress, though,
%that for this more precise line profile fits need to be preformed. 
We also took a look at the actual line widths across the bandpass. For this we restrict the analysis
to fit generic Gaussian line profiles to selected bright lines in order to determine the order of magnitude of
the velocity broadening in the resolved lines. We find that the lines are resolved with very
moderate broadening of about $300\kms$. Specifically we find $369\pm16 \kms$ for Ne X, $279\pm8 \kms$
for Mg XII, $326\pm8\kms$ for Si XIV, $381\pm26 \kms$ for S XVI, and $318\pm52$ for Ar XVIII.
The consistency of these values over a large wavelength range as well as the small uncertainties 
are a reflection of the superb properties of this data set.

\begin{figure*}[t]
\centering
\includegraphics[angle=0,width=18cm]{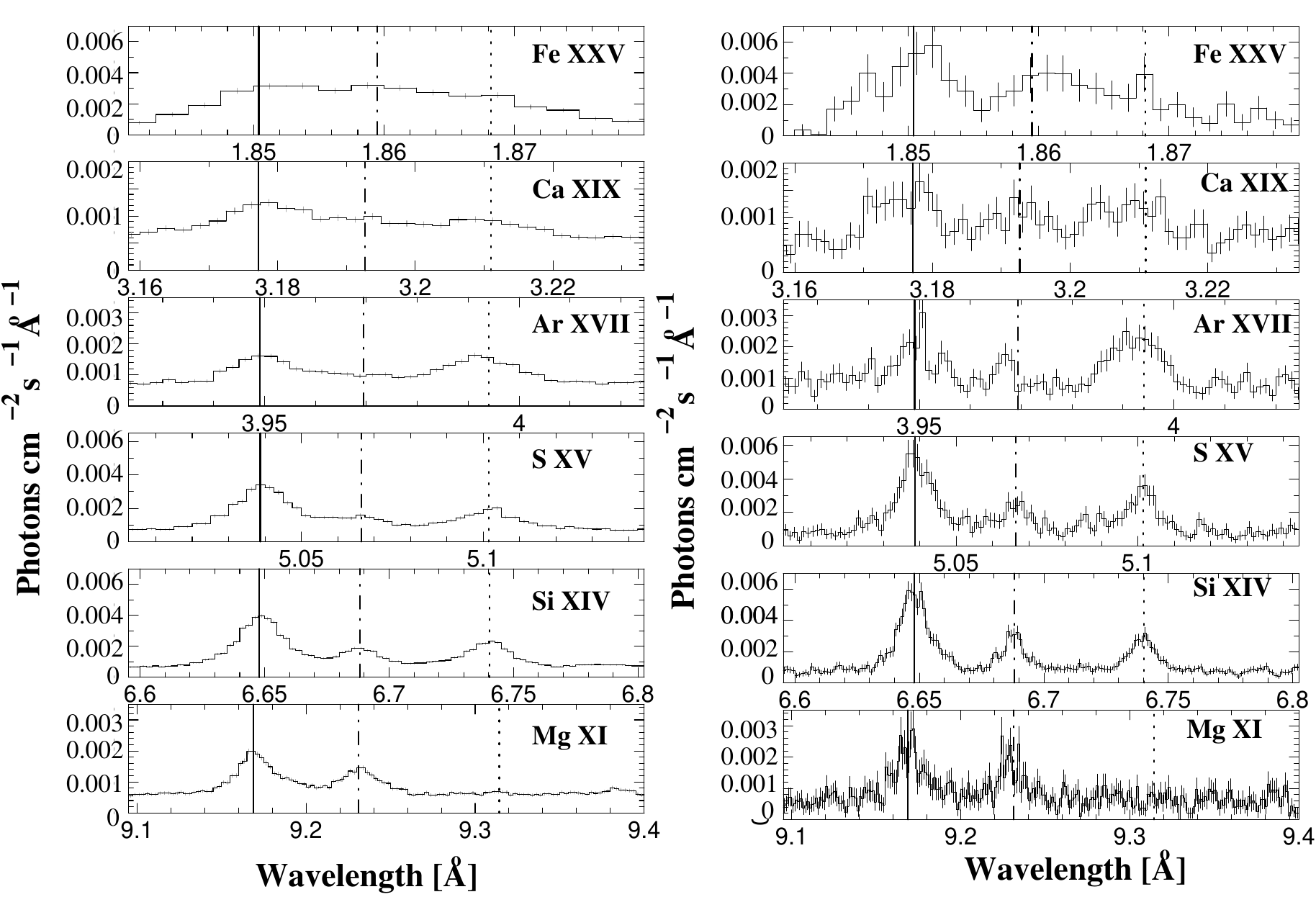}
\caption{Here we show He-like triplets of Mg, Si, S, Ar, Ca and Fe in various orders for $\theta^1$ Ori C. On the left are the
triplets in 1st order, on the right the ones in higher orders, MEG 3rd and HEG 2nd orders. While the 1st orders provide high
signal to noise power, the significantly higher resolving power of the higher grating orders provide much more details. }
\label{fig:helike}
\end{figure*}

The high significance in the emission lines in the 1st order of $\theta^1$ Ori C allows the
analysis of the spectral properties at the highest possible spectral resolution with nearly perfect 
statistics throughout the entire waveband between 1.7 and 23~\AA. One example where these
conditions benefit this analysis are the He-like triplets in this bandpass. Fig.~\ref{fig:helike} on
the left side shows the triplets from Mg, Si, S, Ar, Ca, and Fe at no data binning. The statistical
$1\sigma$ errors are plotted as well but so small that they are not visible. Previous HETG studies
of the source ~\citep{schulz2000, schulz2003, gagne2005} showed that the lines are well resolved with
a FWHM of a few hundred km/s.
%{\em I would not say turbulent broadening. Perhaps Doppler broadening? I would drop the following:}
%{\em show turbulent broadening from the magnetic wind confinement.}

The combination of high resolution 
and good count statistics should prove invaluable for magnetic wind shock model analysis.
However at 1st order resolution spectral details of the triplets start to fade past Si,
i.e. the line components of higher Z triplets
are not fully resolved. Here this long exposure allows the utilization of the higher orders of the
transmission gratings, specifically the MEG 3rd and HEG 2nd orders which features each nearly
10$\%$ of the 1st order efficiency. Fig.~\ref{fig:helike} on the right shows He-like triplets
at this higher resolution. The $1\sigma$ statistical error bars are now clearly visible due to
the reduced efficiency of the higher orders. However, the main triplet components are now
resolved up to Ca and partially at Fe. The resolving power at Mg XI is now 1480, at Si XIII is 1000, at 
s XV is 820, at Ar XVII is 640, at Ca XIX is 515 and at Fe almost 310, which are the
highest resolving powers in He triplets to date.
Resolving He-like triplets is a very powerful tool for analysis. In the case of massive stars
such as $\theta^1$ Ori C He-like line triplet ratios are sensitive tracers of where these lines
originated in the particular wind geometry. The higher Z elements we can resolve the deeper into
the wind geometry we can trace. Low-Z triplets are also powerful diagnostics for accretion, higher-Z
can determine levels of UV exposure in stars, which is generally difficult to measure in the ONC.

\subsubsection{Other Massive Stars}

The three other massive stars in the sample are $\theta^1$ Ori A, $\theta^2$ Ori A and V1230 Ori
(see Appendix \ref{app:spectra} for spectra). 
For the latter two stars the cleaning procedure leaves about 1.5 Msec of remaining exposure,
while for $\theta^1$ Ori A the exposure is 1.2 Msec. This lower exposure is caused by the combination
of this star being very close to $\theta^1$ Ori C and a period of unfortunate roll angle of the 
telescope which caused more confusion of the two stars. In both cases we harvest several
10$^4$ cts in the bandpass between 1.7~\AA{} and 20~\AA. 

There are three more massive stars in the sample for which we could not harvest valid counts 
in the HETG 1st orders. The most prominent example is $\theta^1$ Ori D, which is optically 
supposed to be very close to $\theta^1$ Ori A, but not only do we have a large amount 
of  confusion with other Trapezium stars, the star appears to be also very dim in X-rays,
i.e.\ it is hardly detected even in the 0th order. The similarity of these stars is striking,
as their massive components have very similar mass, and both stars have two low- and intermediate
mass companions (see  Tab~\ref{tab:theta1Ori}). The absence of X-ray detection can have two reasons, 
one is that its spectrum is very soft and suffers from ACIS filter absorption, another is that it is 
inherently X-ray weak. Both explanations are at odds with the appearance of $\theta^1$ Ori A.
Specifically the fact that $\theta^1$ Ori D has lower mass companions but no significant coronal
emissions are detected is quite puzzling.

The other massive stars are $\theta^1$ Ori B and $\theta^2$ Ori B. According to Tab~\ref{tab:theta1Ori},
$\theta^1$ Ori B is a cluster of at least six stars, mostly  of intermediate mass. The cluster is detected
in 0th order but we do no have HETG 1st order spectra. The same is true for $\theta^2$ Ori B, which 
is well detected in 0th order but no significant emissions could be recovered in HETG 1st order.

\begin{deluxetable*}{lcccccccccc}\label{tab:3aped}
\tablecaption{HETG Spectral Parameters of 2 Temperature APED fits (Data available in machine readable table (MRT)):}
\tablehead{
    \colhead{Star}&
    \colhead{N$_{H}$}&
    \colhead{T$_1$}&
    \colhead{T$_2$}&
    \colhead{EM$_1$}&
    \colhead{EM$_2$}&
    \colhead{v$_{Ne}$}&
    \colhead{v$_{Si}$}&
    \colhead{f$_x$}&
    \colhead{L$_x$}&
    \colhead{C$_{\nu}$}
\\
&
    \colhead{(1)}&
    \colhead{(2)}&
    \colhead{(2)}&
    \colhead{(3)}&
    \colhead{(3)}&
    \colhead{(4)}&
    \colhead{(4)}&
    \colhead{(5)}&
    \colhead{(6)}&
    \colhead{}
}
\startdata
%
% 2024.02.02 dph added t1oE info;   vNe vSi are line sigma
$\theta^1\,$Ori~E& 
$1.7\,^{0.1}_{0.1}$ & 
$12.98\,^{0.31}_{0.75}$ &
$40.62\,^{0.80}_{1.40}$ &  
$2.04\,^{0.21}_{0.19}$ &  
$8.05\,^{0.30}_{0.15}$ &
$239\,^{34}_{34}$ &  
$247\,^{34}_{35}$ &  
$53.8\,^{0.3}_{0.3}$ &  
$13.7$&
1.19
\\
MV Ori & 10.10 $^{  0.56 }_{  0.55 }$ & 13.81 $^{  0.65 }_{  0.67 }$ & 90.00 $^{  0.00 }_{ 19.17 }$ &  2.25 $^{  0.17 }_{  0.05 }$ &  0.72 $^{  0.11 }_{  0.04 }$ &  153 $^{   96 }_{  227 }$ &  986 $^{  202 }_{   61 }$ &  4.6 $^{  4.9 }_{  4.4 }$ &  1.03 & 1.15 \\
TU Ori &  9.20 $^{  1.50 }_{  0.43 }$ & 15.67 $^{  2.27 }_{  1.62 }$ & 69.09 $^{ 13.91 }_{ 16.47 }$ &  1.04 $^{  0.16 }_{  0.18 }$ &  0.56 $^{  0.16 }_{  0.15 }$ &  319 $^{  410 }_{  220 }$ &  646 $^{  484 }_{  597 }$ &  2.6 $^{  2.7 }_{  2.4 }$ &  0.56 & 1.22 \\
V2279 Ori &  5.43 $^{  0.49 }_{  0.50 }$ &  9.18 $^{  1.09 }_{  1.02 }$ & 45.61 $^{  3.52 }_{  3.38 }$ &  0.93 $^{  0.17 }_{  0.16 }$ &  1.50 $^{  0.10 }_{  0.09 }$ &  115 $^{  269 }_{  115 }$ &  374 $^{  257 }_{  167 }$ &  4.9 $^{  5.1 }_{  4.7 }$ &  1.03  & 1.16\\
V348 Ori &  2.55 $^{  0.25 }_{  0.23 }$ & 10.27 $^{  0.70 }_{  0.71 }$ & 41.10 $^{  1.66 }_{  1.53 }$ &  0.61 $^{  0.10 }_{  0.09 }$ &  2.80 $^{  0.08 }_{  0.08 }$ &  209 $^{   24 }_{   95 }$ &  258 $^{   98 }_{  258 }$ &  9.4 $^{  9.9 }_{  8.9 }$ &  1.78  & 1.15\\
V1229 Ori &  2.76 $^{  0.28 }_{0.28 }$ &  9.61 $^{  0.78 }_{  1.19 }$ & 35.15 $^{  1.31 }_{  1.53 }$ &  0.56 $^{  0.08 }_{  0.09 }$ &  2.47 $^{  0.07 }_{  0.12 }$ &  153 $^{   58 }_{  123 }$ &  679 $^{  455 }_{  380 }$ &  5.7 $^{  6.0 }_{  5.5 }$ &  1.12  & 1.17\\
V1399 Ori &  3.14 $^{  0.36 }_{  0.24 }$ &  9.70 $^{  0.73 }_{  0.76 }$ & 31.33 $^{  1.16 }_{  1.19 }$ &  0.62 $^{  0.12 }_{  0.09 }$ &  2.20 $^{  0.09 }_{  0.08 }$ &  257 $^{   64 }_{   50 }$ &  268 $^{  121 }_{  189 }$ &  7.3 $^{  7.7 }_{  7.0 }$ &  1.40  & 1.12\\
V2299 Ori & 10.58 $^{  0.84 }_{  0.73 }$ & 16.83 $^{  2.96 }_{  2.71 }$ & 57.82 $^{ 19.51 }_{ 10.68 }$ &  0.81 $^{  0.18 }_{ 0.55 }$ &  1.23 $^{  0.45 }_{  0.17 }$ &  219 $^{   85 }_{  103 }$ &  218 $^{  267 }_{  218 }$ &  4.4 $^{  4.6 }_{  4.2 }$ &  0.94  & 1.20\\
LR Ori &  4.09 $^{  0.82 }_{  0.74 }$ & 12.00 $^{  0.87 }_{  1.55 }$ & 60.00 $^{ 11.00 }_{ 10.53 }$ &  0.57 $^{  0.13 }_{  0.14 }$ &  0.51 $^{  0.25 }_{  0.03 }$ &  213 $^{  112 }_{   99 }$ &  179 $^{  221 }_{  179 }$ &  1.9 $^{  2.0 }_{  1.8 }$ &  0.37  & 1.28\\
2MASS3 &  5.10 $^{  0.84 }_{  0.80 }$ & 14.46 $^{  1.11 }_{  1.32 }$ & 74.60 $^{ 15.40 }_{ 17.00 }$ &  0.36 $^{  0.40 }_{  0.03 }$ &  0.57 $^{  0.09 }_{  0.09 }$ &  153 $^{  122 }_{    50 }$ &    50 $^{  112 }_{  373 }$ &  1.6 $^{  1.7 }_{  1.5 }$ &  0.37  & 1.19\\
MT Ori &  3.38 $^{  0.12 }_{  0.11 }$ & 12.35 $^{  0.78 }_{  0.64 }$ & 40.95 $^{  0.96 }_{  8.17 }$ &  1.37 $^{  0.22 }_{  0.17 }$ &  9.96 $^{  0.17 }_{  0.22 }$ &  195 $^{   27 }_{   24 }$ &  289 $^{   58 }_{   89 }$ & 34.5 $^{ 0.8 }_{ 1.7 }$ &  6.73  & 1.14\\
LU Ori &  4.45 $^{  0.63 }_{  0.64 }$ & 10.96 $^{  0.49 }_{  0.48 }$ & 45.35 $^{  4.58 }_{  3.97 }$ &  0.68 $^{  0.15 }_{  0.14 }$ &  0.77 $^{  0.06 }_{  0.06 }$ &  322 $^{   87 }_{  134 }$ &  470 $^{  213 }_{  181 }$ &  2.6 $^{  2.7 }_{  2.4 }$ &  0.56  & 1.23\\
V1333 Ori &  9.29 $^{  0.58 }_{  0.60 }$ & 12.04 $^{  0.61 }_{  0.75 }$ & 30.39 $^{  2.57 }_{  2.60 }$ &  1.52 $^{  0.25 }_{  0.27 }$ &  1.30 $^{  0.20 }_{  0.15 }$ &  222 $^{   87 }_{ 208 }$ &  636 $^{  635 }_{  260 }$ &  3.3 $^{  3.5 }_{  3.2 }$ &  0.65  & 1.32\\
Par 1842 &  1.77 $^{  0.43 }_{  0.37 }$ & 10.82 $^{  0.93 }_{  1.07 }$ & 36.39 $^{  2.14 }_{  2.19 }$ &  0.45 $^{  0.11 }_{  0.10 }$ &  1.52 $^{  0.09 }_{  0.08 }$ &  216 $^{   31 }_{  138 }$ &  556 $^{  276 }_{  341 }$ &  4.3 $^{  4.5 }_{  4.1 }$ &  0.84 & 1.17 \\
V1330 Ori &  4.95 $^{  0.45 }_{  0.40 }$ & 10.46 $^{  0.74 }_{  0.47 }$ & 43.05 $^{  3.08 }_{ 4.95 }$ &  0.65 $^{  0.13 }_{  0.11 }$ &  1.57 $^{  0.07 }_{  0.08 }$ &  152 $^{  176 }_{  108 }$ &  254 $^{  124 }_{  253 }$ &  5.3 $^{  5.6 }_{  5.0 }$ &  1.03  & 1.14\\
Par 1837 &  5.45 $^{  0.37  }_{  0.84 }$ &  7.22 $^{  1.44 }_{  1.49 }$ & 45.03 $^{  4.66 }_{  5.42 }$ &  0.35 $^{  0.21 }_{  0.09 }$ &  0.52 $^{  0.06 }_{  0.04 }$ &  321 $^{  140 }_{  138 }$ &  597 $^{  288 }_{  326 }$ &  1.5 $^{  1.6 }_{  1.4 }$ &  0.28  & 1.34\\
Par 1895 &  0.05 $^{  0.27 }_{  0.04 }$ & 13.13 $^{  1.93 }_{  1.31 }$ & 64.33 $^{ 15.71 }_{  9.72 }$ &  0.18 $^{  0.05 }_{  0.04 }$ &  0.39 $^{  0.04 }_{  0.05 }$ &  318 $^{  290 }_{  156 }$ &  253 $^{   99 }_{   89 }$ &  1.5 $^{  1.5 }_{  1.4 }$ &  0.28  & 0.95\\
V1279 Ori &  2.02 $^{  0.58 }_{  0.40 }$ &  9.58 $^{  0.92 }_{  1.08 }$ & 38.34 $^{  2.67 }_{  2.52 }$ &  0.26 $^{  0.10 }_{  0.07 }$ &  0.92 $^{  0.05 }_{  0.07 }$ &  279 $^{   55 }_{  132 }$ &  247 $^{  101 }_{  199 }$ &  2.7 $^{  2.8 }_{  2.5 }$ &  0.47  & 1.20\\
V491 Ori & 16.21 $^{  0.91 }_{  0.50 }$ &  -- & 43.40 $^{  2.23 }_{  3.08 }$ &  -- &  2.69 $^{  0.06 }_{  0.10 }$ &  --  &  400 $^{  358 }_{  220 }$ &  7.7 $^{  8.1 }_{  7.3 }$ &  1.68  & 1.28\\
Par 1839 &  2.99 $^{  0.86 }_{  0.84 }$ & 11.67 $^{  1.00 }_{  1.22 }$ & 81.22 $^{  8.78 }_{ 11.71 }$ &  0.48 $^{  0.11 }_{  0.11 }$ &  0.43 $^{  0.06 }_{  0.03 }$ &  269 $^{   88 }_{  164 }$ &  278 $^{  200 }_{  277 }$ &  1.9 $^{  2.0 }_{  1.8 }$ &  0.37 & 1.25 \\
LQ Ori &  0.29 $^{  0.20 }_{  0.23 }$ & 10.52 $^{  0.36 }_{  0.31 }$ & 34.28 $^{  2.16 }_{  1.32 }$ &  0.68 $^{  0.11 }_{  0.07 }$ &  1.89 $^{  0.15 }_{  0.30 }$ &  213 $^{   31 }_{   29 }$ &  221 $^{  160 }_{  220 }$ &  5.0 $^{  5.3 }_{  4.8 }$ &  0.94  & 1.16\\
V1326 Ori &  3.19 $^{  0.41 }_{  0.44 }$ &  6.04 $^{  0.55 }_{  0.50 }$ & 29.46 $^{  1.56 }_{  1.21 }$ &  0.98 $^{  0.21 }_{  0.18 }$ &  1.27 $^{  0.06 }_{  0.06 }$ &  191 $^{   39 }_{   59 }$ &  301 $^{  188 }_{  194 }$ &  2.7 $^{  2.8 }_{  2.6 }$ &  0.56 & 1.51 \\
COUP 1023 &  5.56 $^{  1.34 }_{  1.13 }$ & 18.96 $^{  3.09 }_{  2.95 }$ & 78.00 $^{ 12.00 }_{ 14.96 }$ &  0.60 $^{  0.13 }_{  0.13 }$ &  0.28 $^{  0.11 }_{  0.04 }$ &   86 $^{   83 }_{   78 }$ &  243 $^{  193 }_{  145 }$ &  1.7 $^{  1.8 }_{  1.6 }$ &  0.37  & 1.29\\
V495 Ori &  5.24 $^{  0.72 }_{  0.69 }$ & 11.61 $^{  1.47 }_{  1.01 }$ & 69.02 $^{ 14.58 }_{  9.43 }$ &  0.51 $^{  0.12 }_{  0.11 }$ &  0.66 $^{  0.06 }_{  0.07 }$ &  257 $^{  127 }_{  133 }$ &  307 $^{  165 }_{  143 }$ &  3.0 $^{  3.2 }_{  2.9 }$ &  0.65  & 1.26\\
V1228 Ori &  3.04 $^{  0.80 }_{  1.54 }$ &  9.18 $^{  0.78 }_{  0.65 }$ & 37.33 $^{  5.33 }_{  2.67 }$ &  0.48 $^{  0.17 }_{  0.10 }$ &  0.62 $^{  0.05 }_{  0.08 }$ &  135 $^{   74 }_{   74 }$ &  359 $^{   50 }_{  354 }$ &  1.7 $^{  1.8 }_{  1.6 }$ &  0.37  & 1.30\\
V1501 Ori &  3.41 $^{  0.82 }_{  0.69 }$ & 12.19 $^{  0.99 }_{  1.29 }$ & 42.42 $^{  4.09 }_{  5.42 }$ &  0.59 $^{  0.18 }_{  0.15 }$ &  0.79 $^{  0.11 }_{  0.07 }$ &  301 $^{  105 }_{   97 }$ &  384 $^{  180 }_{  307 }$ &  2.5 $^{  2.7 }_{  2.4 }$ &  0.56  & 1.31\\
V1496 Ori &  3.27 $^{  0.83 }_{  0.84 }$ & 13.00 $^{  2.11 }_{  1.44 }$ & 65.90 $^{ 18.16 }_{ 10.35 }$ &  0.27 $^{  0.09 }_{  0.07 }$ &  0.41 $^{  0.05 }_{  0.05 }$ &  210 $^{  556 }_{  186 }$ &   36 $^{  286 }_{   31 }$ &  1.6 $^{  1.7 }_{  1.6 }$ &  0.37  & 1.29\\
2MASS1 & 14.49 $^{  0.82 }_{  0.82 }$ & 12.07 $^{  1.61 }_{  1.32 }$ & 47.93 $^{  9.85 }_{  6.32 }$ &  1.27 $^{  0.27 }_{  0.28 }$ &  1.08 $^{  0.18 }_{  0.18 }$ &  -- &  770 $^{  246 }_{  293 }$ &  3.6 $^{  3.8 }_{  3.4 }$ &  0.84 & 1.25 \\
COUP 450 & 30.95 $^{  0.78 }_{  0.78 }$ &  -- & 34.92 $^{  1.55 }_{  1.34 }$ &  -- &  5.58 $^{  0.12 }_{  0.26 }$ &    -- &  460 $^{  603 }_{  302 }$ & 11.2 $^{ 0.5 }_{ 0.6 }$ &  3.09  & 1.15 \\
Par 1936 & 16.88 $^{  1.94 }_{  1.76 }$ & 13.28 $^{  1.54 }_{  1.42 }$ & 83.00 $^{  7.00 }_{  7.05 }$ &  0.80 $^{  0.21 }_{  0.16 }$ &  0.26 $^{  0.03 }_{  0.02 }$ &    --  &  356 $^{ 1089 }_{  356 }$ &  1.3 $^{  1.4 }_{  1.2 }$ &  0.28 & 1.32 \\
COUP 662 & 21.52 $^{  1.66 }_{  1.60 }$ &  -- & 89.00 $^{  1.00 }_{  1.00 }$ &  0.00 $^{  0.00 }_{  0.00 }$ &  0.62 $^{  0.02 }_{  0.02 }$ &    -- &  364 $^{ 729 }_{  359 }$ &  2.6 $^{  2.7 }_{  2.5 }$ &  0.56  & 1.37\\
V1398 Ori &  5.01 $^{  1.05 }_{  1.24 }$ & 12.89 $^{  1.31 }_{  1.00 }$ & 79.00 $^{ 11.00 }_{ 11.05 }$ &  0.50 $^{  0.12 }_{  0.16 }$ &  0.36 $^{  0.06 }_{  0.02 }$ &  197 $^{  121 }_{  195 }$ &  415 $^{  314 }_{  290 }$ &  1.7 $^{  1.7 }_{  1.6 }$ &  0.37  & 1.32\\
\enddata
\tablecomments{(1) 10$^{21}$ cm$^{-2}$
(2) 10$^6$ K
(3) 10$^{54}$ cm$^{-3}$
(4) km s$^{-1}$
(5) 10$^{-13}$ erg cm$^{-2}$ s$^{-1}$
(6) 10$^{31}$ erg s$^{-1}$ \\
N$_H$ = column density, T = X-ray temperature, EM = emission measure, v = line width from individual fits to Ne and Si, f$_x$ = X-ray flux,
L$_x$ = X-ray luminosity at 396.5 pc, C$_{\nu}$ = Cash statistic of 2 APED broadband fit }
\end{deluxetable*}

\subsection{Intermediate- and Low-Mass Stars}

There are 11 stars of masses between 1.5 \Msun and 3 \Msun in the sample, which we designate as intermediate
mass stars and 20 stars below 1.5 \Msun, which we designate as low-mass stars
(see Appendix \ref{app:spectra} for spectra). The mass designation is somewhat 
arbitrary but helps in the discussion of their properties. In the analysis we treat them similar as
coronal sources and apply the same model to their data. This model consists of the standard soft background,
column density and two APED temperature components. 
The spectra have a large range in terms of statistical quality from very low to very high levels. Consequently for all spectral fits we
use the \emph{Cash} statistical concept~\citep{cash1979} that allows for properly treating data bins with low statistics by the use of a 
maximum likelihood ratio test. We dynamically binned the spectra to make sure we preserve maximum spectral 
resolution and have non-zero count data bins. In \emph{ISIS} we then can fit multiple APED functions with common abundance 
and column density values. We conducted the model fits in two steps. In a first step we fit the spectra with all
parameters free. This fit should already generate an acceptable reduced Cash statistic C$_{\nu}$. However, in this overview 
analysis we are not interested in all the details and we fixed the APED abundance values to the fit result and 
in a second step we computed 90$\%$ confidence limits for the 
absorption column N$_H$, the involved temperatures  kT$_i$, where \emph{i} is the APED component index, 
and the emission measures EM$_i$ of each component. This second step improves the
Cash statistic by roughly 10$\%$. More detailed analyses involving abundances should be done in 
follow-up studies within the framework of a differential emission measure analysis as described in
\citet{huenemoerder2003}. We also kept the turbulent velocity $v_t$ (in a Gaussian line profile with 
the width defined by $\sigma =\sqrt{\frac{2\mathrm{k}T}{A m_h} + v_t^2}$ where $\mathrm{k}$ is the Boltzmann 
constant, $T$ is the temperature of the component, $A$ is the atomic number and $m_h$ is the hydrogen mass)
free to be fitted but after pre-screening of 
all the data we constrained them to values between 100 \kms and 500 \kms for the broadband fit, which helped stabilize the fit procedure.
%{\bf Note that this line widths represent the bulk turbulent velocity of each temperature component.}

In order to further characterize the actual line widths we performed individual line fits on the Ne X and Si XIV
lines in all sources where they were detected. We simply applied Gaussian functions to determine the line
widths. The resulting velocities were converted from the $\sigma$ width of the Gaussian line and
are therefore slightly different from the APED global turbulent velocities.
The final results of these fits are shown in Table~\ref{tab:3aped}. The broadband fits produced 
Cash statistics between 0.95 and 1.51, which are also listed in Table~\ref{tab:3aped}.
 
\subsubsection{Surface Flux}

The global fits result in X-ray fluxes of a few 10$^{-13}$ \ergcm for all sources except MT Ori,
which is an order of magnitude brighter. In order to determine what we call surface flux we calculate
the source luminosity from the measured unabsorbed flux and divide by the surface area of the 
star. The radius of each star is calculated from the bolometric luminosity and the effective 
surface temperature, which are measured quantities and listed in the standard COUP tables.

\begin{figure}
    \centering
    \includegraphics[width=0.8\columnwidth]{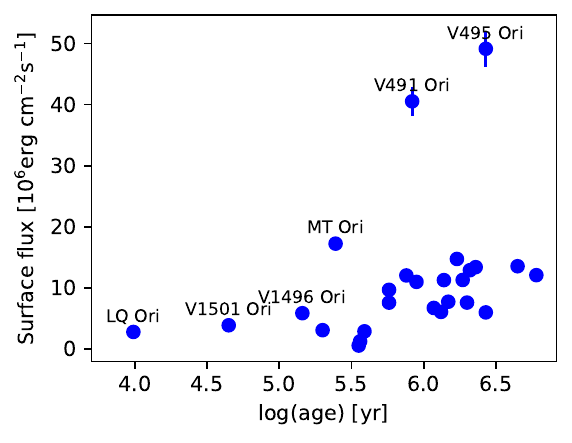}
    \caption{The surface flux plotted against the modeled age of the ONC stars
    The ages are taken from the COUP tables, the surface flux is the source luminosity divided by the stellar surface area. 
    The latter was determined from 
    the bolometric luminosity and the effective surface temperature, both also
    from the COUP tables.
    An interactive version of this figure is available in the online journal with the ability to zoom, pan, and display name and additional information for each source.}
    \label{fig:agesflux}
\end{figure}

In Fig.~\ref{fig:agesflux} we plot the surface flux versus the age of the
cluster source as listed in Tab.~\ref{tab:master}. These ages are also taken from the COUP tables and
even though not very well known they allow global order of magnitude comparisons. The COUP radii are also
subject to systematical uncertainties and we added a 10$\%$ contribution to the uncertainty of the surface
flux. 
%As was done in the \cite{schulz2015} paper we added two young active coronal sources to the plot,
%SU Aur and V982 Ori, where the ages of a few Myr are a bit better determined. 
The plot shows that similar age stars have similar surface fluxes. There may be a possible trend of increasing (coronal) 
X-ray surface brightness with PMS age. Such a trend would seem to be consistent with studies of the evolutionary behavior of 
TTS X-ray emission dating at back to \citet{kastner1997} in the TW Hydra Association. 
There are two exceptions. V495 Ori exhibited a giant flare that lasted for a week;
V491 Ori is a highly absorbed persistent source that will need special attention.

\subsubsection{Absorption Column Densities}
\label{sect:nhav}

\begin{figure}
    \centering
    \includegraphics[width=0.8\columnwidth]{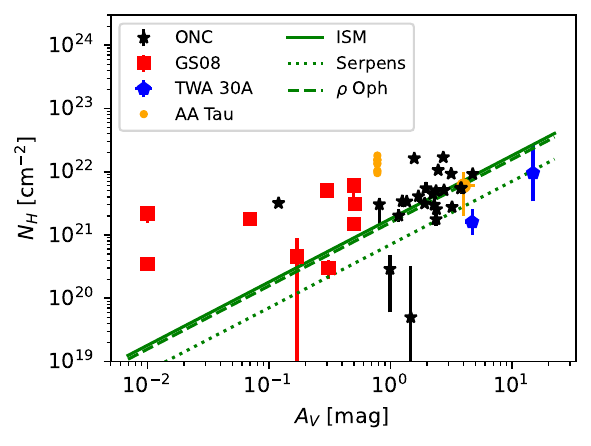}
    \caption{The N$_H$ from the APED fits plotted versus the N$_H$ determined from
    optical extinction $A_V$
    in comparison to AA~Tau (the small dots without error bars are measurements before the dimming) and TWA~30A. Red squares are from the the sample from \citet[][GS08]{guenther2008}. Green lines show the $N_\mathrm{H}/A_V$ ratio observed in the ISM and the average value for two other star forming regions. Data sources are given in section~\ref{sect:nhav}. Only for AA~Tau and TWA~30A extinction and absorption data are contemporaneous, while all other cases rely on  optical and X-ray data taken non-contemporaneously.
    An interactive version of this figure is available in the online journal with the ability to zoom, pan, and display name and additional information for each source.
    Clicking on the legend entries mutes/unmutes the data for better visibility.}
    \label{fig:nhav}
\end{figure}

The global fits resulted in column densities $N_H$ between a few times 10$^{21}$ cm$^{-2}$ and 
a few times 10$^{22}$ cm$^{-2}$. The largest column was observed in COUP 450 with 
1.3$\times 10^{22}$ cm$^{-2}$. LQ Ori exhibits the lowest column consistent with a value below
10$^{20}$ cm$^{-2}$. The column density towards the ONC is estimated to be $\sim 2.3\times10^{21}$ cm$^{-2}$
(see discussion in \citealt{schulz2015}) which implies most of the excess absorption observed is likely intrinsic to the stellar systems.
In Fig.~\ref{fig:nhav} we plot the measured X-ray absorption column versus the
optical extinction.   

The figure also shows other young stars from the literature for comparison. The sample of \citet{guenther2008} concentrates 
on stars that are observed with high-resolution X-ray spectroscopy similar to our sample from the ONC. The 
figure also
displays two stars where the absorbing column density and the optical extinction have been observed to 
change with time, 
in particular in \object{TWA 30A} \citep{principe2016} and \object{AA Tau} \citep{grosso2007, 
schneider2015}. Green 
lines indicate $N_\mathrm{H}/A_V$ ratios from the ISM and two star forming regions from \citet{vuong2003}; 
for the ONC 
those authors have only a very small sample with large uncertainties that appears compatible with the ISM.

To provide an independent means of estimating N$_H$, we compared the flux in the Ne X alpha line with the 
flux in the Ne X 
beta line.  The Ne X lines are relatively strong in the spectra of our sources and the wavelength separation 
of the alpha 
and beta lines is adequate to estimate N$_H$.  
A two-temperature APED model was used for the continuum in each case and the emission lines were fit with 
Gaussian profiles.  
The ratio was used to interpolate the N$_H$ transmission curves. The Ne-based $N_H$ values are consistent 
with the ones from the APED fits.

\subsubsection{Coronal Temperatures}

Table~\ref{tab:3aped} shows all the APED temperatures of the spectral fits. Most spectra required
two APED components with moderate absorption. About half a dozen sources are so
absorbed that we only detect one hot component. The sources with low or moderate temperatures produced a moderately hot
APED component of 6 to 19 MK. The temperatures of APED components are determined by the observed
relative line strengths within an ion species and the strength of the underlying continuum. The 
uncertainties of this temperature component are relatively small indicating it is well determined
specifically due a high number of contributing lines. In that respect the spread in temperature 
between the ONC stars is likely real. 
 
\begin{figure}
    \centering
    \includegraphics[width=0.8\columnwidth]{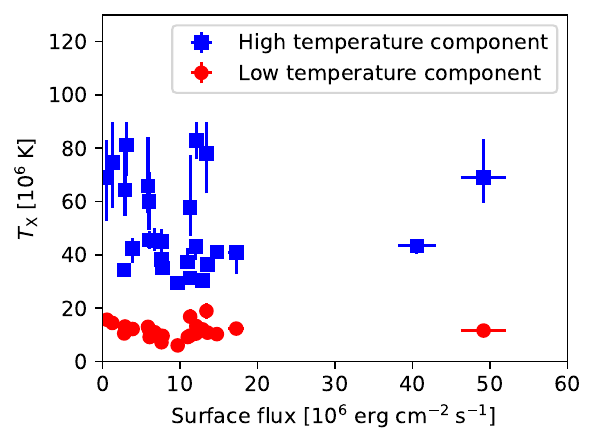}
    \caption{The coronal temperatures from the APED fits versus the surface
    flux.
    An interactive version of this figure is available in the online journal with the ability to zoom, pan, and display name and additional information for each source.}
    \label{fig:kt}
\end{figure}

Figure ~\ref{fig:kt} plots all temperatures against surface flux. The very hot component not only shows quite
a large scatter between 30 MK and 90 MK, but likely a bifurcation of values. It shows the presence 
of two temperature regimes, one between 30 and 50 MK, and a very hot one between 60 MK and 90 MK.
While in the case of the hot components there are a few supporting lines from Si, S, Ar and Ca, the
very hot component at best has line contributions from Fe XXV and Fe XXVI but is mostly defined by
the continuum. Of the highly absorbed stars there is only one, COUP 662, that exhibited an 
extremely high temperature component at 89 MK. At such high temperatures no lines will be detected
as the plasma is completely ionized. Another interesting case is V495 Ori, which is bright in only two observations
and exhibits a giant flare. Its shows a moderate and a very hot component of 69 MK indicating that
sources with very hot components likely engage in heavy flaring. 

It is also important to consider the underlying emission measure contributions. For the 
two components we measure values between a few times 10$^{53}$ cm$^{-3}$ and 10$^{54}$ cm$^{-3}$.
This shows that the ensemble of coronal stars exhibit fairly consistent properties.
These are, except for MT Ori, slightly smaller than the ones determined in the early observations
~\citep{schulz2015}, but not by much. However, there are some significant trends with respect to 
X-ray temperature. The first is that on average the emission measures of the low temperature
component ($\sim$ 10 MK) is about a factor 2-3 smaller than that  of the hot component ($\sim$ 40 MK).
This is not the case for the very hot component ($>$ 60 MK) which is similar or even lower in value than the
one associated with the low temperature component. Thus it appears that all three X-ray temperature
regimes possess distinct properties with respect to their coronal nature with respect to emission volume
and maybe even plasma densities.

\begin{figure}
    \centering
    \includegraphics[width=0.8\columnwidth]{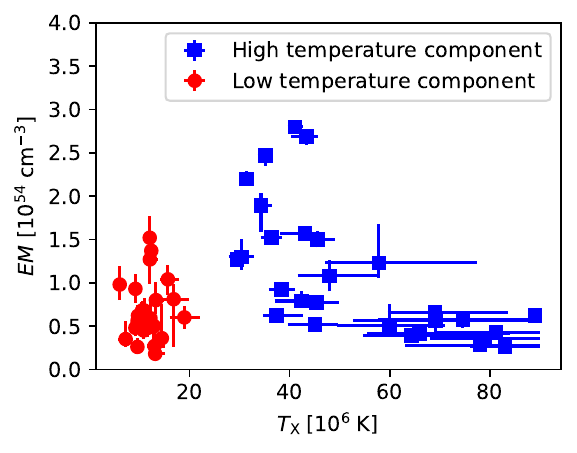}
    \caption{The emission measures from the APED fits plotted versus the temperatures from the fits. This figure is zoomed in to avoid large values in MT~Ori and V450~Ori.
    An interactive version of this figure is available in the online journal with the ability to zoom, pan, and display name and additional information for each source.}
    \label{fig:em}
\end{figure}

One item we did not pursue in this global coronal analysis  is a more detailed study of abundances, which
should be done in more detailed followup studies of this coronal sample. However we did perform a fit of APED abundance values,
which can be useful already. However, when we determine  a set of average values we need to optimize the sample.
For example, for highly absorbed sources values for Ne and Mg are more unreliable because only a few weak lines might exist.
Similarly we might exclude high Z element abundances from the subset of very high temperature components because
lines are weak and/or likely only some Fe K lines exist. In calculating the average values for the remaining sources
we also drop the highest and lowest values to remove some bias where the fit was unable to make a sensible 
determination. The average abundance distribution for the coronal fits then yields the following
values with respect to solar (\citet{anders1989}): Ne (1.52+/-0.70), Mg (0.18+/-0.16), Si (0.20+/-0.13), S (0.28+/-0.21),
Ar (0.59+/-0.45), Ca (0.26+/-0.25), Fe (0.14+/-0.18). The +/- values are not uncertainties but the variance from the average
in the sample. These abundance ranges are consistent with the abundances stated in \citet{schulz2015} (see also
\citet{maggio2007} for COUP results). 

\subsubsection{Line Broadening}

%The broadband fits were conducted with unbroadend line contributions, i.e.  APED lines were 
%treated as delta functions with no thermal and turbulent contributions. This was quite 
%warranted as the continua were dominating the fit and thermal and turbulent contributions
%are expected to be relatively small in coronal sources. It also guarantees that specifically
%in the hotter components, where lines are weak and absent, line widths do not run away 
%during the fits as dilute the continuum fit. 

\begin{figure*}
    \centering
    \includegraphics[width=0.8\columnwidth]{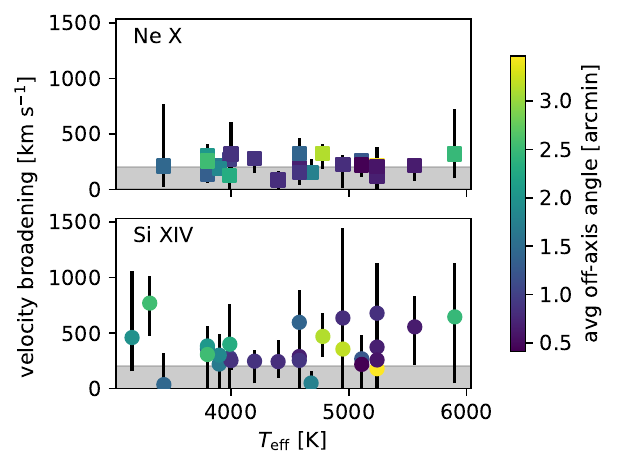}
    \includegraphics[width=0.8\columnwidth]{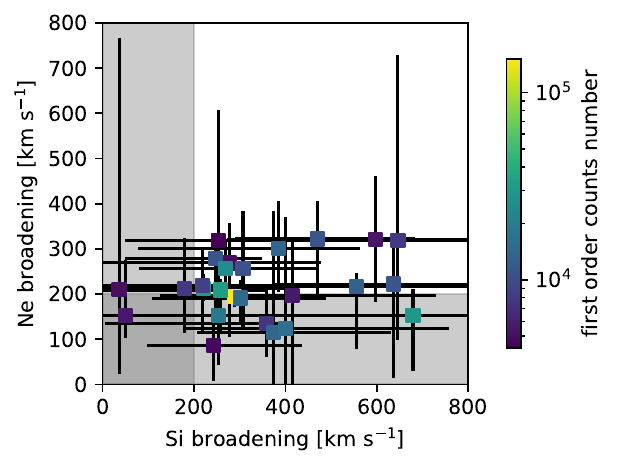}
    \caption{The measured line widths from the single line fits. Gray boxes show the regions where the measured line width cannot be distinguished from instrumental broadening alone.
    \emph{left:} Velocity broadening vs. effective temperature. The symbol color indicates how far off axis the source is located, averaged over all observations.
    \emph{right:} Comparing the velocity broadening in Ne and Si directly. The symbol color indicates the number of counts in the zeroth order for each source.
    An interactive version of this figure is available in the online journal with the ability to zoom, pan, and display name and additional information for each source.
    Clicking on the legend entries mutes/unmutes the data for better visibility.}
    \label{fig:velocities}
\end{figure*}

We also performed separate line fits to the bright lines
in the spectrum. The best cases were the Ne X and the Si XIV lines, both H-like single line
systems. We ignored the spin-orbit coupling and fitted these lines
with single Gaussian line functions at the appropriate wavelength. Here we also have to worry about the spatial distribution
of sources. The stars in Tab.~\ref{tab:master} distribute around the aim-point within about 
3 arcmin radius. The HETG instrument can tolerate zeroth orders to about 2 arcmin off-axis and not suffer
degradation of spectral resolution. This means that about 25$\%$ of the stars in  Tab.~\ref{tab:master}
will suffer some form of spectral degradation. We plotted all line fits
and color coded the off-axis information (left panel of Fig.~\ref{fig:velocities}) and number of counts in the first order (right panel). Stronger sources are generally measured better, but the flux in the relevant lines also depends on the spectral shape. The figure shows a general trend where source with broadened Ne lines also have broadened Si lines, though the error bars are also compatible with no measurable broadening for most sources.

\begin{figure}[t]
  \centering\leavevmode
  \includegraphics[width=0.95\columnwidth]{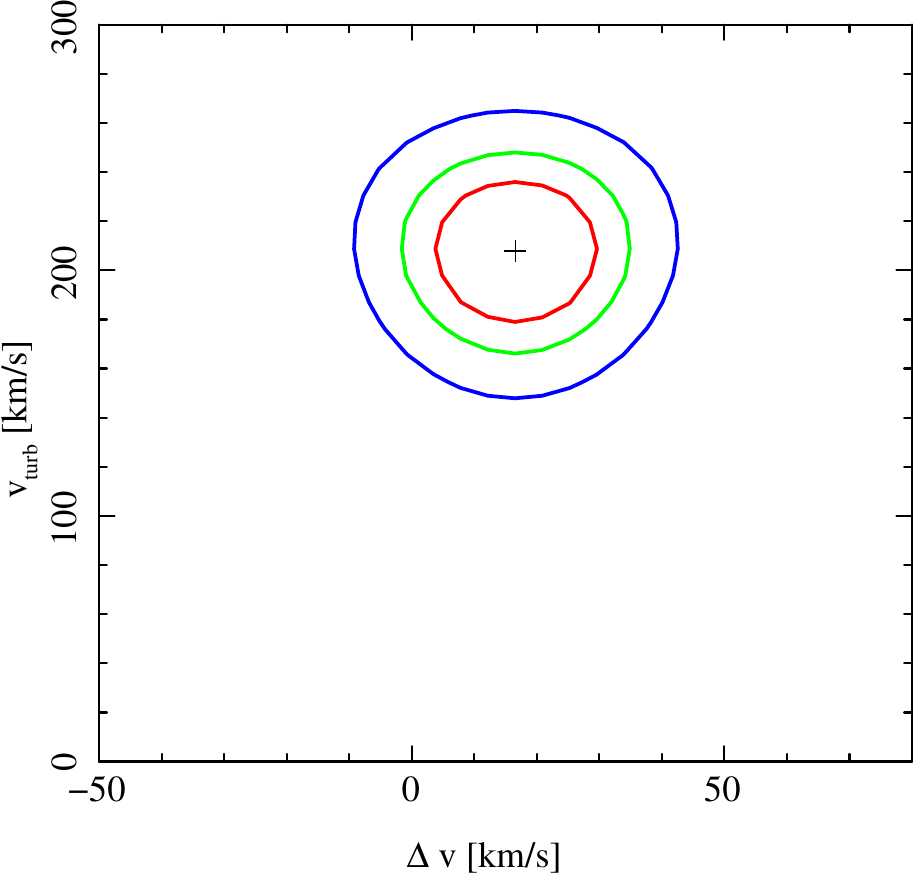}
\caption{The confidence contours for the turbulent broadening term against Doppler shift for a plasma model fit to the $8$--$14\mang$ region of MT Ori.  Contours are for $68\%$, $90\%$, and $99\%$ limits.}
\label{fig:mtorivturb}
\end{figure}

%%% mt ori vturb; dph 2023.07.27
To further quantify line broadening, since we do not expect it in
typical coronal sources, we took one case to investigate in more
detail.  MT Ori has well detected broadening in \eli{Ne}{10}.  We
started with the two-temperature plasma model (see
Table~\ref{tab:3aped}) and allowed the turbulent broadening term and
the redshift to be free parameters and re-fit the merged spectrum over
the $8$--$14\mang$ region where there are many lines from Mg, Ne, and
Fe. We also let the normalization float (but tied the ratio), but kept
the two temperatures frozen.  In addition, we allowed relative
abundances of Mg, Ne, and Fe to be free.  In this way, we implicitly
include all blending implicit in the model, account for thermal
broadening, and determine any excess broadening required to fit the
spectrum.   This confirms the result found for fitting individual
features. We show the confidence contours of the excess broadening
against the Doppler shift in Figure~\ref{fig:mtorivturb}, and contours
are closed.  This is a barely resolved result --- if the broadening
were a bit lower ($v_\mathrm{turb}\gtrsim100\kms$), then the contours
would likely be unbounded on the lower limit.  We suspect that
broadening in this case could be due to orbital motions in a binary
system.

\subsubsection{$\theta^1\,$Ori~E}

\toneE is a spectroscopic binary with a $9.9\,$ day period in which both components,
each a G-type giant, have an intermediate mass of about $2.8\,M_\odot$\citep{morales2012}.
The basic characteristics were reviewed by
\citet{huenemoerder2009}, along with a detailed analysis of the
HETG spectrum. We now have an effective exposure of about $1.5\,$Ms,
compared to the previous $260\ks$. Due to detector efficiency reduction and
source confusion, the largest exposure gains are in the short
wavelength region, below $10\mang$ and we fully realize the expected
increase in signal-to-noise ratio of $2$ or more.  This will allow us to
put better constraints on the highest temperature plasma through the
continuum emission and the emission from the H- and He-like ions of
Si, S, Ar, Ca, and Fe.  Here we provide an overview of the improved
spectrum, with a look at an approximate plasma model, variability, and
line profiles.

\begin{figure*}[ht]
  \centering\leavevmode
  \includegraphics[width=1.8\columnwidth]{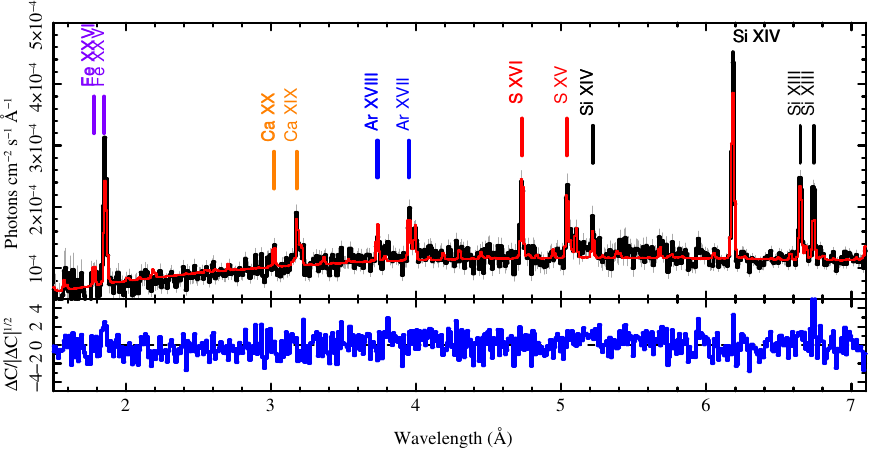}
\caption{The short-wavelength region HETGS spectrum of
  \toneE, having an effective exposure of $1.5\,$Ms. The prominent H-
  and He-like emission lines are labeled.  The flux spectrum is shown
  in black, the model in red, and residuals in the lower panel.  Line label colors are arbitrary.}
\label{fig:t1oeshort}
\end{figure*}

\begin{figure*}[t]
  \centering\leavevmode
  \includegraphics[width=1.8\columnwidth]{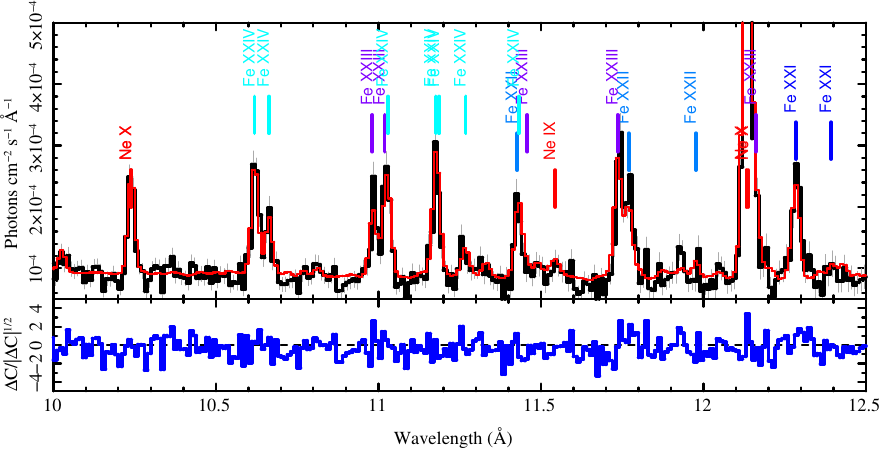}
\caption{The $10$--$13\mang$ region HETGS spectrum of
  \toneE, which is important for establishing the relative Fe and Ni
  abundances, given the emission measure model.  The fluxed spectrum is
shown in black, the model in red, and below are the residuals.
Prominent emission from \eli{Fe}{21} to \eli{Fe}{24} are labeled, as
well as some neon lines (label colors are arbitrary).}
\label{fig:t1oelong}
\end{figure*}

A three-temperature APED model provided an overall characterization of this high brilliance
spectrum, but as we noticed for \toneC, there were large 
residuals that could not be eliminated with few-temperature-component models.
We thus adopted a broken powerlaw emission measure distribution model
which approximates the line-based emission-measure reconstruction of \citet{huenemoerder2009}.
The model parameters are the normalization, the temperature of maximum emission measure, 
and powerlaw slopes below and above that temperature, and relative elemental abundances;
fitted values are given in Table~\ref{tbl:t1oemodel}. Uncertainties for the emission measure
shape were determined from a Monte-Carlo evaluation, with relative
abundances frozen.  The Fe and Ni values were determined post-facto
from confidence levels determined using only the $10$--$13\mang$
region, which has many Fe lines and the brightest Ni lines.  The oxygen
abundance uncertainty was scaled from the flux uncertainty, and is the
most uncertain value due to the low counts in that region, due both to
line-of-sight absorption and detector contamination.  Portions
of the spectra and models are shown in Figure~\ref{fig:t1oeshort} and
\ref{fig:t1oelong}.

\begin{deluxetable}{lc}
  \tablecaption{Broken Powerlaw Emission Measure Model Parameters\label{tbl:t1oemodel}}
  \tablewidth{4in}
  \tablehead{
    \colhead{Parameter} &
    \colhead{Value}
  }
  \startdata
   Norm& $6.3\times10^{-3}$ ($1.0\times10^{-4}$) [$\mathrm{cm^{-5}}$] \\
   $T_\mathrm{max}$& 26.3 (2.4) MK \\
   $\alpha$& $0.9$ ($0.1$)\\
   $\beta$& $-2.5$ ($0.2$) \\
    O&  $0.22$ ($0.08$)\\
    Ne& $0.82$ ($0.08:$)\\
    Mg& $0.30$ ($0.03:$)\\
    Si& $0.22$ ($0.02:$)\\
    S&  $0.25$ ($0.04:$)\\
    Ar& $0.58$ ($0.1:$)\\
    Ca& $0.84$ ($0.2:$)\\
    Fe& $0.17$ ($0.01$)\\
    Ni& $0.11$ ($0.06$)\\
  \enddata
  \tablecomments{Model parameters, for an emission model defined by
    $EM(T) = Norm * (T/T_\mathrm{max})^{a(T)}$; $a(T<T_\mathrm{max}) =
    \alpha$; $a(T \ge T_\mathrm{max})=\beta$.   Elemental abundances are given
    relative by number to the fiducial values of \citet{anders1989}. The
    emission measure and normalization are related in the usual
    scaling: $EM = 10^{14} \times Norm / (4\pi
    d^2)\,[\mathrm{cm}^{-3}]$.
  Abundance uncertainties not formally evaluated, but estimated from
  counts are designated with a ``:''.}
\end{deluxetable} 

The plasma model fits include a ``turbulent'' velocity term and a
redshift. Emission lines were also fit individually with Gaussian
profiles. The lines in the merged spectrum showed significant excess
broadening (in addition to instrumental or thermal terms), having
about $400\kms$ full-width-half-maximum with an uncertainty of
$50\kms$ (corresponding to $v_\mathrm{turb}\approx200\pm30\kms$).  The
maximum orbital radial velocity separation is about $160\kms$.  Since
the spectrum fit was merged over all observations, we expect there to
be some width due to orbital dynamics.  However, the measured width is
somewhat larger than expected from photospheric radial velocities alone.   
The mean profile Doppler shifts are consistent with $0.0\pm30\kms$ (not accounting for heliocentric motion).   
The values are consistent with \citet{huenemoerder2009}, but have smaller
uncertainties.  The widths and offsets definitely need further
scrutiny, especially relative to orbital phase.

\begin{figure}[th]
  \centering\leavevmode
  \includegraphics[width=0.95\columnwidth]{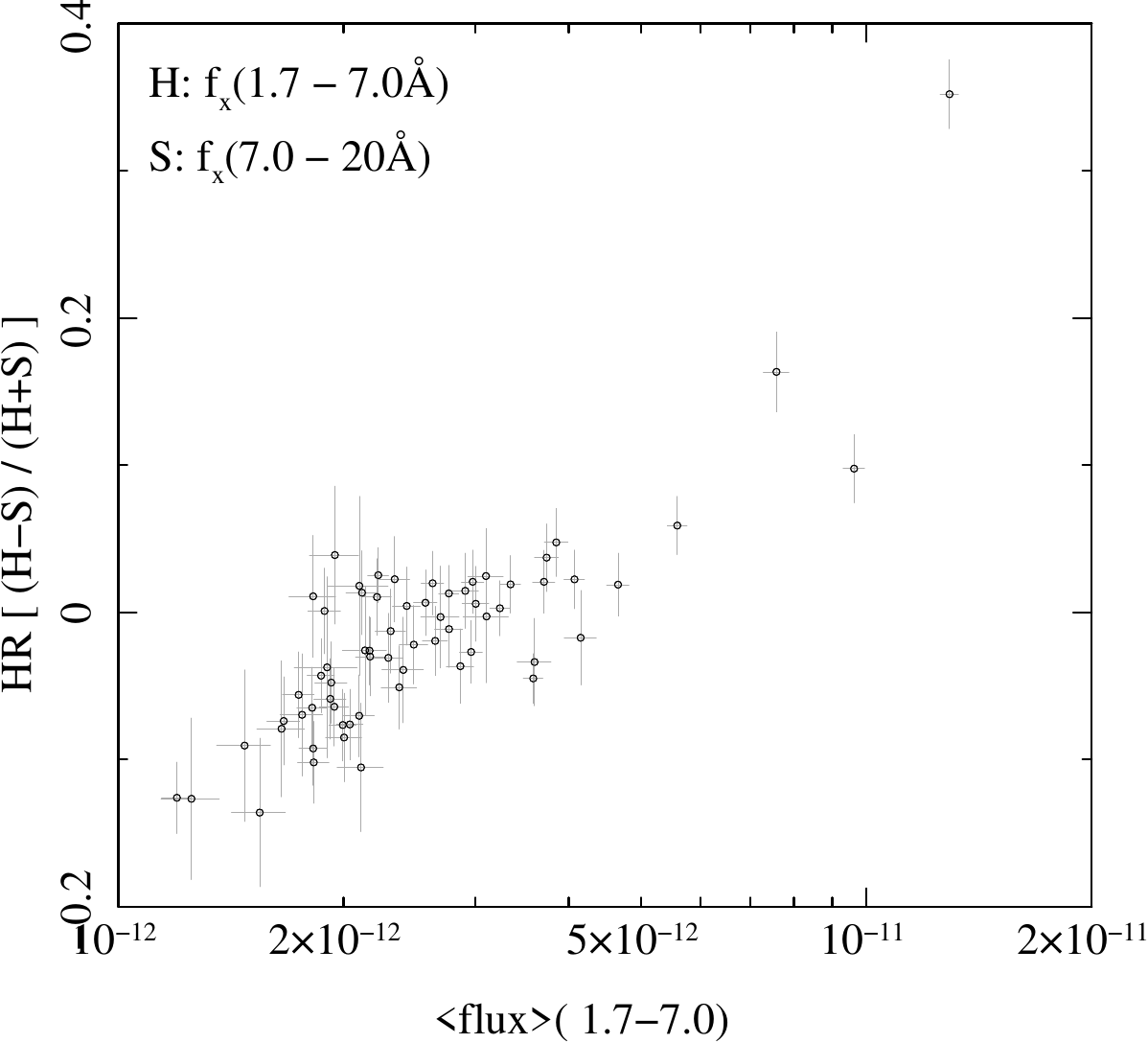}
  \caption{ The hardness ratio vs.\ hard flux for \toneE. Each point
    represents a single observation ID.  A hardness increasing
    directly with flux is a characteristic of stellar coronal flares
    (magnetic reconnection events). Flux is in units of $\mathrm{erg\,cm^{-2}\,s^{-1}}$}.
\label{fig:t1oehardness}
\end{figure}

With this deeper exposure of \toneE we have significantly improved
diagnostics from the $2$--$7\mang$ region, specifically from the
emission lines of Si, S, Ar, and Ca. The broken
powerlaw emission measure model, though, may be too simple, since the
model seems to under-predict \eli{Fe}{25}, as seen in the residuals in
Figure~\ref{fig:t1oeshort}.   

\toneE is highly variable.  As a broad overview of this, we fit the mean flux in a hard band
($1.7$--$7.0\mang$) and in a soft band ($7.0$--$20.0\mang$) and formed
a hardness-ratio, $HR = (H-S)/(H+S)$, where $H$ and $S$ refer the the
hard and soft band fluxes.
in Figure~\ref{fig:t1oehardness} we plot
the $HR$ against $H$ for each individual \chan observation.
This shows over an order of magnitude range in $H$, in a direct
correlation with $HR$.  The flux-hardness trend, is likely due to 
coronal magnetic flare events. This is consistent with  one of the defining characteristics of 
stellar coronal (magnetic) flares, that they are hotter and brighter.

The abundance of Ne seems significantly larger than determined by
\citet{huenemoerder2009}, either due to flaring, or could be due
to emission measure distribution structure, which will require more careful evaluation using
reconstruction using line fluxes, or via exploration
of more complex emission measure models.  Abundances of Ca, Ar, and Ni
are consistent with upper limits of previously determined values, but
now much better constrained.

\section{Summary and Outlook}
\label{sect:summary}
This data set was designed to provide the first collection of high resolution X-ray spectra of a 
very young massive stellar cluster. We were able to harvest about three dozen of high resolution X-ray 
spectra from young massive, intermediate mass, and low mass stars with sufficient statistical properties 
to determine spectral fluxes, coronal temperatures, line widths, line ratios, and abundances. This
data set now provides a unique base of high resolution X-ray spectra of some of the youngest stars known.
The ONC cluster study provides common initial conditions for all extracted objects:
stars are chemically similar, they have young ages, a common ISM evolution and are exposed to fairly
similar global extinction. This first extraction is designed to provide average properties
of extracted stars, it does not yet allow to extract time slices needed for flare studies, for example.

The sample of extracted HETG spectra includes four massive stars. The most prominent star is 
$\theta^1$ Ori C ~\citep{schulz2000, schulz2003, gagne2005} with over 10$^6$ of total counts in
1st order providing for high S/N at the provided oversampling of each HETG resolution element which 
will allow for high brilliance line profile studies and weak line searches (see Gagne et al.\ 2024, in prep.).
One of the most intriguing outcome of the long exposure for this star is the potential use of higher
order grating data, which in this case resolves high Z He-like triplets of Mg, Si, S, Ar Ca,
and Fe with unprecedented high resolution. Plasma density and UV pumping studies should be
highly beneficial for future high resolution missions. $\theta^2$ Ori C is the second most massive
star in the sample and here the survey added only about $10^4$ counts to the previously existing 
data as published in ~\citep{schulz2006, mitschang2011}. The new data should primarily improve the study of 
Mg and Si lines. The zero order data shows high variability in the source indicating
that the star system did engage in flaring activity as reported in ~\citet{schulz2006} and
while the HETG 1st order covers only a fraction of this activity it should prove essential
in the in depth flare analysis. Another interesting
but also unfortunate outcome of the survey is the almost complete absence of $\theta^1$ Ori D in
HETG 1st order and also a surprising weakness in 0th order. The latter is likely a result of the
fact that softer X-rays are blocked by detector contamination. More interesting is the collection
of over 7$\times 10^4$ counts in $\theta^1$ Ori A, a massive trapezium star that is with respect
to primary spectral type as well as number of companions
very similar to $\theta^1$ Ori D, a fact that certainly needs further study.
The fifth massive star in the sample is V1230 Ori, which is not part of the Orion Trapezium and 
farther away in the ONC. Since the survey could not produce any 1st order spectra for 
$\theta^1$ Ori B and $\theta^2$ Ori B, the 2.4$\times 10^4$ counts in HETG 1st order are
the only data to study later B-type massive stars.

The survey also produced over 30 HETG 1st order spectra of intermediate mass and low-mass
CTTS which at their current evolutionary state should exhibit accretion and coronal signatures.
At a canonical age of the ONC of around 1 Myr and older we expect mostly the latter.
The \toneE binary and MV Ori are the most massive among the stars in this sample.
The deeper exposure of \toneE improves the determination of the high-temperature emission measure and elemental abundances 
through the well-detected emission lines of Si, S, Ar, Ca, and Fe in the $1.8$--$7\mang$ spectral region.  The longer exposure
also better quantifies the variability as typical of coronal flares.  Future work is needed to improve the emission
measure distribution, since it probably has more structure than 
the model adopted here, to study the distribution 
at different emission levels to help model flaring structures, and to phase resolve line-shifts and broadening to further
model the emission from each stellar binary component.

We analyzed all the non-massive stars with a two temperature coronal plasma model to characterize their
global coronal properties. From these fits we determined X-ray fluxes between 1.3 $\times 10^{-13}$ \ergcm
and 3.4 $\times 10^{-12}$ \ergcm with the bulk of the fluxes trending more to the low end of
these limits. Most ONC stars are even fainter. 
It is important to note that more exposure would not result in additional sources with successfully extracted
first order spectra as source confusion becomes more dominant. Table~\ref{tab:master} shows that some extraction
efficiencies are already well below 0.5 and final exposures well below 50$\%$ indicating that spectral extraction
becomes very inefficient. In that respect  this survey is going to the limit of what the high resolution
gratings can achieve in a crowded cluster field. 

From the extracted sample we can see that if we 
describe X-ray activity in terms of surface flux, then Fig.~\ref{fig:agesflux} might show that 
activity increases with age in CTTS, even though not as strongly as was suggested in 
\citet{schulz2015}. The surface fluxes of the bulk of the ONC stars appear quite similar to 
other CTTS stars. What is striking in these global fits is the distribution of coronal temperatures.
A large number of ONC stars can be described by a bi-modal temperature distribution, where one
temperature is around 10 MK, the other one more around 40 MK. This is what \citet{schulz2015} observed
in the six bright ONC CTTS and is no surprise. It is observed in many other CTTS outside the ONC,
such as TW Hya \citep{kastner2002}, HD 9880 \citep{kastner2004}, and BP Tau \citep{robrade2006} to
mention a few of the many we know today. Here we see these common properties in almost
three dozen T Tauri stars of a single cluster. What is new is that there is a subsample of sources where the 
high temperature component is more like 60 MK and higher, something that is not expected under 
normal coronal conditions. This definitely requires further study. It is interesting to note
that the emission measures of the two normal temperature components distribute somewhat similar to 
what was projected in ~\citet{schulz2015} but not as extreme, the average between high and low
temperatures differ more  like 2.5 instead of the factor 3 to 6. However, it should be noted
that in the cases of the very high temperatures, the emission measures are systematically low
indicating that here we may deal with high plasma densities and low volumes.
We should add that very high temperatures have been reported with \asca \citep{tsuboi1998} and 
the COUP project ~\citep{getman2005, maggio2007}. Here we confirm the high temperatures with
a high resolution dispersive device.
%However, we stress that such high temperatures 
%detected with CCD devices are prone to systematical uncertainties caused by photon pileup.

%\subsection{Possible contribution by accretion}
CTTS are also characterized by active accretion and in some nearby stars with low absorption, 
accretion signatures are seen prominently in the grating spectra: The line ratios in the He-like 
triplets of \ion{O}{7} and \ion{Ne}{9} have unusually low forbidden to intercombination (f/i) line 
ratios, that can only be explained by high densities in the emission region 
\citep{kastner2002,brickhouse2010}. The observed densities are higher than seen in the corona 
and do not correlate with flares, thus a natural explanation is for the emission to come from 
the cooling flow behind the accretion shock. Unfortunately, the data presented here cannot be
used to test for this as the high contamination on the ACIS camera makes those lines inaccessible 
to us. The other signature of accretion seems to be an excess of soft plasma when comparing 
accreting and non-accreting CTTS \citep{robrade2007,telleschi2007}, which could 
again be a direct signature of the post-shock cooling flow or an indirect effect where the presence of accretion 
columns cools or distorts the fields in the corona \citep{schneider2018}. Again, the low sensitivity of ACIS in 
our observations and the high absorbing column densities for many objects in the ONC make it hard to test for
this conclusively. 

%\subsection{Absorption Column Densities}
The X-ray absorbing column density and the optical/IR extinction (or ``reddening'') probe different 
aspects of the material in the line-of-sight. The optical extinction is typically expressed as dimming 
in a certain band, e.g.\ $A_V$, and it is caused by small dust grains. The X-ray absorption is dominated 
by inner-shell absorption of heavy elements with  contribution from H and He; this absorption occurs 
both in gas and small dust grains. Only large grains that block all energies of X-ray light do not 
change the shape of the observed X-ray spectrum; instead they cause grey absorption that just reduces 
the overall intensity. The X-ray absorbing column density is measured as $N_\mathrm{H}$, the equivalent 
hydrogen column density that would cause the observed absorption for some standard set of elemental 
abundances. A naive interpretation of the $N_\mathrm{H}/A_V$ ratio is that this measures 
the gas-to-dust ratio averaged over the line-of-sight. However, grain growth and non-standard abundances also 
influence the measured ratio and might be different in the accretion columns, the disk atmosphere, the cloud
material, and the ISM between the ONC and Earth.

One promising approach is to study time variability, where the time scale can give us a hint in which region the 
absorber is located. \citet{principe2016} observed a change in $N_\mathrm{H}$ over a month, but with constant 
$N_\mathrm{H}/A_V$ ratio, in \object{TWA 30}. This star is seen nearly edge-on, so we are looking through some 
layer of the disk, with different column density at different times or locations, but constant dust grain 
properties. In \object{AA Tau}, \citet{grosso2007} observed repeated changes of $N_\mathrm{H}$ over 
the 8-day rotation period consistent with a wedge of the inner disk rotating in and out of view; this inner
part of the disk appears gas rich, while an outer ($R > 1$~au) dimming indicates ISM like material 
\citep{schneider2015}. Another prominent example is \object{RW Aur} \citep{guenther2018} which
showed an increase in $N_\mathrm{H}$ by a factor $>100$ over time scales of months to years, clearly related
to major changes in the disk structure. In our general analysis in the ONC, we do not have the time information 
as in these examples, but we can look at the properties of the sample. Figure~\ref{fig:nhav} shows ONC sources 
both above and below the line of an ISM-like $N_\mathrm{H}/A_V$ ratio. While there are certain systematics 
in the measurement of both $N_\mathrm{H}$ and $A_V$, this spread is likely real and represents the different 
viewing geometries. If the line-of-sight passes though a structure close to the star, within the dust 
sublimation radius, $N_\mathrm{H}/A_V$ is large. For a star seen at high inclination angle, the structure 
could be a polar accretion column, while for stars at lower inclination, the inner disk might contribute. 
On the other hand, stars seen through the outer disk might have more evolved dust grains, leading to 
low $N_\mathrm{H}/A_V$ values. Since disks are dynamic, this processed dust can be lifted into higher layers 
of the disk and might be in the line-of-sight, even we do not view the star through the disk mid-plane.

Of the 45 stars in this study, 33 are known variables, 7 are suspected variables, 4 have not been identified 
as variable, and one was excluded due to pileup in the zeroth order.  These statistics will allow us to carry out 
a time-resolved analysis of a significant number of flares in a population of cool stars of approximately the 
same age.  For several of the identified flares, high resolution spectra can be obtained starting before the 
flare begins to be visible in the X-rays and continue through the end of the X-ray emission of the flare.  
Such an analysis technique is rare in the study of flares due to their unpredictable nature.  
The flare spectra will be analyzed to determine spectral changes during the flares.

The percent of obsids that are probably or definitely variable for each source ranges from 0\% to 36\%.
Of course, the longer the exposure time of an obsid, the greater the possibility of detecting variability. 
But the statistics for each source include the same set of obsids (except those with known zeroth order 
confusion) of the same exposure time, so the variability percent is relevant and should be considered 
in concert with the presence of flares and periodicity. Light curves produced by the glvary tool 
for each obsid are being evaluated to verify the timing and duration of flares detected by the statistical method.

\vspace{5mm}

%\begin{acknowledgments}
This project was funded by Chandra grant GO0-21015A and  by NASA through the Smithsonian Astrophysical Observatory (SAO) 
contract SV3-73016 to MIT for Support of the Chandra X-Ray Center (CXC) and Science Instruments. CXC is operated by SAO 
for and on behalf of NASA under contract NAS8-03060.  N.S.~would
like to thank MKI postdoctoral associate Jun Yang for participation and comments. J.N.~acknowledges the assistance of Thomas Firnhaber (University of Kansas) in the zeroth order variability analysis.
The research of T.P.~was partly supported by the Deutsche Forschungsgemeinschaft (DFG, German Research Foundation) 
in the research unit \textit{Transition Discs} (Ref no.~325594231 - FOR~2634/2 - TE 1024/2-1) and 
in the  Excellence
Cluster \textit{ORIGINS} - EXC 2094 - 390783311. T.P.~would like to thank the LMU PhD students S.~Flaischlen and
C.~G\"oppl for assistance in the preparation of data tables for this project. The team would also like to thank Wayne Waldron for
helpful comments early in the project.
%\end{acknowledgments}

\bibliography{orion_vlp}{}

\begin{thebibliography}{}
\expandafter\ifx\csname natexlab\endcsname\relax\def\natexlab#1{#1}\fi

\bibitem[{{Anders} \& {Grevesse}(1989)}]{anders1989}
{Anders}, E., \& {Grevesse}, N. 1989, \gca, 53, 197

\bibitem[{{Argiroffi} {et~al.}(2007){Argiroffi}, {Maggio}, \&
  {Peres}}]{argiroffi2007}
{Argiroffi}, C., {Maggio}, A., \& {Peres}, G. 2007, \aap, 465, L5

\bibitem[{{Argiroffi} {et~al.}(2012){Argiroffi}, {Maggio}, {Montmerle},
  {Huenemoerder}, {Alecian}, {Audard}, {Bouvier}, {Damiani}, {Donati},
  {Gregory}, {G{\"u}del}, {Hussain}, {Kastner}, \& {Sacco}}]{argiroffi2012}
{Argiroffi}, C., {Maggio}, A., {Montmerle}, T., {et~al.} 2012, \apj, 752, 100

\bibitem[{{Bally} {et~al.}(2000){Bally}, {O'Dell}, \&
  {McCaughrean}}]{bally2000}
{Bally}, J., {O'Dell}, C.~R., \& {McCaughrean}, M.~J. 2000, \aj, 119, 2919

\bibitem[{{Brickhouse} {et~al.}(2010){Brickhouse}, {Cranmer}, {Dupree}, {Luna},
  \& {Wolk}}]{brickhouse2010}
{Brickhouse}, N.~S., {Cranmer}, S.~R., {Dupree}, A.~K., {Luna}, G.~J.~M., \&
  {Wolk}, S. 2010, \apj, 710, 1835

\bibitem[{{Broos} {et~al.}(2007){Broos}, {Feigelson}, {Townsley}, {Getman},
  {Wang}, {Garmire}, {Jiang}, \& {Tsuboi}}]{broos2007}
{Broos}, P.~S., {Feigelson}, E.~D., {Townsley}, L.~K., {et~al.} 2007, \apjs,
  169, 353

\bibitem[{{Broos} {et~al.}(2011){Broos}, {Townsley}, {Feigelson}, {Getman},
  {Garmire}, {Preibisch}, {Smith}, {Babler}, {Hodgkin}, {Indebetouw}, {Irwin},
  {King}, {Lewis}, {Majewski}, {McCaughrean}, {Meade}, \&
  {Zinnecker}}]{broos2011}
{Broos}, P.~S., {Townsley}, L.~K., {Feigelson}, E.~D., {et~al.} 2011, \apjs,
  194, 2

\bibitem[{{Canizares} {et~al.}(2000){Canizares}, {Huenemoerder}, {Davis},
  {Dewey}, {Flanagan}, {Houck}, {Markert}, {Marshall}, {Schattenburg},
  {Schulz}, {Wise}, {Drake}, \& {Brickhouse}}]{canizares2000}
{Canizares}, C.~R., {Huenemoerder}, D.~P., {Davis}, D.~S., {et~al.} 2000,
  \apjl, 539, L41

\bibitem[{{Cash}(1979)}]{cash1979}
{Cash}, W. 1979, \apj, 228, 939

\bibitem[{{Crowther} {et~al.}(2022){Crowther}, {Broos}, {Townsley}, {Pollock},
  {Tehrani}, \& {Gagn{\'e}}}]{crowther2022}
{Crowther}, P.~A., {Broos}, P.~S., {Townsley}, L.~K., {et~al.} 2022, \mnras,
  515, 4130

\bibitem[{{Da Rio} {et~al.}(2010){Da Rio}, {Robberto}, {Soderblom}, {Panagia},
  {Hillenbrand}, {Palla}, \& {Stassun}}]{dario2010}
{Da Rio}, N., {Robberto}, M., {Soderblom}, D.~R., {et~al.} 2010, \apj, 722,
  1092

\bibitem[{{den Boggende} {et~al.}(1978){den Boggende}, {Mewe}, {Gronenschild},
  {Heise}, \& {Grindlay}}]{denboggende1978}
{den Boggende}, A.~J.~F., {Mewe}, R., {Gronenschild}, E.~H.~B.~M., {Heise}, J.,
  \& {Grindlay}, J.~E. 1978, \aap, 62, 1

\bibitem[{{Espaillat} {et~al.}(2021){Espaillat}, {Robinson}, {Romanova},
  {Thanathibodee}, {Wendeborn}, {Calvet}, {Reynolds}, \&
  {Muzerolle}}]{espaillat2021}
{Espaillat}, C.~C., {Robinson}, C.~E., {Romanova}, M.~M., {et~al.} 2021, \nat,
  597, 41

\bibitem[{{Fabricius} {et~al.}(2021){Fabricius}, {Luri}, {Arenou}, {Babusiaux},
  {Helmi}, {Muraveva}, {Reyl{\'e}}, {Spoto}, {Vallenari}, {Antoja}, {Balbinot},
  {Barache}, {Bauchet}, {Bragaglia}, {Busonero}, {Cantat-Gaudin}, {Carrasco},
  {Diakit{\'e}}, {Fabrizio}, {Figueras}, {Garcia-Gutierrez}, {Garofalo},
  {Jordi}, {Kervella}, {Khanna}, {Leclerc}, {Licata}, {Lambert}, {Marrese},
  {Masip}, {Ramos}, {Robichon}, {Robin}, {Romero-G{\'o}mez}, {Rubele}, \&
  {Weiler}}]{fabricius2021}
{Fabricius}, C., {Luri}, X., {Arenou}, F., {et~al.} 2021, \aap, 649, A5

\bibitem[{{Feigelson} \& {Decampli}(1981)}]{feigelson1981}
{Feigelson}, E.~D., \& {Decampli}, W.~M. 1981, \apjl, 243, L89

\bibitem[{{Feigelson} {et~al.}(2005){Feigelson}, {Getman}, {Townsley},
  {Garmire}, {Preibisch}, {Grosso}, {Montmerle}, {Muench}, \&
  {McCaughrean}}]{feigelson2005}
{Feigelson}, E.~D., {Getman}, K., {Townsley}, L., {et~al.} 2005, \apjs, 160,
  379

\bibitem[{{Feigelson} {et~al.}(2011){Feigelson}, {Getman}, {Townsley}, {Broos},
  {Povich}, {Garmire}, {King}, {Montmerle}, {Preibisch}, {Smith}, {Stassun},
  {Wang}, {Wolk}, \& {Zinnecker}}]{feigelson2011}
{Feigelson}, E.~D., {Getman}, K.~V., {Townsley}, L.~K., {et~al.} 2011, \apjs,
  194, 9

\bibitem[{{Fruscione} {et~al.}(2006){Fruscione}, {McDowell}, {Allen},
  {Brickhouse}, {Burke}, {Davis}, {Durham}, {Elvis}, {Galle}, {Harris},
  {Huenemoerder}, {Houck}, {Ishibashi}, {Karovska}, {Nicastro}, {Noble},
  {Nowak}, {Primini}, {Siemiginowska}, {Smith}, \& {Wise}}]{fruscione2006}
{Fruscione}, A., {McDowell}, J.~C., {Allen}, G.~E., {et~al.} 2006, in Presented
  at the Society of Photo-Optical Instrumentation Engineers (SPIE) Conference,
  Vol. 6270, SPIE Conference Series

\bibitem[{{Gagne} {et~al.}(1995){Gagne}, {Caillault}, \&
  {Stauffer}}]{gagne1995}
{Gagne}, M., {Caillault}, J.-P., \& {Stauffer}, J.~R. 1995, \apj, 445, 280

\bibitem[{{Gagn{\'e}} {et~al.}(2005){Gagn{\'e}}, {Oksala}, {Cohen}, {Tonnesen},
  {ud-Doula}, {Owocki}, {Townsend}, \& {MacFarlane}}]{gagne2005}
{Gagn{\'e}}, M., {Oksala}, M.~E., {Cohen}, D.~H., {et~al.} 2005, \apj, 628, 986

\bibitem[{{Gagn{\'e}} {et~al.}(2011){Gagn{\'e}}, {Fehon}, {Savoy}, {Cohen},
  {Townsley}, {Broos}, {Povich}, {Corcoran}, {Walborn}, {Remage Evans},
  {Moffat}, {Naz{\'e}}, \& {Oskinova}}]{gagne2011}
{Gagn{\'e}}, M., {Fehon}, G., {Savoy}, M.~R., {et~al.} 2011, \apjs, 194, 5

\bibitem[{{Getman} {et~al.}(2005){Getman}, {Flaccomio}, {Broos}, {Grosso},
  {Tsujimoto}, {Townsley}, {Garmire}, {Kastner}, {Li}, {Harnden}, {Wolk},
  {Murray}, {Lada}, {Muench}, {McCaughrean}, {Meeus}, {Damiani}, {Micela},
  {Sciortino}, {Bally}, {Hillenbrand}, {Herbst}, {Preibisch}, \&
  {Feigelson}}]{getman2005}
{Getman}, K.~V., {Flaccomio}, E., {Broos}, P.~S., {et~al.} 2005, \apjs, 160,
  319

\bibitem[{{Giacconi} {et~al.}(1972){Giacconi}, {Murray}, {Gursky}, {Kellogg},
  {Schreier}, \& {Tananbaum}}]{giacconi1972}
{Giacconi}, R., {Murray}, S., {Gursky}, H., {et~al.} 1972, \apj, 178, 281

\bibitem[{{GRAVITY Collaboration} {et~al.}(2018){GRAVITY Collaboration},
  {Karl}, {Pfuhl}, {Eisenhauer}, {Genzel}, {Grellmann}, {Habibi}, {Abuter},
  {Accardo}, {Amorim}, {Anugu}, {{\'A}vila}, {Benisty}, {Berger}, {Blind},
  {Bonnet}, {Bourget}, {Brandner}, {Brast}, {Buron}, {Caratti O Garatti},
  {Chapron}, {Cl{\'e}net}, {Collin}, {Coud{\'e} Du Foresto}, {de Wit}, {de
  Zeeuw}, {Deen}, {Delplancke-Str{\"o}bele}, {Dembet}, {Derie}, {Dexter},
  {Duvert}, {Ebert}, {Eckart}, {Esselborn}, {F{\'e}dou}, {Finger}, {Garcia},
  {Garcia Dabo}, {Garcia Lopez}, {Gao}, {Gendron}, {Gillessen}, {Gont{\'e}},
  {Gordo}, {Gr{\"o}zinger}, {Guajardo}, {Guieu}, {Haguenauer}, {Hans},
  {Haubois}, {Haug}, {Hau{\ss}mann}, {Henning}, {Hippler}, {Horrobin}, {Huber},
  {Hubert}, {Hubin}, {Jakob}, {Jochum}, {Jocou}, {Kaufer}, {Kellner},
  {Kendrew}, {Kern}, {Kervella}, {Kiekebusch}, {Klein}, {K{\"o}hler}, {Kolb},
  {Kulas}, {Lacour}, {Lapeyr{\`e}re}, {Lazareff}, {Le Bouquin}, {L{\'e}na},
  {Lenzen}, {L{\'e}v{\^e}que}, {Lin}, {Lippa}, {Magnard}, {Mehrgan},
  {M{\'e}rand}, {Moulin}, {M{\"u}ller}, {M{\"u}ller}, {Neumann}, {Oberti},
  {Ott}, {Pallanca}, {Panduro}, {Pasquini}, {Paumard}, {Percheron}, {Perraut},
  {Perrin}, {Pfl{\"u}ger}, {Duc}, {Plewa}, {Popovic}, {Rabien}, {Ram{\'\i}rez},
  {Ramos}, {Rau}, {Riquelme}, {Rodr{\'\i}guez-Coira}, {Rohloff}, {Rosales},
  {Rousset}, {Sanchez-Bermudez}, {Scheithauer}, {Sch{\"o}ller}, {Schuhler},
  {Spyromilio}, {Straub}, {Straubmeier}, {Sturm}, {Suarez}, {Tristram},
  {Ventura}, {Vincent}, {Waisberg}, {Wank}, {Widmann}, {Wieprecht}, {Wiest},
  {Wiezorrek}, {Wittkowski}, {Woillez}, {Wolff}, {Yazici}, {Ziegler}, \&
  {Zins}}]{gravity2018}
{GRAVITY Collaboration}, {Karl}, M., {Pfuhl}, O., {et~al.} 2018, \aap, 620,
  A116

\bibitem[{{Gregory} \& {Loredo}(1992)}]{gregory1992}
{Gregory}, P.~C., \& {Loredo}, T.~J. 1992, \apj, 398, 146

\bibitem[{{Grellmann} {et~al.}(2013){Grellmann}, {Preibisch}, {Ratzka},
  {Kraus}, {Helminiak}, \& {Zinnecker}}]{grellmann2013}
{Grellmann}, R., {Preibisch}, T., {Ratzka}, T., {et~al.} 2013, \aap, 550, A82

\bibitem[{{Grosso} {et~al.}(2007){Grosso}, {Bouvier}, {Montmerle},
  {Fern{\'a}ndez}, {Grankin}, \& {Zapatero Osorio}}]{grosso2007}
{Grosso}, N., {Bouvier}, J., {Montmerle}, T., {et~al.} 2007, \aap, 475, 607

\bibitem[{{G{\"u}del} {et~al.}(2011){G{\"u}del}, {Audard}, {Bacciotti}, {Bary},
  {Briggs}, {Cabrit}, {Carmona}, {Codella}, {Dougados}, {Eisl{\"o}ffel},
  {Gueth}, {G{\"u}nther}, {Herczeg}, {Kundurthy}, {Matt}, {Mutel}, {Ray},
  {Schmitt}, {Schneider}, {Skinner}, \& {van Boekel}}]{guedel2011}
{G{\"u}del}, M., {Audard}, M., {Bacciotti}, F., {et~al.} 2011, in Astronomical
  Society of the Pacific Conference Series, Vol. 448, 16th Cambridge Workshop
  on Cool Stars, Stellar Systems, and the Sun, ed. C.~{Johns-Krull}, M.~K.
  {Browning}, \& A.~A. {West}, 617

\bibitem[{{G{\"u}nther} {et~al.}(2006){G{\"u}nther}, {Liefke}, {Schmitt},
  {Robrade}, \& {Ness}}]{guenther2006}
{G{\"u}nther}, H.~M., {Liefke}, C., {Schmitt}, J.~H.~M.~M., {Robrade}, J., \&
  {Ness}, J.-U. 2006, \aap, 459, L29

\bibitem[{{G{\"u}nther} \& {Schmitt}(2008)}]{guenther2008}
{G{\"u}nther}, H.~M., \& {Schmitt}, J.~H.~M.~M. 2008, \aap, 481, 735

\bibitem[{{G{\"u}nther} {et~al.}(2007){G{\"u}nther}, {Schmitt}, {Robrade}, \&
  {Liefke}}]{guenther2007}
{G{\"u}nther}, H.~M., {Schmitt}, J.~H.~M.~M., {Robrade}, J., \& {Liefke}, C.
  2007, \aap, 466, 1111

\bibitem[{{G{\"u}nther} {et~al.}(2018){G{\"u}nther}, {Birnstiel},
  {Huenemoerder}, {Principe}, {Schneider}, {Wolk}, {Dubois}, {Logie}, {Rau}, \&
  {Vanaverbeke}}]{guenther2018}
{G{\"u}nther}, H.~M., {Birnstiel}, T., {Huenemoerder}, D.~P., {et~al.} 2018,
  \aj, 156, 56

\bibitem[{{Hartmann} {et~al.}(2016){Hartmann}, {Herczeg}, \&
  {Calvet}}]{hartmann2016}
{Hartmann}, L., {Herczeg}, G., \& {Calvet}, N. 2016, \araa, 54, 135

\bibitem[{{Herbig} \& {Griffin}(2006)}]{herbig2006}
{Herbig}, G.~H., \& {Griffin}, R.~F. 2006, \aj, 132, 1763

\bibitem[{{Hillenbrand}(1997)}]{hillenbrand1997}
{Hillenbrand}, L.~A. 1997, \aj, 113, 1733

\bibitem[{{Hillenbrand} {et~al.}(2013){Hillenbrand}, {Hoffer}, \&
  {Herczeg}}]{hillenbrand2013}
{Hillenbrand}, L.~A., {Hoffer}, A.~S., \& {Herczeg}, G.~J. 2013, \aj, 146, 85

\bibitem[{Houck \& Denicola(2000)}]{houck2000}
Houck, J.~C., \& Denicola, L.~A. 2000, in Astronomical {{Society}} of the
  {{Pacific Conference Series}}, Vol. 216, Astronomical {{Data Analysis
  Software}} and {{Systems IX}}, ed. N.~Manset, C.~Veillet, \& D.~Crabtree, 591

\bibitem[{{Huenemoerder} {et~al.}(2003){Huenemoerder}, {Canizares}, {Drake}, \&
  {Sanz-Forcada}}]{huenemoerder2003}
{Huenemoerder}, D.~P., {Canizares}, C.~R., {Drake}, J.~J., \& {Sanz-Forcada},
  J. 2003, \apj, 595, 1131

\bibitem[{{Huenemoerder} {et~al.}(2007){Huenemoerder}, {Kastner}, {Testa},
  {Schulz}, \& {Weintraub}}]{huenemoerder2007}
{Huenemoerder}, D.~P., {Kastner}, J.~H., {Testa}, P., {Schulz}, N.~S., \&
  {Weintraub}, D.~A. 2007, \apj, 671, 592

\bibitem[{{Huenemoerder} {et~al.}(2009){Huenemoerder}, {Schulz}, {Testa},
  {Kesich}, \& {Canizares}}]{huenemoerder2009}
{Huenemoerder}, D.~P., {Schulz}, N.~S., {Testa}, P., {Kesich}, A., \&
  {Canizares}, C.~R. 2009, \apj, 707, 942

\bibitem[{{Karl} {et~al.}(2018){Karl}, {Pfuhl}, {Eisenhauer}, {Genzel}, \&
  {Grellmann}}]{karl2018}
{Karl}, M., {Pfuhl}, O., {Eisenhauer}, F., {Genzel}, R., \& {Grellmann}, R.
  e.~a. 2018, \aap, 620, A116

\bibitem[{{Kastner} {et~al.}(2005){Kastner}, {Franz}, {Grosso}, {Bally},
  {McCaughrean}, {Getman}, {Feigelson}, \& {Schulz}}]{kastner2005}
{Kastner}, J.~H., {Franz}, G., {Grosso}, N., {et~al.} 2005, \apjs, 160, 511

\bibitem[{{Kastner} {et~al.}(2004){Kastner}, {Huenemoerder}, {Schulz},
  {Canizares}, {Li}, \& {Weintraub}}]{kastner2004}
{Kastner}, J.~H., {Huenemoerder}, D.~P., {Schulz}, N.~S., {et~al.} 2004, \apjl,
  605, L49

\bibitem[{{Kastner} {et~al.}(2002){Kastner}, {Huenemoerder}, {Schulz},
  {Canizares}, \& {Weintraub}}]{kastner2002}
{Kastner}, J.~H., {Huenemoerder}, D.~P., {Schulz}, N.~S., {Canizares}, C.~R.,
  \& {Weintraub}, D.~A. 2002, \apj, 567, 434

\bibitem[{{Kastner} {et~al.}(1997){Kastner}, {Zuckerman}, {Weintraub}, \&
  {Forveille}}]{kastner1997}
{Kastner}, J.~H., {Zuckerman}, B., {Weintraub}, D.~A., \& {Forveille}, T. 1997,
  Science, 277, 67

\bibitem[{{Kounkel} {et~al.}(2017){Kounkel}, {Hartmann}, {Loinard},
  {Ortiz-Le{\'o}n}, {Mioduszewski}, {Rodr{\'\i}guez}, {Dzib}, {Torres}, {Pech},
  {Galli}, {Rivera}, {Boden}, {Evans}, {Brice{\~n}o}, \& {Tobin}}]{kounke2017}
{Kounkel}, M., {Hartmann}, L., {Loinard}, L., {et~al.} 2017, \apj, 834, 142

\bibitem[{{Kraus} {et~al.}(2009){Kraus}, {Weigelt}, {Balega}, {Docobo},
  {Hofmann}, {Preibisch}, {Schertl}, {Tamazian}, {Driebe}, {Ohnaka}, {Petrov},
  {Sch{\"o}ller}, \& {Smith}}]{kraus2009}
{Kraus}, S., {Weigelt}, G., {Balega}, Y.~Y., {et~al.} 2009, \aap, 497, 195

\bibitem[{{Kuhn} {et~al.}(2019){Kuhn}, {Hillenbrand}, {Sills}, {Feigelson}, \&
  {Getman}}]{kuhn2019}
{Kuhn}, M.~A., {Hillenbrand}, L.~A., {Sills}, A., {Feigelson}, E.~D., \&
  {Getman}, K.~V. 2019, \apj, 870, 32

\bibitem[{{Lamzin}(1998)}]{lamzin1998}
{Lamzin}, S.~A. 1998, Astronomy Reports, 42, 322

\bibitem[{Lindegren(2018)}]{lindegren2018}
Lindegren, L. 2018, gAIA-C3-TN-LU-LL-124

\bibitem[{{Lindegren} {et~al.}(2021){Lindegren}, {Bastian}, {Biermann},
  {Bombrun}, {de Torres}, {Gerlach}, {Geyer}, {Hern{\'a}ndez}, {Hilger},
  {Hobbs}, {Klioner}, {Lammers}, {McMillan}, {Ramos-Lerate},
  {Steidelm{\"u}ller}, {Stephenson}, \& {van Leeuwen}}]{lindegren2021}
{Lindegren}, L., {Bastian}, U., {Biermann}, M., {et~al.} 2021, \aap, 649, A4

\bibitem[{{Maggio} {et~al.}(2007){Maggio}, {Flaccomio}, {Favata}, {Micela},
  {Sciortino}, {Feigelson}, \& {Getman}}]{maggio2007}
{Maggio}, A., {Flaccomio}, E., {Favata}, F., {et~al.} 2007, \apj, 660, 1462

\bibitem[{{Ma{\'\i}z Apell{\'a}niz} {et~al.}(2022){Ma{\'\i}z Apell{\'a}niz},
  {Barb{\'a}}, {Fern{\'a}ndez Aranda}, {Pantaleoni Gonz{\'a}lez}, {Crespo
  Bellido}, {Sota}, \& {Alfaro}}]{maizapellaniz2022}
{Ma{\'\i}z Apell{\'a}niz}, J., {Barb{\'a}}, R.~H., {Fern{\'a}ndez Aranda}, R.,
  {et~al.} 2022, \aap, 657, A131

\bibitem[{{Manara} {et~al.}(2012){Manara}, {Robberto}, {Da Rio}, {Lodato},
  {Hillenbrand}, {Stassun}, \& {Soderblom}}]{manara2012}
{Manara}, C.~F., {Robberto}, M., {Da Rio}, N., {et~al.} 2012, \apj, 755, 154

\bibitem[{{Menten} {et~al.}(2007){Menten}, {Reid}, {Forbrich}, \&
  {Brunthaler}}]{menten2007}
{Menten}, K.~M., {Reid}, M.~J., {Forbrich}, J., \& {Brunthaler}, A. 2007, \aap,
  474, 515

\bibitem[{{Mitschang} {et~al.}(2011){Mitschang}, {Schulz}, {Huenemoerder},
  {Nichols}, \& {Testa}}]{mitschang2011}
{Mitschang}, A.~W., {Schulz}, N.~S., {Huenemoerder}, D.~P., {Nichols}, J.~S.,
  \& {Testa}, P. 2011, \apj, 734, 14

\bibitem[{{Morales-Calder{\'o}n} {et~al.}(2012){Morales-Calder{\'o}n},
  {Stauffer}, {Stassun}, {Vrba}, {Prato}, {Hillenbrand}, {Terebey}, {Covey},
  {Rebull}, {Terndrup}, {Gutermuth}, {Song}, {Plavchan}, {Carpenter},
  {Marchis}, {Garc{\'\i}a}, {Margheim}, {Luhman}, {Angione}, \&
  {Irwin}}]{morales2012}
{Morales-Calder{\'o}n}, M., {Stauffer}, J.~R., {Stassun}, K.~G., {et~al.} 2012,
  \apj, 753, 149

\bibitem[{{Muench} {et~al.}(2002){Muench}, {Lada}, {Lada}, \&
  {Alves}}]{muench2002}
{Muench}, A.~A., {Lada}, E.~A., {Lada}, C.~J., \& {Alves}, J. 2002, \apj, 573,
  366

\bibitem[{{Olivares} {et~al.}(2020){Olivares}, {Sarro}, {Bouy}, {Miret-Roig},
  {Casamiquela}, {Galli}, {Berihuete}, \& {Tarricq}}]{olivares2020}
{Olivares}, J., {Sarro}, L.~M., {Bouy}, H., {et~al.} 2020, \aap, 644, A7

\bibitem[{{Orlando} {et~al.}(2010){Orlando}, {Sacco}, {Argiroffi}, {Reale},
  {Peres}, \& {Maggio}}]{orlando2010}
{Orlando}, S., {Sacco}, G.~G., {Argiroffi}, C., {et~al.} 2010, \aap, 510, A71

\bibitem[{{Orlando} {et~al.}(2013){Orlando}, {Bonito}, {Argiroffi}, {Reale},
  {Peres}, {Miceli}, {Matsakos}, {Stehl{\'e}}, {Ibgui}, {de Sa}, {Chi{\`e}ze},
  \& {Lanz}}]{orlando2013}
{Orlando}, S., {Bonito}, R., {Argiroffi}, C., {et~al.} 2013, \aap, 559, A127

\bibitem[{{Petr} {et~al.}(1998){Petr}, {Coud{\'e} du Foresto}, {Beckwith},
  {Richichi}, \& {McCaughrean}}]{petr1998}
{Petr}, M.~G., {Coud{\'e} du Foresto}, V., {Beckwith}, S. V.~W., {Richichi},
  A., \& {McCaughrean}, M.~J. 1998, \apj, 500, 825

\bibitem[{{Pillitteri} {et~al.}(2013){Pillitteri}, {Wolk}, {Megeath}, {Allen},
  {Bally}, {Gagn{\'e}}, {Gutermuth}, {Hartmann}, {Micela}, {Myers}, {Oliveira},
  {Sciortino}, {Walter}, {Rebull}, \& {Stauffer}}]{pillitteri2013}
{Pillitteri}, I., {Wolk}, S.~J., {Megeath}, S.~T., {et~al.} 2013, \apj, 768, 99

\bibitem[{{Preibisch} {et~al.}(1999){Preibisch}, {Balega}, {Hofmann},
  {Weigelt}, \& {Zinnecker}}]{preibisch1999}
{Preibisch}, T., {Balega}, Y., {Hofmann}, K.-H., {Weigelt}, G., \& {Zinnecker},
  H. 1999, \na, 4, 531

\bibitem[{{Preibisch} {et~al.}(2011){Preibisch}, {Hodgkin}, {Irwin}, {Lewis},
  {King}, {McCaughrean}, {Zinnecker}, {Townsley}, \& {Broos}}]{preibisch2011}
{Preibisch}, T., {Hodgkin}, S., {Irwin}, M., {et~al.} 2011, \apjs, 194, 10

\bibitem[{{Principe} {et~al.}(2016){Principe}, {Sacco}, {Kastner}, {Stelzer},
  \& {Alcal{\'a}}}]{principe2016}
{Principe}, D.~A., {Sacco}, G., {Kastner}, J.~H., {Stelzer}, B., \&
  {Alcal{\'a}}, J.~M. 2016, \mnras, 459, 2097

\bibitem[{{Reale} {et~al.}(2014){Reale}, {Orlando}, {Testa}, {Landi}, \&
  {Schrijver}}]{reale2014}
{Reale}, F., {Orlando}, S., {Testa}, P., {Landi}, E., \& {Schrijver}, C.~J.
  2014, \apjl, 797, L5

\bibitem[{{Reale} {et~al.}(2013){Reale}, {Orlando}, {Testa}, {Peres}, {Landi},
  \& {Schrijver}}]{reale2013}
{Reale}, F., {Orlando}, S., {Testa}, P., {et~al.} 2013, Science, 341, 251

\bibitem[{{Robberto} {et~al.}(2010){Robberto}, {Soderblom}, {Scandariato},
  {Smith}, {Da Rio}, {Pagano}, \& {Spezzi}}]{robberto2010}
{Robberto}, M., {Soderblom}, D.~R., {Scandariato}, G., {et~al.} 2010, \aj, 139,
  950

\bibitem[{{Robrade} \& {Schmitt}(2006)}]{robrade2006}
{Robrade}, J., \& {Schmitt}, J.~H.~M.~M. 2006, \aap, 449, 737

\bibitem[{{Robrade} \& {Schmitt}(2007)}]{robrade2007}
---. 2007, \aap, 473, 229

\bibitem[{{Rots}(2012)}]{rots2012}
{Rots}, A.~H. 2012, in New Horizons in Time Domain Astronomy, ed. E.~{Griffin},
  R.~{Hanisch}, \& R.~{Seaman}, Vol. 285, 402--403

\bibitem[{{Rzaev} {et~al.}(2021){Rzaev}, {Shimansky}, \& {Kolbin}}]{rzaev2021}
{Rzaev}, A.~K., {Shimansky}, V.~V., \& {Kolbin}, A.~I. 2021, \mnras, 504, 3787

\bibitem[{{Sacco} {et~al.}(2010){Sacco}, {Orlando}, {Argiroffi}, {Maggio},
  {Peres}, {Reale}, \& {Curran}}]{sacco2010}
{Sacco}, G.~G., {Orlando}, S., {Argiroffi}, C., {et~al.} 2010, \aap, 522, A55

\bibitem[{{Schmitt} {et~al.}(2005){Schmitt}, {Robrade}, {Ness}, {Favata}, \&
  {Stelzer}}]{schmitt2005}
{Schmitt}, J.~H.~M.~M., {Robrade}, J., {Ness}, J.-U., {Favata}, F., \&
  {Stelzer}, B. 2005, \aap, 432, L35

\bibitem[{{Schneider} {et~al.}(2015){Schneider}, {France}, {G{\"u}nther},
  {Herczeg}, {Robrade}, {Bouvier}, {McJunkin}, \& {Schmitt}}]{schneider2015}
{Schneider}, P.~C., {France}, K., {G{\"u}nther}, H.~M., {et~al.} 2015, \aap,
  584, A51

\bibitem[{{Schneider} {et~al.}(2018){Schneider}, {G{\"u}nther}, {Robrade},
  {Schmitt}, \& {G{\"u}del}}]{schneider2018}
{Schneider}, P.~C., {G{\"u}nther}, H.~M., {Robrade}, J., {Schmitt},
  J.~H.~M.~M., \& {G{\"u}del}, M. 2018, \aap, 618, A55

\bibitem[{{Schneider} {et~al.}(2022){Schneider}, {G{\"u}nther}, \&
  {Ustamujic}}]{schneider2022}
{Schneider}, P.~C., {G{\"u}nther}, H.~M., \& {Ustamujic}, S. 2022, arXiv
  e-prints, arXiv:2207.06886

\bibitem[{{Schulz} {et~al.}(2003){Schulz}, {Canizares}, {Huenemoerder}, \&
  {Tibbets}}]{schulz2003}
{Schulz}, N.~S., {Canizares}, C., {Huenemoerder}, D., \& {Tibbets}, K. 2003,
  \apj, 595, 365

\bibitem[{{Schulz} {et~al.}(2000){Schulz}, {Canizares}, {Huenemoerder}, \&
  {Lee}}]{schulz2000}
{Schulz}, N.~S., {Canizares}, C.~R., {Huenemoerder}, D., \& {Lee}, J.~C. 2000,
  \apjl, 545, L135

\bibitem[{{Schulz} {et~al.}(2015){Schulz}, {Huenemoerder}, {G{\"u}nther},
  {Testa}, \& {Canizares}}]{schulz2015}
{Schulz}, N.~S., {Huenemoerder}, D.~P., {G{\"u}nther}, M., {Testa}, P., \&
  {Canizares}, C.~R. 2015, \apj, 810, 55

\bibitem[{{Schulz} {et~al.}(2006){Schulz}, {Testa}, {Huenemoerder},
  {Ishibashi}, \& {Canizares}}]{schulz2006}
{Schulz}, N.~S., {Testa}, P., {Huenemoerder}, D.~P., {Ishibashi}, K., \&
  {Canizares}, C.~R. 2006, \apj, 653, 636

\bibitem[{{Skrutskie} {et~al.}(2006){Skrutskie}, {Cutri}, {Stiening},
  {Weinberg}, {Schneider}, {Carpenter}, {Beichman}, {Capps}, {Chester},
  {Elias}, {Huchra}, {Liebert}, {Lonsdale}, {Monet}, {Price}, {Seitzer},
  {Jarrett}, {Kirkpatrick}, {Gizis}, {Howard}, {Evans}, {Fowler}, {Fullmer},
  {Hurt}, {Light}, {Kopan}, {Marsh}, {McCallon}, {Tam}, {Van Dyk}, \&
  {Wheelock}}]{skrutskie2006}
{Skrutskie}, M.~F., {Cutri}, R.~M., {Stiening}, R., {et~al.} 2006, \aj, 131,
  1163

\bibitem[{{Telleschi} {et~al.}(2007){Telleschi}, {G{\"u}del}, {Briggs},
  {Audard}, \& {Palla}}]{telleschi2007}
{Telleschi}, A., {G{\"u}del}, M., {Briggs}, K.~R., {Audard}, M., \& {Palla}, F.
  2007, \aap, 468, 425

\bibitem[{{Testa} {et~al.}(2004){Testa}, {Drake}, \& {Peres}}]{testa2004}
{Testa}, P., {Drake}, J.~J., \& {Peres}, G. 2004, \apj, 617, 508

\bibitem[{{Townsley} {et~al.}(2006){Townsley}, {Broos}, {Feigelson}, {Brandl},
  {Chu}, {Garmire}, \& {Pavlov}}]{townsley2006}
{Townsley}, L.~K., {Broos}, P.~S., {Feigelson}, E.~D., {et~al.} 2006, \aj, 131,
  2140

\bibitem[{{Townsley} {et~al.}(2011){Townsley}, {Broos}, {Corcoran},
  {Feigelson}, {Gagn{\'e}}, {Montmerle}, {Oey}, {Smith}, {Garmire}, {Getman},
  {Povich}, {Remage Evans}, {Naz{\'e}}, {Parkin}, {Preibisch}, {Wang}, {Wolk},
  {Chu}, {Cohen}, {Gruendl}, {Hamaguchi}, {King}, {Mac Low}, {McCaughrean},
  {Moffat}, {Oskinova}, {Pittard}, {Stassun}, {ud-Doula}, {Walborn}, {Waldron},
  {Churchwell}, {Nichols}, {Owocki}, \& {Schulz}}]{townsley2011}
{Townsley}, L.~K., {Broos}, P.~S., {Corcoran}, M.~F., {et~al.} 2011, \apjs,
  194, 1

\bibitem[{{Tsuboi} {et~al.}(1998){Tsuboi}, {Koyama}, {Murakami}, {Hayashi},
  {Skinner}, \& {Ueno}}]{tsuboi1998}
{Tsuboi}, Y., {Koyama}, K., {Murakami}, H., {et~al.} 1998, \apj, 503, 894

\bibitem[{{Vilhu}(1984)}]{vilhu1984}
{Vilhu}, O. 1984, \aap, 133, 117

\bibitem[{{Vilhu} \& {Walter}(1987)}]{vilhu1987}
{Vilhu}, O., \& {Walter}, F.~M. 1987, \apj, 321, 958

\bibitem[{{Vuong} {et~al.}(2003){Vuong}, {Montmerle}, {Grosso}, {Feigelson},
  {Verstraete}, \& {Ozawa}}]{vuong2003}
{Vuong}, M.~H., {Montmerle}, T., {Grosso}, N., {et~al.} 2003, \aap, 408, 581

\bibitem[{{Waldron} {et~al.}(2004){Waldron}, {Cassinelli}, {Miller},
  {MacFarlane}, \& {Reiter}}]{waldron2004}
{Waldron}, W.~L., {Cassinelli}, J.~P., {Miller}, N.~A., {MacFarlane}, J.~J., \&
  {Reiter}, J.~C. 2004, \apj, 616, 542

\bibitem[{{Wang} {et~al.}(2008){Wang}, {Townsley}, {Feigelson}, {Broos},
  {Getman}, {Rom{\'a}n-Z{\'u}{\~n}iga}, \& {Lada}}]{Wang2008}
{Wang}, J., {Townsley}, L.~K., {Feigelson}, E.~D., {et~al.} 2008, \apj, 675,
  464

\bibitem[{{Wang} {et~al.}(2007){Wang}, {Townsley}, {Feigelson}, {Getman},
  {Broos}, {Garmire}, \& {Tsujimoto}}]{wang2007}
---. 2007, \apjs, 168, 100

\bibitem[{{Wolk} {et~al.}(2006){Wolk}, {Spitzbart}, {Bourke}, \&
  {Alves}}]{wolk2006}
{Wolk}, S.~J., {Spitzbart}, B.~D., {Bourke}, T.~L., \& {Alves}, J. 2006, \aj,
  132, 1100

\bibitem[{{Wright} {et~al.}(2011){Wright}, {Drake}, {Mamajek}, \&
  {Henry}}]{Wright2011}
{Wright}, N.~J., {Drake}, J.~J., {Mamajek}, E.~E., \& {Henry}, G.~W. 2011,
  \apj, 743, 48

\end{thebibliography}
\bibliographystyle{aasjournal}

\clearpage

\appendix
% \begin{appendices}

\section{Lightcurves \label{app:light}}

%\clearpage
This appendix contains the concatenated, time-ordered zeroth order light curves for each source in Tab. \ref{tab:master}. The data have been binned at 1 ksec and  time gaps between observations have been eliminated.  Time on the x-axis is cumulative observing time since the beginning of the first observation.  Data for obsids where confusion affects the zeroth order count rate have been eliminated in the plots.  Therefore each point on each plot is derived from a non-confused zeroth order. The vertical dotted line in each plot indicates the significant time gap of about 12 years in observations.  The name of each source is in blue in the upper right-hand corner of each plot.  An example plot for LQ Ori is included in the text (Fig. \ref{plt:lq}) and not repeated in this appendix.
\begin{figure*}[t]
\centering
\includegraphics[angle=0,width=16.2cm]{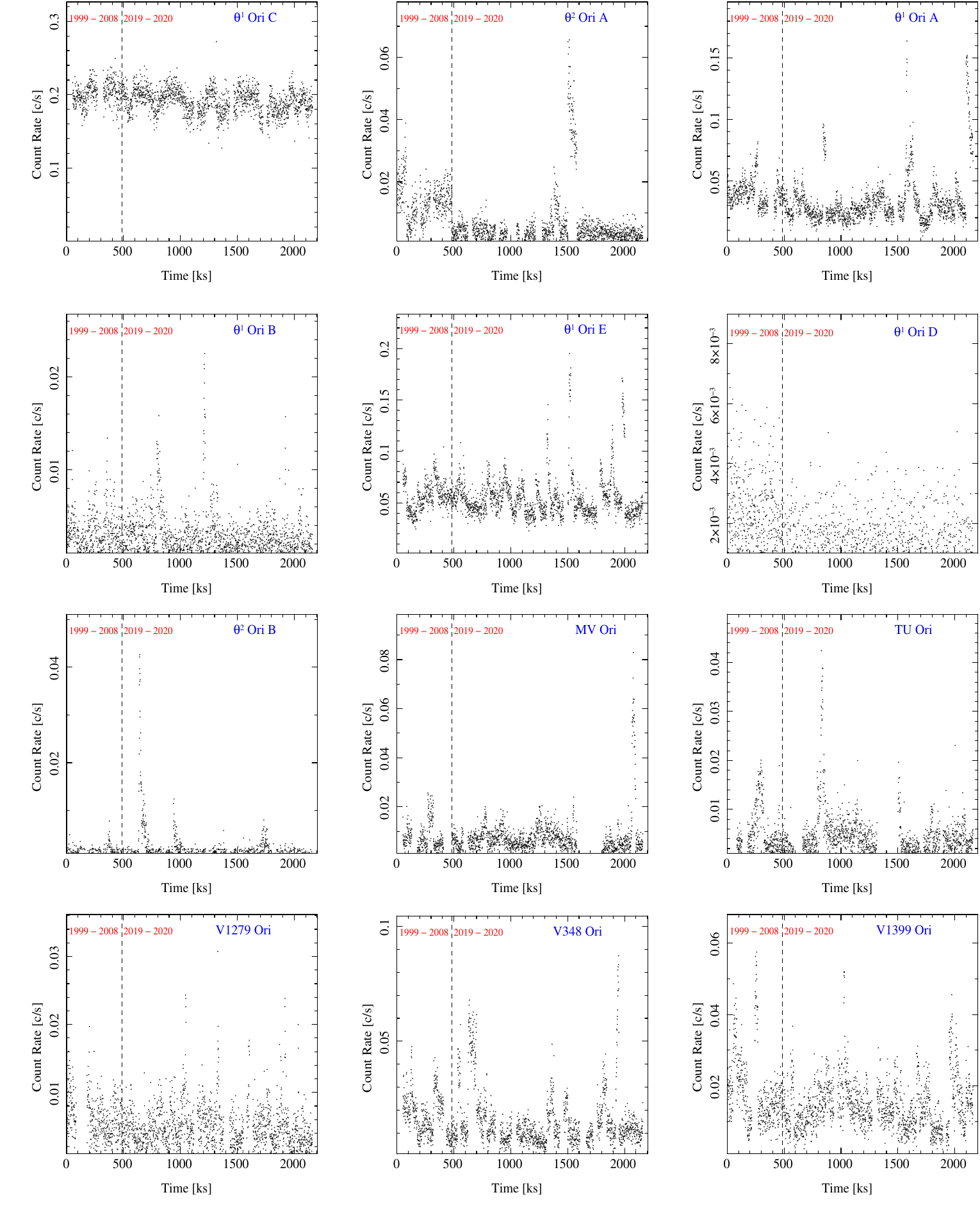}
%\caption{Merged zero order image over the entire exposure using a three color (rgb) scheme reflecting the stars energy spectra. }
\label{fig:joy}
\end{figure*}

\begin{figure*}
\centering
\includegraphics[angle=0,width=16.2cm]{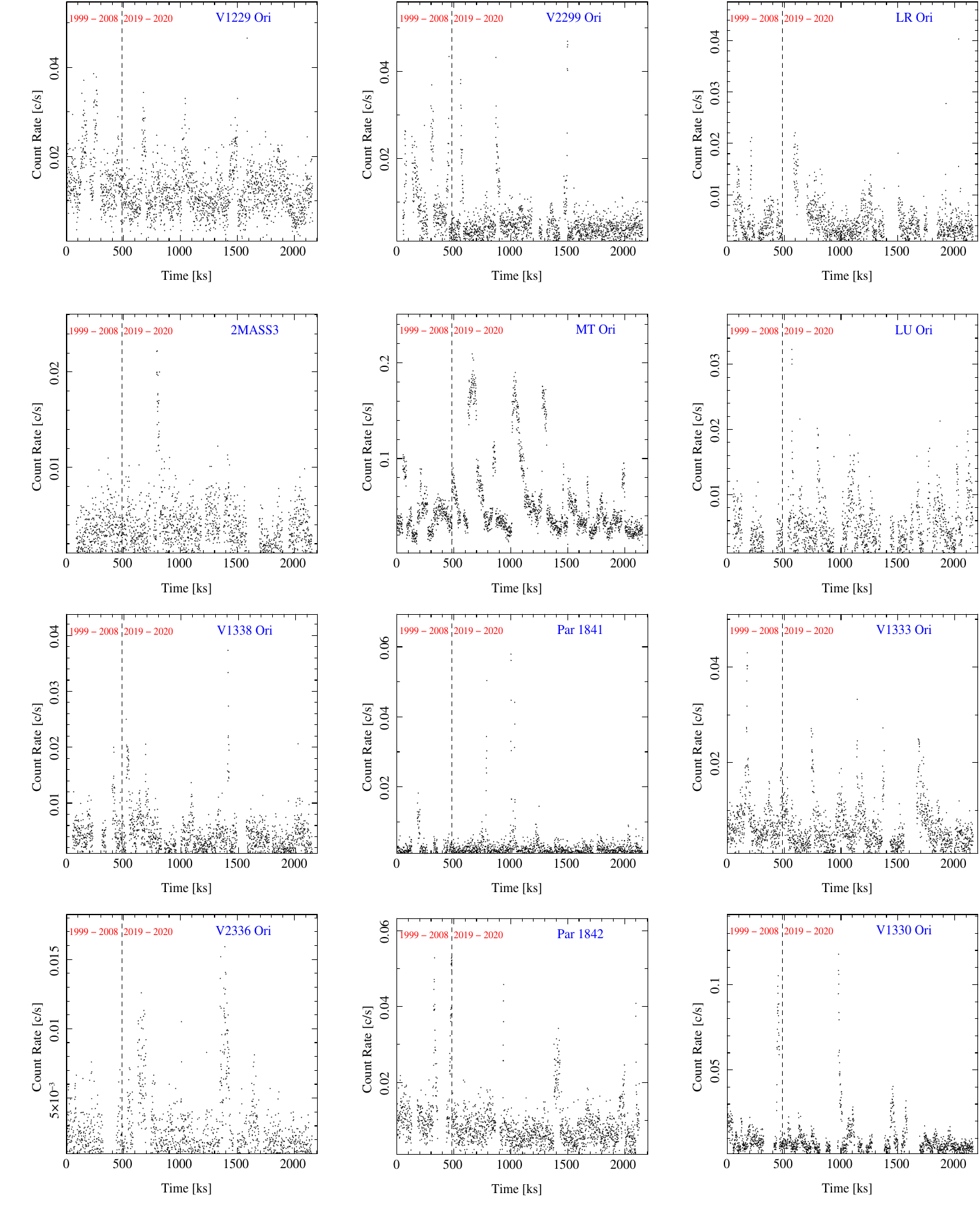}
%\caption{Merged zero order image over the entire exposure using a three color (rgb) scheme reflecting the stars energy spectra. }
\label{fig:joy2}
\end{figure*}

\begin{figure*}
\centering
\includegraphics[angle=0,width=16.2cm]{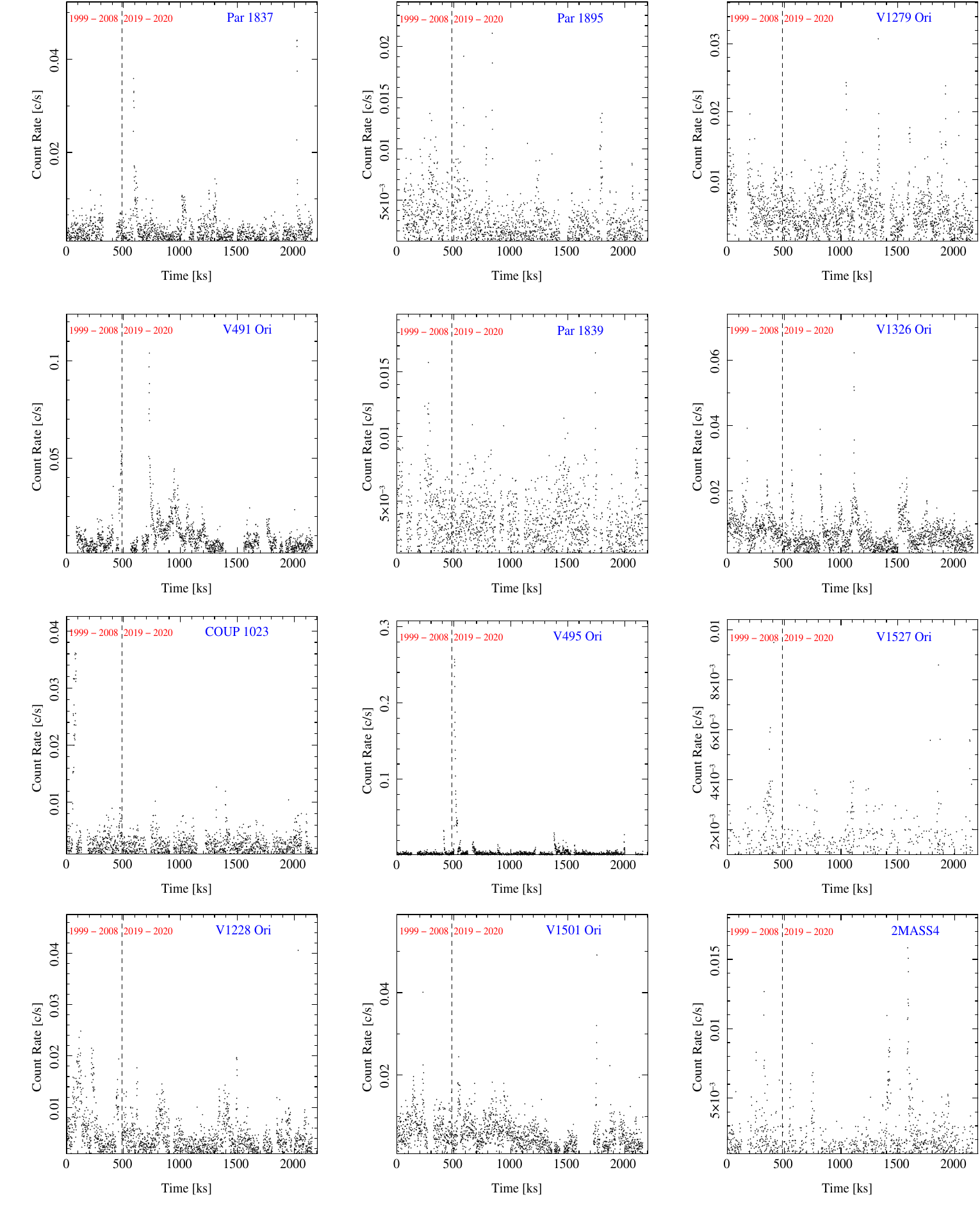}
%\caption{Merged zero order image over the entire exposure using a three color (rgb) scheme reflecting the stars energy spectra. }
\label{fig:joy3}
\end{figure*}

\begin{figure*}
\centering
\includegraphics[angle=0,width=16.2cm]{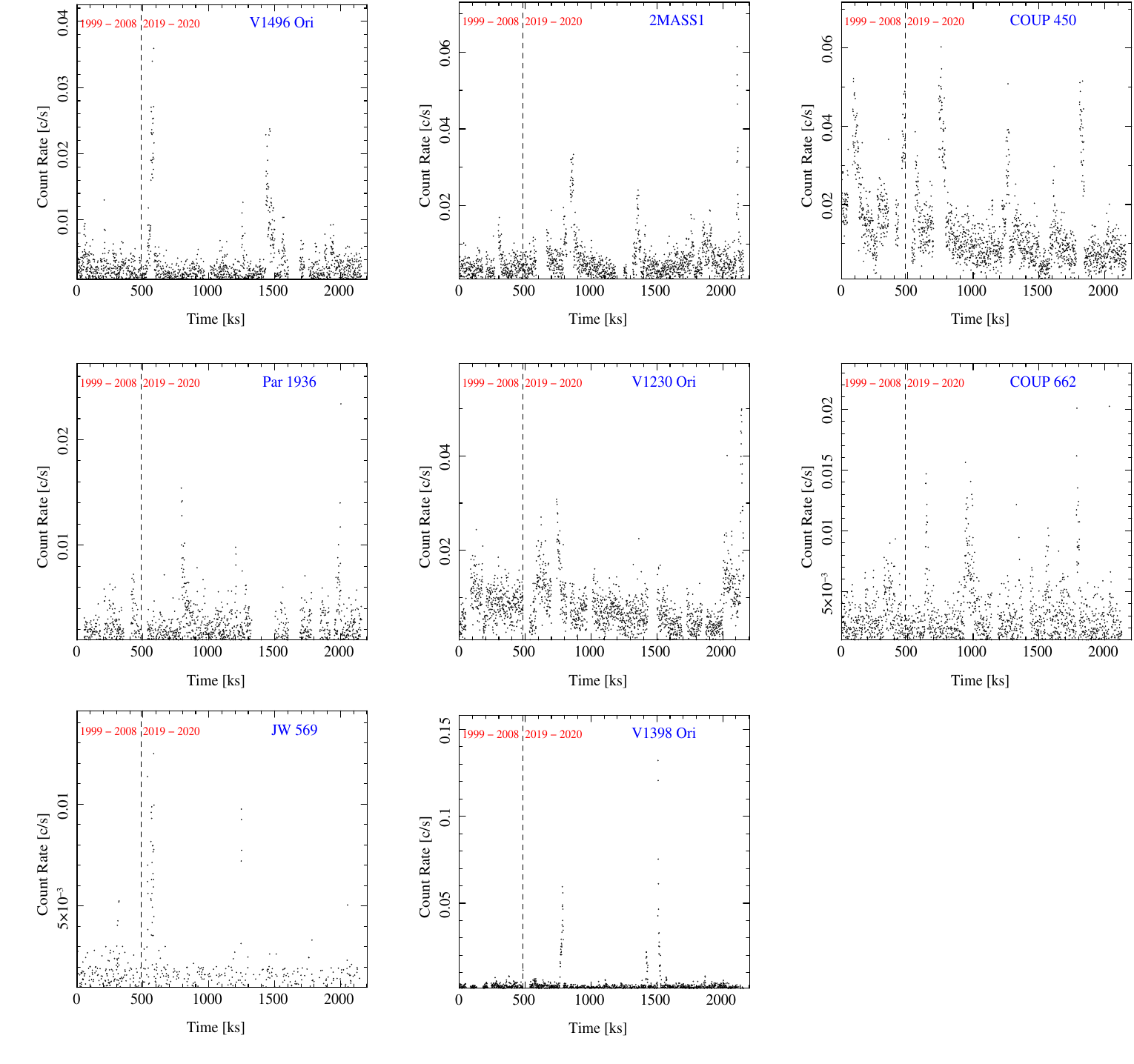}
%\caption{Merged zero order image over the entire exposure using a three color (rgb) scheme reflecting the stars energy spectra. }
\label{fig:joy4}
\end{figure*}

\clearpage
\section{HETG Spectra \label{app:spectra}}

In this appendix we provide plots of all the HETG spectra we extracted so far for this study.
Here we do not include the HETG spectra we already showed in the main part of the paper, i.e. $\theta^1$ Ori C, 
$\theta^1$ Ori E, and COUP 450. Merged HEG and MEG spectra are binned to at minimum of $0.01\mang$. Some low-signal 
spectral regions have coarser binning. We note, this was done for plotting purposes only.
We also added line identification where it deemed appropriate. This was done by visual inspection. At first
all lines that were identified in the APED database are labelled. However in some cases we added identifications
in cases were lines that should have been detected were not there. The first three panels are the other massive 
stars, $\theta^1$ Ori A, $\theta^2$ Ori A. and V1230 Ori. The remaining panels are spectra from extracted 
intermediate and low-mass stars. The HETG 1st order data can be downloaded from the \emph{Chandra} archive
~\url{https://cxc.harvard.edu/cda/contributedsets.html} and \emph{Zenodo}~\url{https://doi.org/10.5281/zenodo.10853416} .

\begin{figure*}[b]
\centering
\includegraphics[angle=0,width=16.2cm]{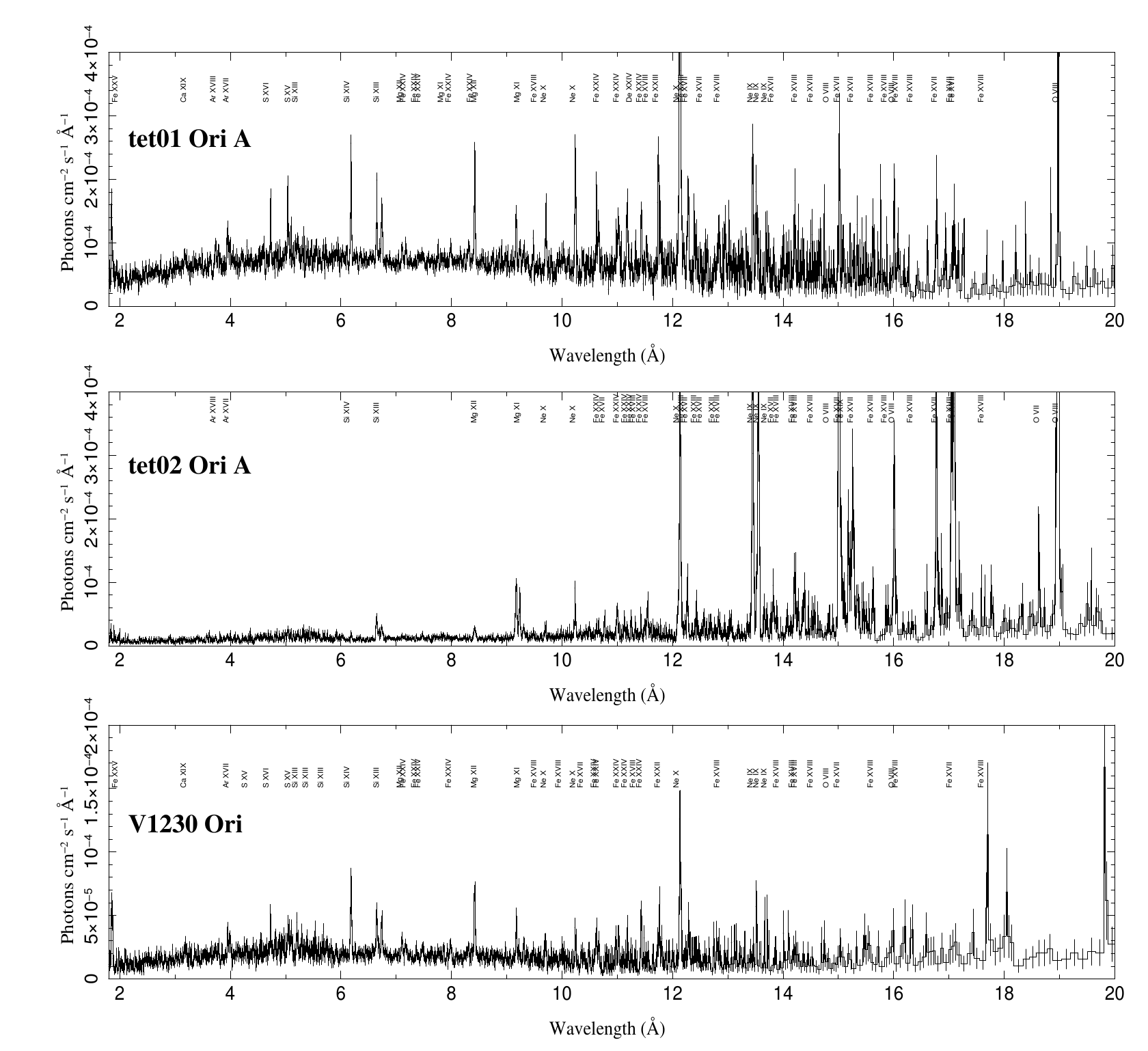}
%\caption{Merged zero order image over the entire exposure using a three color (rgb) scheme reflecting the stars energy spectra. }
\label{fig:massive}
\end{figure*}

\begin{figure*}[t]
\centering
\includegraphics[angle=0,width=16.2cm]{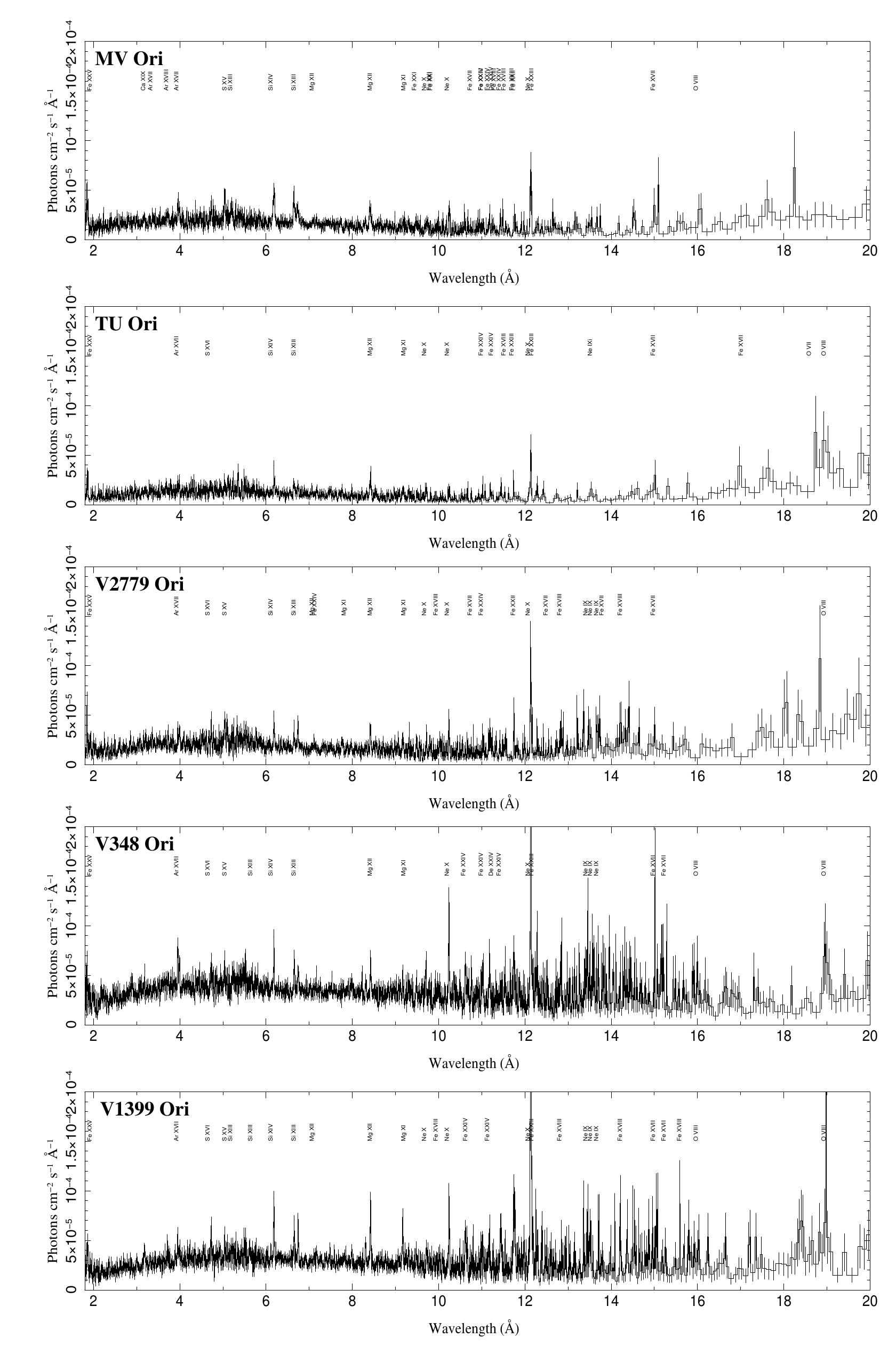}
%\caption{Merged zero order image over the entire exposure using a three color (rgb) scheme reflecting the stars energy spectra. }
\label{fig:aped21}
\end{figure*}

\begin{figure*}[t]
\centering
\includegraphics[angle=0,width=16.2cm]{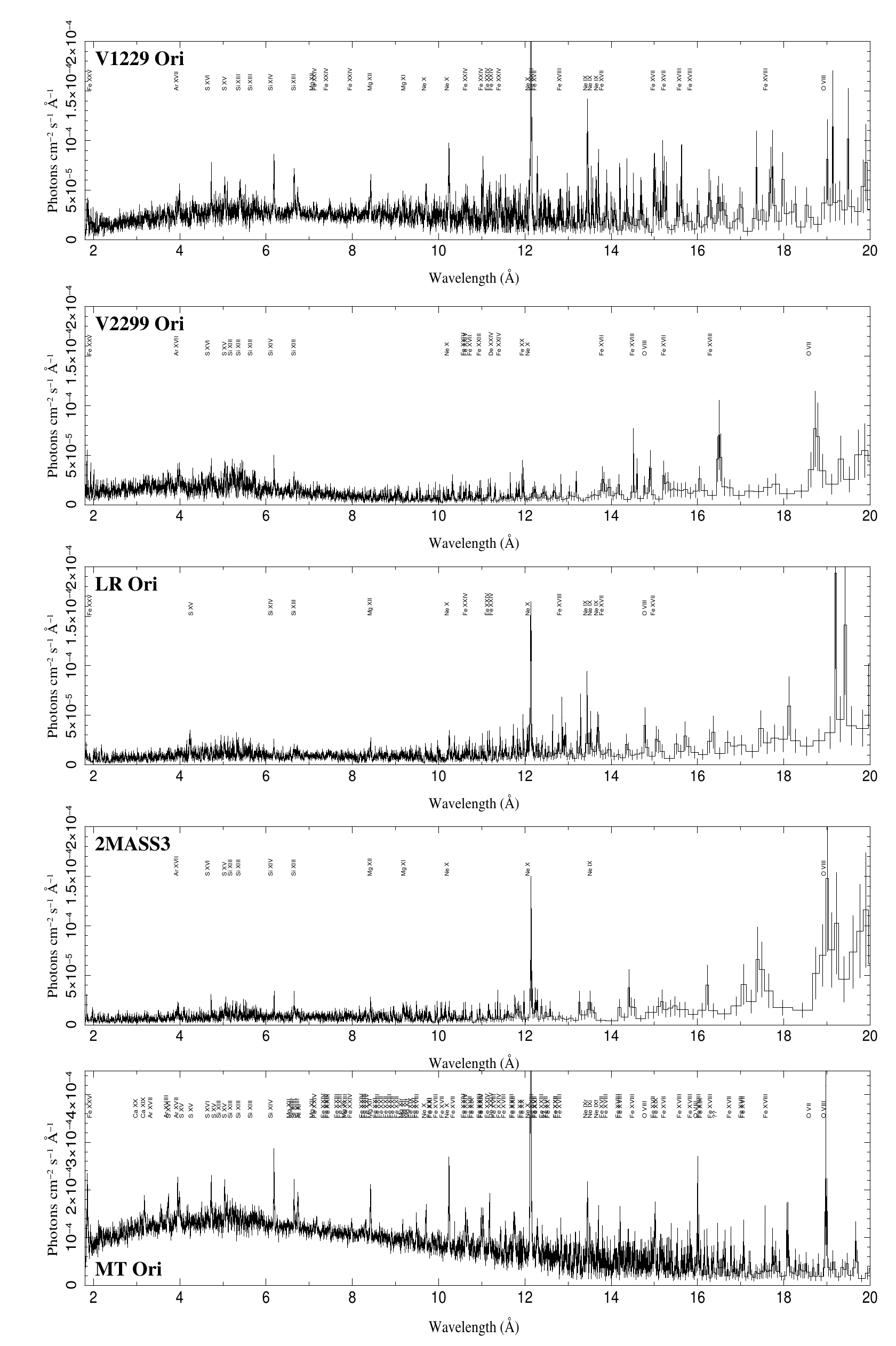}
%\caption{Merged zero order image over the entire exposure using a three color (rgb) scheme reflecting the stars energy spectra. }
\label{fig:aped22}
\end{figure*}

\begin{figure*}[t]
\centering
\includegraphics[angle=0,width=16.2cm]{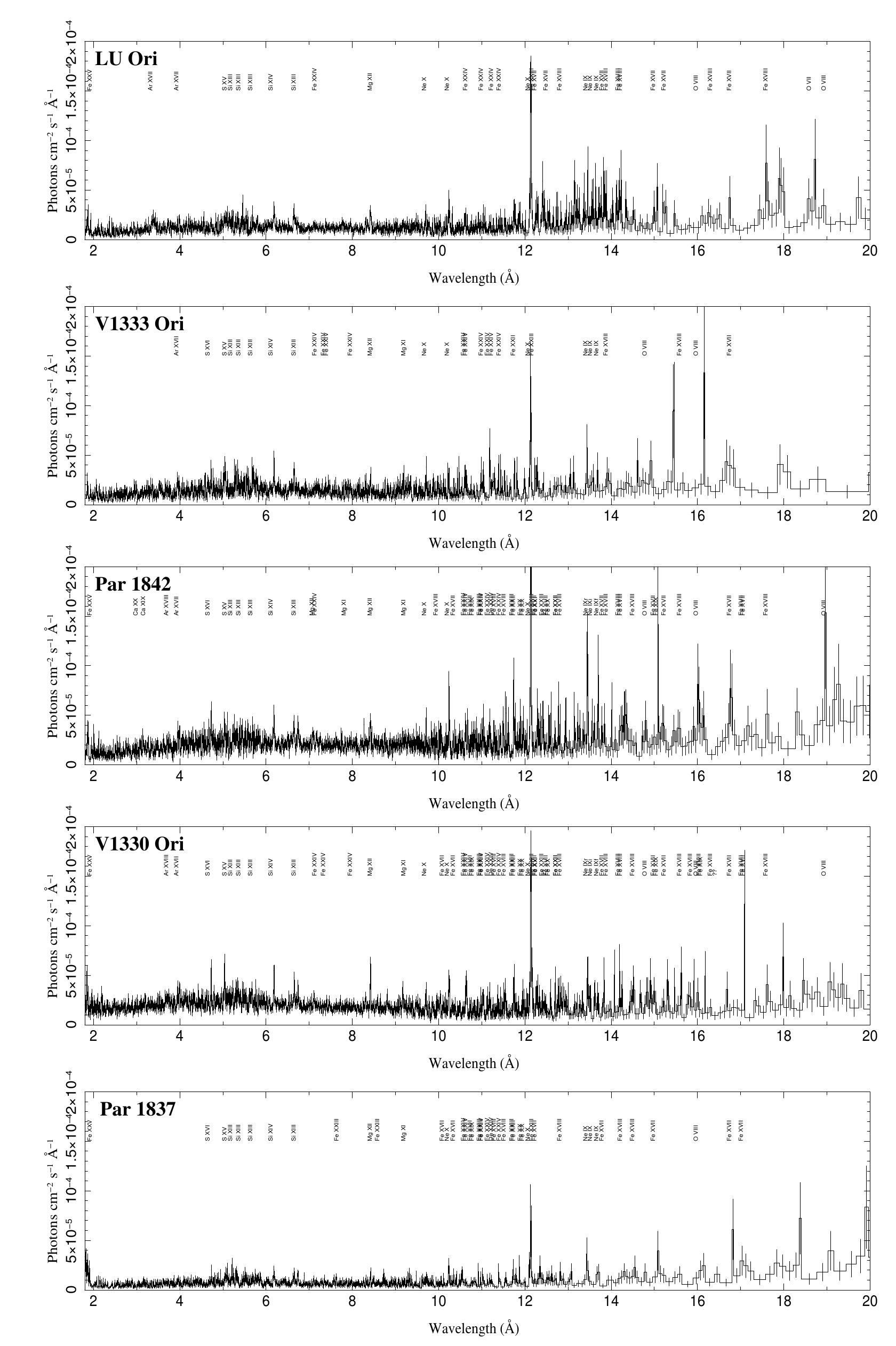}
%\caption{Merged zero order image over the entire exposure using a three color (rgb) scheme reflecting the stars energy spectra. }
\label{fig:aped23}
\end{figure*}

\begin{figure*}[t]
\centering
\includegraphics[angle=0,width=16.2cm]{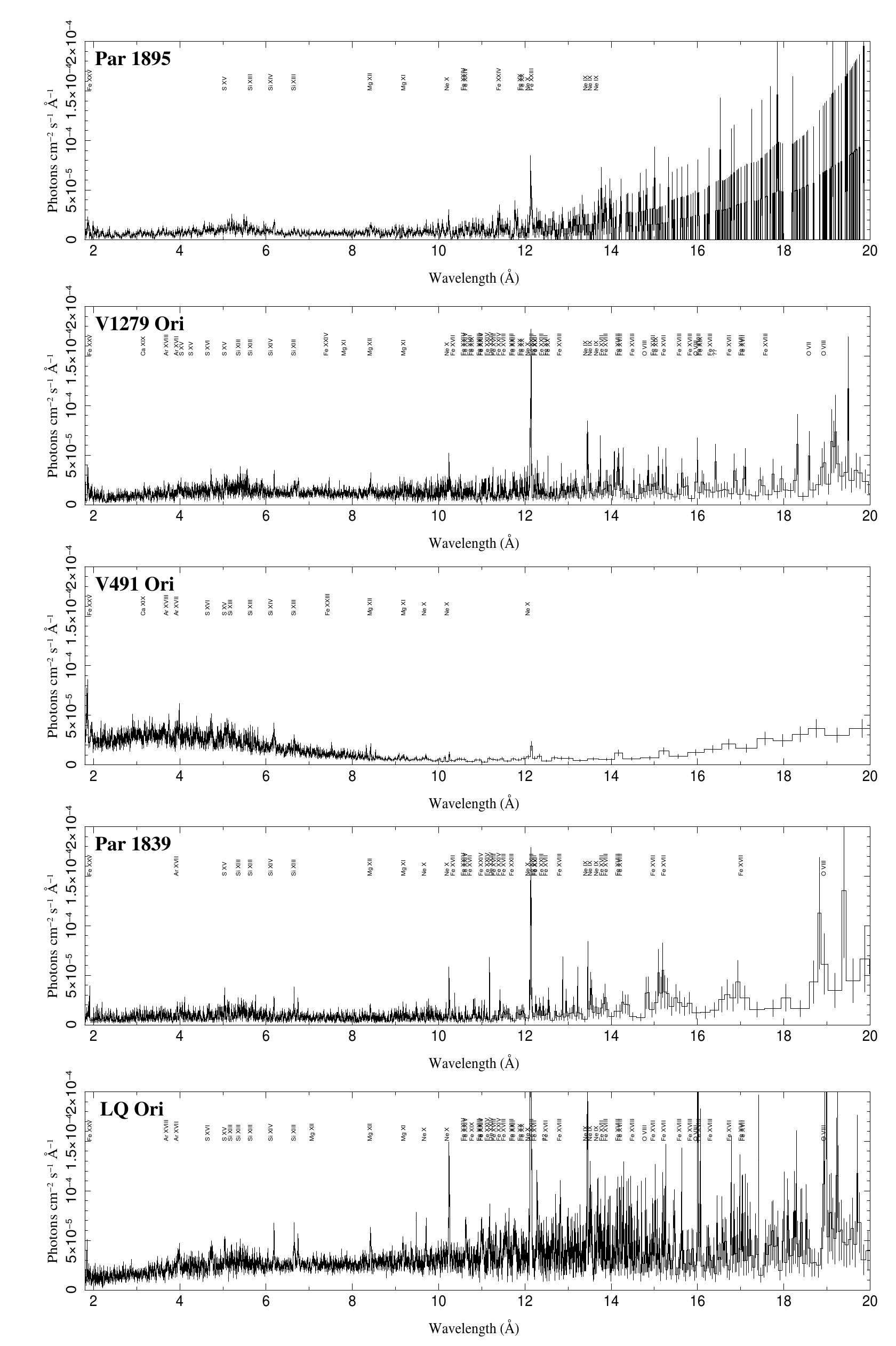}
%\caption{Merged zero order image over the entire exposure using a three color (rgb) scheme reflecting the stars energy spectra. }
\label{fig:aped24}
\end{figure*}

\begin{figure*}[t]
\centering
\includegraphics[angle=0,width=16.2cm]{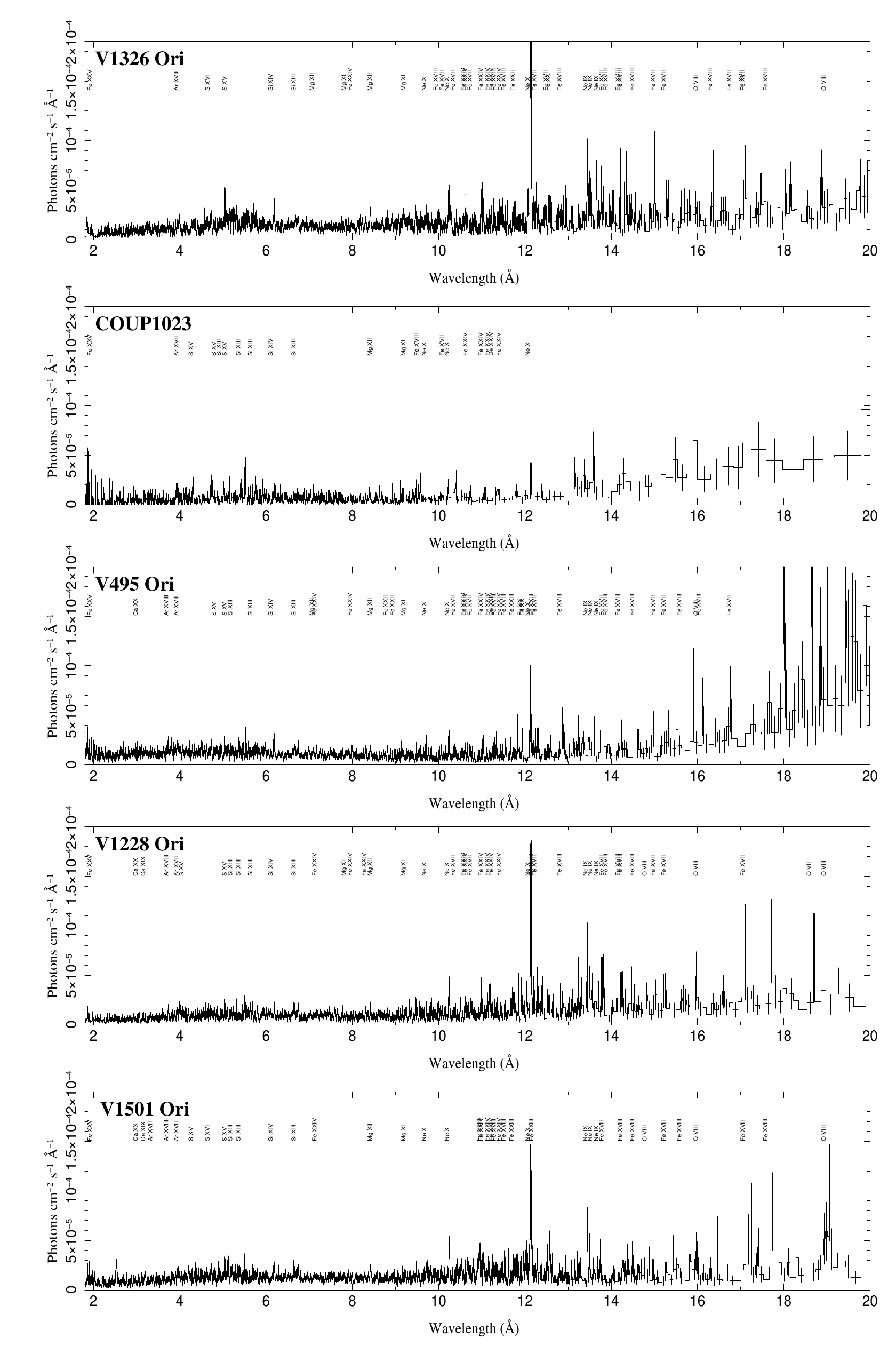}
%\caption{Merged zero order image over the entire exposure using a three color (rgb) scheme reflecting the stars energy spectra. }
\label{fig:aped25}
\end{figure*}

\begin{figure*}[t]
\centering
\includegraphics[angle=0,width=16.2cm]{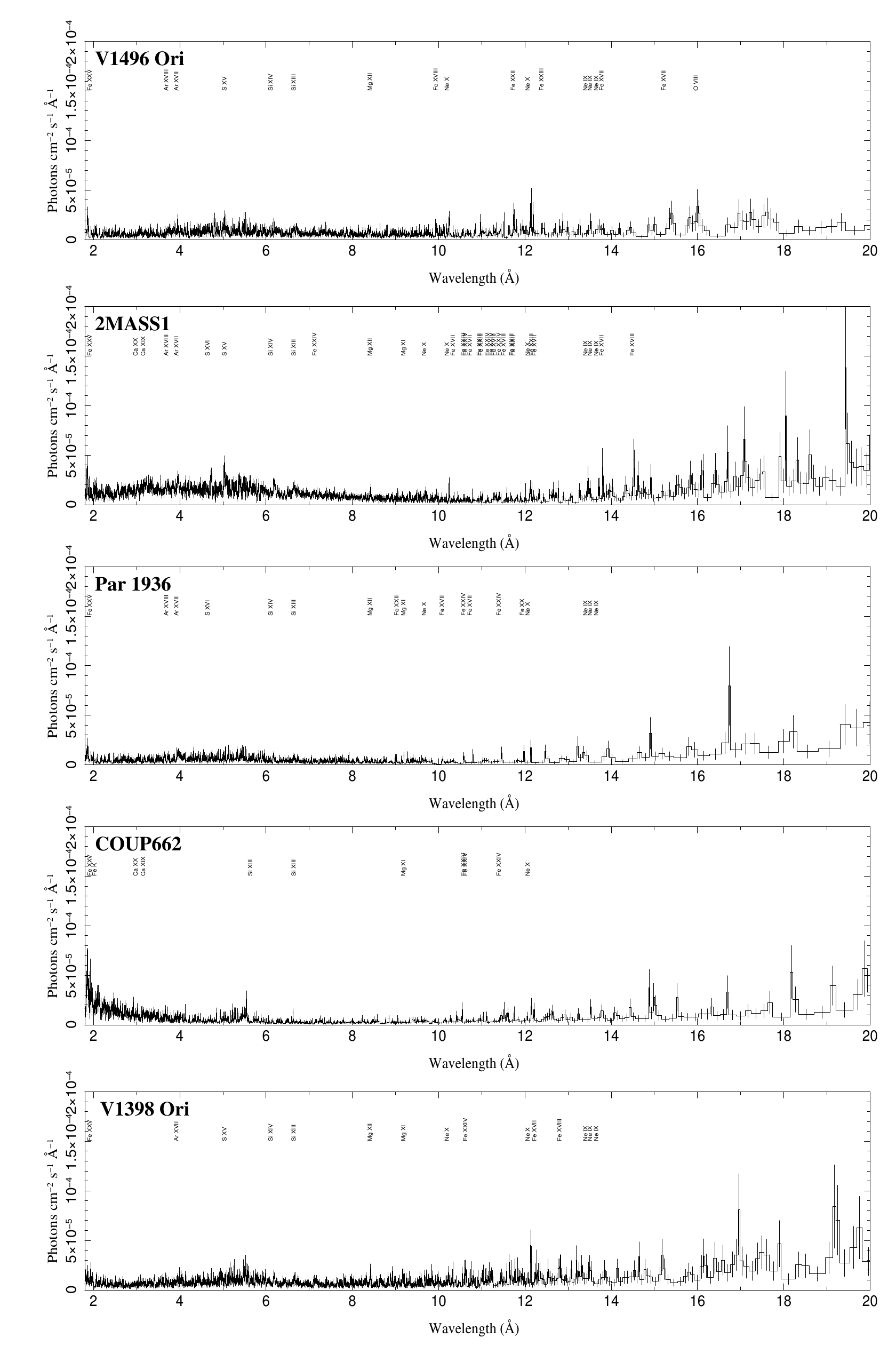}
%\caption{Merged zero order image over the entire exposure using a three color (rgb) scheme reflecting the stars energy spectra. }
\label{fig:aped26}
\end{figure*}

%\end{appendices}

\end{document}